\documentclass[aps,prl,reprint,groupedaddress,showpacs,amsmath,amssymb,twocolumn,floatfix]{revtex4-1}
\usepackage{graphicx}
\usepackage{bm}
\usepackage{color}
\usepackage[normalem]{ulem}
\usepackage{hyperref}
\usepackage{cleveref}
\usepackage{amsmath}

\usepackage{tabularx}

\hypersetup{
    unicode=false,          
    pdftoolbar=true,        
    pdfmenubar=true,        
    pdffitwindow=false,     
    pdfstartview={FitH},    
    pdftitle={Exciton solid},    
    pdfauthor={},     
    pdfsubject={},   
    pdfcreator={},   
    pdfproducer={}, 
    pdfkeywords={} {} {}, 
    pdfnewwindow=true,      
    colorlinks=true,       
    linkcolor=magenta, 
    citecolor=blue,        
    filecolor=magenta,      
    urlcolor=blue           
} 
\usepackage{cleveref}

\newcommand{\nutotal}{$\nu_{total}$ }	
\newcommand{\dnu}{$\Delta \nu$ }

\newcommand{\dlb}{$d/\ell_B$ }
\newcommand{\dle}{$d/\ell_e$ }

\newcommand{\DT}{$\Delta T$ }

\newcommand{\dragxx}{$R_{xx}^{drag}$ }
\newcommand{\dragxy}{$R_{xy}^{drag}$ }

\begin{document}

\title{Observation of a Superfluid-to-insulator Transition of Bilayer Excitons }

\author{Yihang Zeng$^{1}$$^{\ast}$}
\author{Dihao Sun$^{1}$$^{\ast}$}
\author{Naiyuan J. Zhang$^{2}$}
\author{Ron Q. Nguyen$^{2}$}
\author{Qianhui Shi$^{1}$}
\author{A. Okounkova$^{1}$}
\author{K. Watanabe$^{3}$}
\author{T. Taniguchi$^{4}$}
\author{J. Hone$^{5}$}
\author{C.R. Dean$^{1}$$^{\dag}$}
\author{J.I.A. Li$^{6}$$^{\dag}$}

\affiliation{$^{1}$Department of Physics, Columbia University, New York, NY 10027, USA}
\affiliation{$^{2}$Department of Physics, Brown University, Providence, RI 02912, USA}
\affiliation{$^{3}$Research Center for Electronic and Optical Materials, National Institute for Materials Science, 1-1 Namiki, Tsukuba 305-0044, Japan}
\affiliation{$^{4}$Research Center for Materials Nanoarchitectonics, National Institute for Materials Science,  1-1 Namiki, Tsukuba 305-0044, Japan}
\affiliation{$^{5}$Department of Mechanical Engineering, Columbia University, New York, NY 10027, USA}
\affiliation{$^{6}$Department of Physics, University of Texas at Austin, Austin, TX 78712, USA}
\affiliation{$^{\ast}$These authors contributed equally in this work.}
\affiliation{$^{\dag}$ Email: cdean@phys.columbia.edu, jia.li@austin.utexas.edu}

\date{\today}



\maketitle

\textbf{
One of the most spectacular properties associated with Bose-Einstein condensation (BEC) is superfluidity in which the system exhibits zero viscosity and flows without dissipation.  The superfluid phase has been observed in wide ranging Bosonic systems spanning naturally occurring quantum fluids, such as liquid helium, to engineered platforms such as bilayer excitons and cold atom systems~\cite{Allen1938He4,Kapitza1938He4,Anderson1995BEC,Davis1995BEC,Kel.04,Kel.05,Tutuc.04,Wiersma.04,Nandi.12,Eisenstein2014review,Li.17a,Liu.17a,Liu2022crossover,Burg2018strongly,Shi2020,Shi2022}. Theoretical works have proposed that interactions could drive the BEC ground state into another exotic phase that simultaneously exhibits properties of both a crystalline solid and a superfluid - termed a supersolid  ~\cite{Fisher1989BoseMott,Penrose1956supersolid,Andreev1969supersolid,Leggett1970supersolid,Meisel1992supersolid}.  Identifying a material system, however, that hosts the predicted BEC solid phase, driven purely by interactions and without imposing an external lattice potential, has remained elusive~\cite{Tanzi2019AMO,Chomaz2019AMO,Bottcher2019AMO,Norcia2021AMO,Conti2023supersolid}.  
Here we report observation of a superfluid to insulator transition in the layer-imbalanced regime of bilayer magneto-excitons.  Mapping the transport behavior of the bilayer condensate as a function of density and temperature, suggests that the insulating phase is an ordered state of dilute excitons, stabilized by dipole interactions.  The insulator melts into a recovered superfluid upon increasing the temperature, which could indicate that the low temperature solid is also a quantum coherent phase.}

\begin{figure*}
\includegraphics[width=0.8\linewidth]{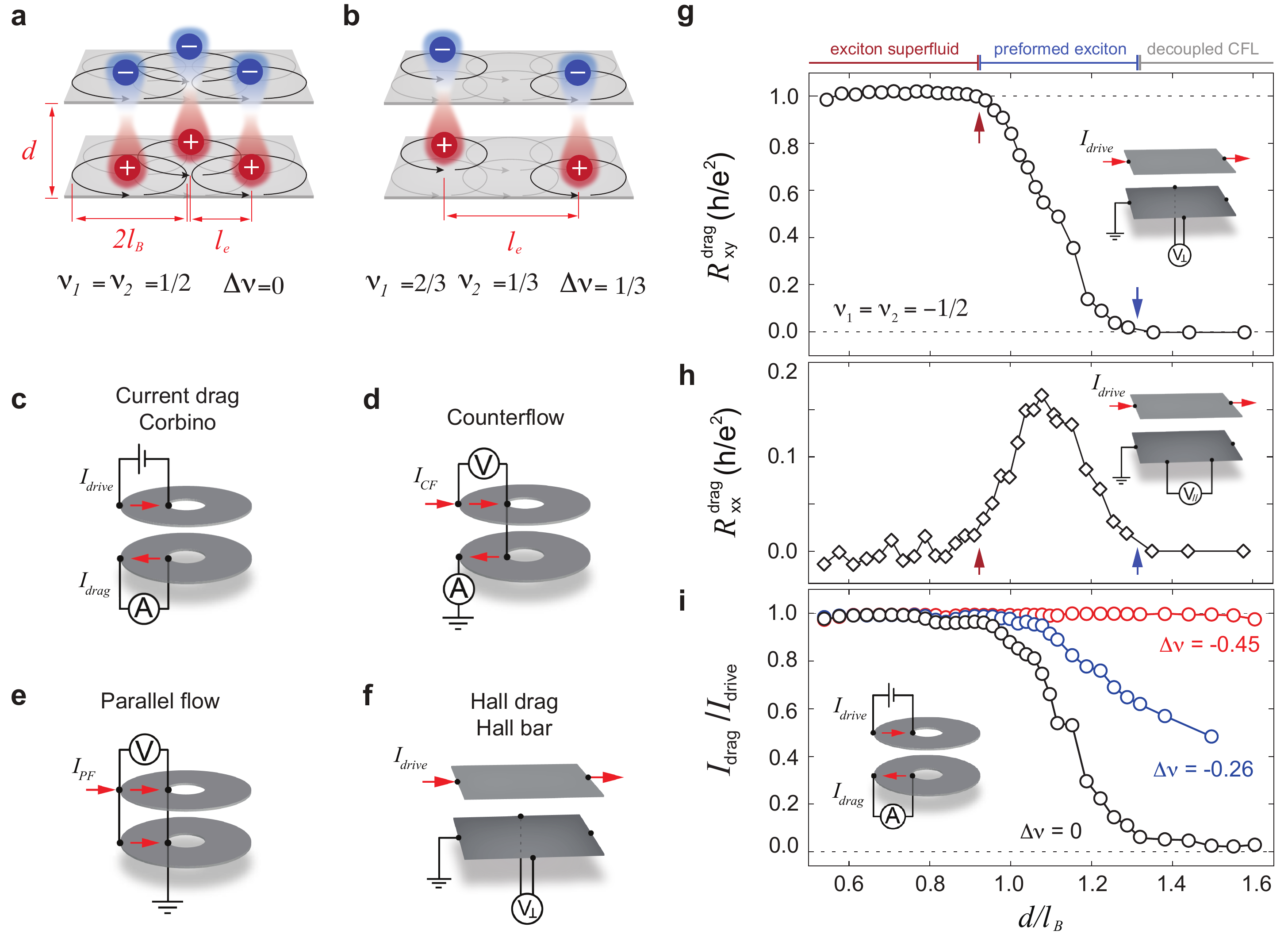}
\caption{\label{fig1} {\bf{Transport signatures of the exciton condensate.}} 
(a-b) Schematics diagram of interlayer excitons in a quantum Hall bilayer system at $\nu_{total}=1$. $\ell_B$ denotes the magnetic length, whereas $\ell_e$ corresponds to the average spacing between excitons. (a) The ratio of $\ell_e/\ell_B$ is minimized under the layer balanced condition $\nu_1 = \nu_2 = 1/2$. (b) The effective exciton density is reduced in the presence of layer imbalance, giving rise to an enhanced $\ell_e/\ell_B$.  (e-f) Schematic diagram for the sample geometry and measurement configuration of various transport methods. (c) Current drag, (d) counterflow, (e) parallel flow measurements are performed in samples with Corbino-shape, whereas (f) Hall drag measurement is performed in samples with a hall bar shape. 
(g) Hall drag, (h) longitudinal drag, and (i) Current drag ratio $I_{drag}/I_{drive}$ as a function of effective interlayer separation \dlb\. Black circles and diamonds denote  measurements under the layer-balanced condition \dnu $=0$, and blue and red circles are measured with an imbalance between the Landau level fillings across two layers. The regimes of exciton superfluid, preformed exciton, and decoupled composite Fermi liquid (CFL) at \dnu $=0$ are marked near the top axis based.
All measurements are measured at $T = 0.3$ K in samples with the interlayer separation of $d = 7.4$ nm.
}
\end{figure*} 

\begin{figure*}
\includegraphics[width=1\linewidth]{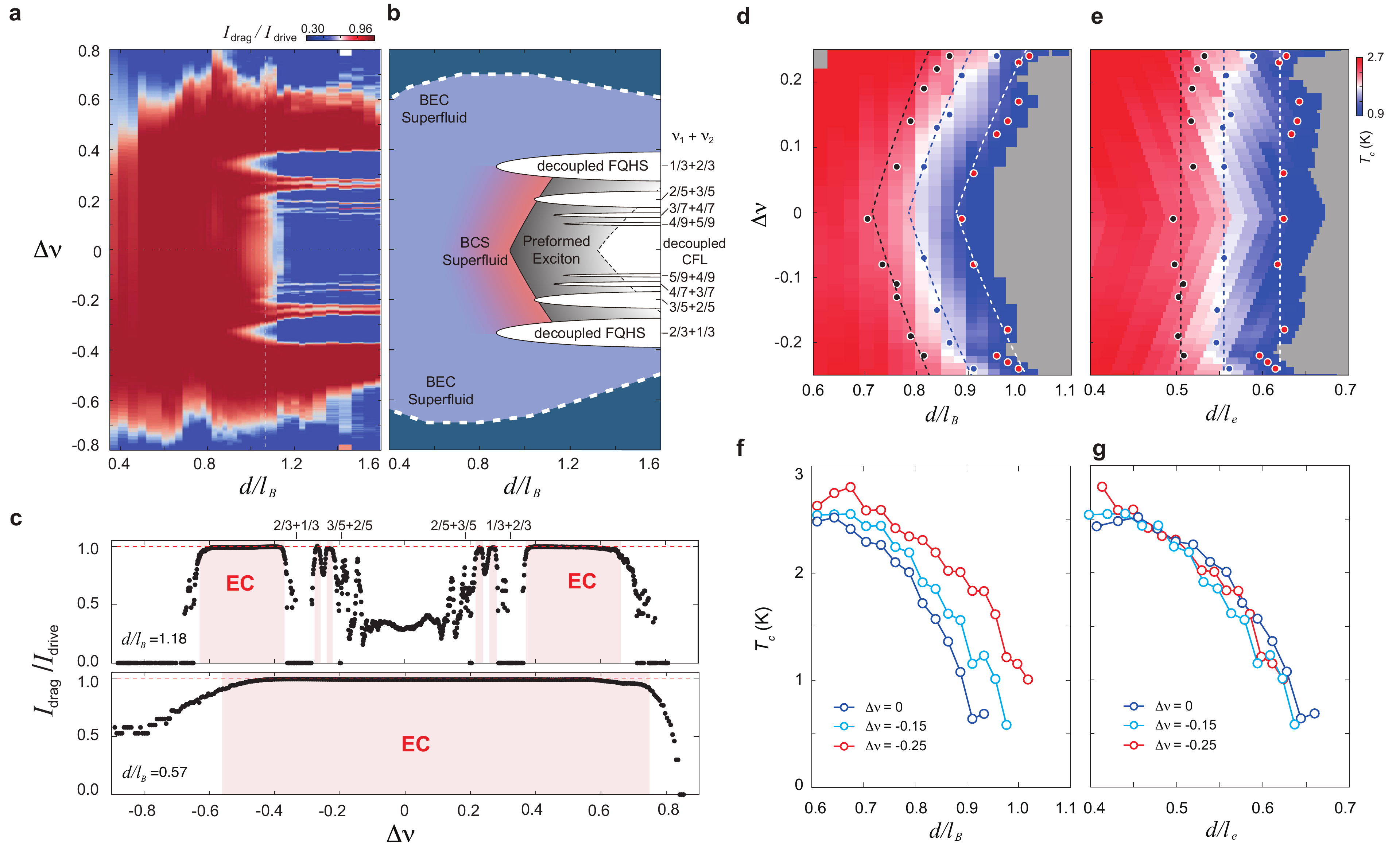}
\caption{\label{fig2} {\bf{Tuning inter-exciton spacing $\ell_e$ with layer imbalance.}} (a) Current drag ratio $I_{drag}/I_{drive}$ as a function of layer-imbalance \dnu\ and effective interlayer separation \dlb. The exciton condensate phase with perfect drag ratio $I_{drag}/I_{drive}=1$ is shown as red in the chosen color scale. The layer-decoupled phases exhibit zero drag ratio, which are shown as blue. (b) Schematic diagram of the \dnu-\dlb\ map. Layer-decoupled fractional quantum Hall states are labeled based on the LL filling of each layer. For instance, $1/3+2/3$ indicates the Laughlin state at $\nu_1=1/3$ and $\nu_2 = 2/3$. Regimes of preformed exciton and decoupled CFL are identified based on drag responses shown in Fig.~1g-i. 
(c) Current drag ratio $I_{drag}/I_{drive}$ as a function of \dnu\ measured at \dlb $=1.18$ (top panel) and \dlb $=0.57$ (bottom panel). Red stripes indicate regions of perfect drag response.  Minima in the upper panel correspond the FQHE sequence labelled in (b). LL filling of each layer is marked near the top axis.  $T_c$ as a function of (d) \dnu\ and \dlb, (e) \dnu\ and \dle.  Constant values of $T_c$ are marked by black, blue and red circles.  $T_c$ as a function of (f) \dlb\ and (g) \dle\ measured at different interlayer density imbalance $\Delta\nu$. $T_{c}$ is operationally defined as the temperature at which Hall drag response equals $98\%$ of the quantized plateau, \dragxy $=0.98h/e^2$ (also see Fig.~\ref{Tc}).  }
\end{figure*} 


Bilayer excitons provide a unique opportunity to study correlated bosonic states in a solid-state system. By spatially confining electrons and holes to closely spaced, but electrically isolated, two-dimensional (2D) layers, a system of long-lived excitons are realized that can undergo
Bose-Einstein condensation to a quantum coherent ground state\cite{Lozovik1975exciton,Pogrebinsky1977exciton,Blatt1962exciton}. Varying the thickness of the 2D material and the thickness of the layer spacer in combination with the field effect gating provides broad experimental control over the relevant energy scales\cite{Li.17a,Liu2022crossover,Liu.17a,Li2019pairing,Liu2019interlayer,Zhang2025exciton}, making the condensate widely tunable. Moreover, a combination of parallel and counterflow measurement geometries allows the transport properties of the condensate to be electrically accessed, despite the exciton having a zero net charge\cite{Li.17a,Liu.17a,Liu2022crossover,Li2019pairing,Liu2019interlayer,Zhang2025exciton}. 

Bilayer exciton condensates have been implemented in a variety of material platforms, including GaAs quantum wells, graphene heterostructures, and transition metal dichalcogendide bilayers ~\cite{Kel.04,Kel.05,Tutuc.04,Wiersma.04,Nandi.12,Eisenstein2014review,Li.17a,Liu.17a,Liu2022crossover,Burg2018strongly,Shi2022,nguyen_perfectdrag2025,Qi_perfectdrag2025}. Within these experimental studies, the condensate ground state has almost exclusively been identified as
a superfluid phase. However, theoretically it has been suggested that long-range dipole-dipole interactions in the bilayer structure yield a rich phase diagram that can include multiple crystalline phases in addition to the superfluid\cite{Chester1970supersolid,Penrose1956supersolid,Andreev1969supersolid,Leggett1970supersolid,Meisel1992supersolid, Vu2023excitonsolid,Hu2024excitonsolid,Chui2020excitonsolid,Yoshioka1990excitonsolid,Joglekar2006excitonsolid,Zarenia2017excitonsolid,De2002excitonsolid,Chen1991excitonsolid,Yang2001excitonsolid,Conti2023supersolid,Boning2011supersolid,Szymanski1994wigner,Astrakharchik2007dipole}.  Particularly intriguing is the prediction that an interaction driven exciton supersolid- an exotic phase of matter that simultaneously exhibits properties of both a solid and superfluid\cite{Leggett1970supersolid,Chester1970supersolid,Andreev1969supersolid,Meisel1992supersolid,Penrose1956supersolid,Chen1991excitonsolid,Conti2023supersolid,Joglekar2006excitonsolid,Vu2023excitonsolid} – may be stabilized within experimentally realizable parameters.  To date, no experimental evidence of a transition from the exciton superfluid to an exciton solid has been reported, and the question of whether the associated supersolid phase can be realized, or even how it may be identified, remains an open experimental challenge. 


Here we address this challenge by studying graphene double layers in the Quantum Hall regime at $\nu=1$ total filling fraction.  We demonstrate the first ever observation of an exciton superfluid-to-insulator phase transition, which we induce by tuning the exciton density at fixed temperature or  by varying the temperature at fixed density.  By mapping the phase boundary in the density-temperature space we conclude that the insulating state is likely to be ordered and coherent, and therefore a possible exciton supersolid.

\hspace{\linewidth}

\noindent\textbf{Quantum Hall bilayers}\\
In symmetrically doped quantum Hall bilayers, electrons in a partially filled Landau level (LL) in one layer couple with  empty states in the other layer, forming interalyer excitons (Fig. 1a). The interlayer attraction is inversely proportional to the layer spacing, $E_{inter} \sim 1/d$, whereas the intralayer repulsion is characterized by the magnetic length, $E_{intra} \sim 1/\ell_{B}$, with $\ell_{B}= \sqrt{\hbar/eB}$.  The  condensate regime is defined by the ratio of these two competing energy scales,  $E_{intra}/E_{inter} \propto d/\ell_B$. A key advantage of magneto-excitons is that the dimensionless parameter $d/l_{B}$ is tunable with magnetic field, providing a way to adjust the exciton pairing strength in a device with fixed layer separation.  Studies of quantum Hall bilayers have confirmed that the magneto-exciton condensate onsets for $d/l_{B}\lesssim1$~\cite{Kel.04,Kel.05,Tutuc.04,Wiersma.04,Nandi.12,Eisenstein2014review,Li.17a,Liu.17a,Liu2022crossover,Burg2018strongly,Shi2020,Shi2022,Liu2022crossover}, with a crossover from weak-pairing to strong-pairing superfluid transition observed as $d/l_{B}$ is tuned from 1 to 0~\cite{Liu2022crossover}. 

Calculations for zero-field bilayer excitons indicate that the condensate can theoretically host both fluid and solid phases.  The phase boundary is determined by an interplay between the $E_{intera}/E_{inter}$ ratio and the so called interaction paramater, $r_{s}=E_{intra}/E_{kinetic}$, defined as the ratio between the exciton repulsion and available kinetic energy\cite{Joglekar2006excitonsolid,Conti2023supersolid}. In the quantum Hall regime $r_{s}$ is ill-defined, since the kinetic energy is quenched within a Landau level. One might consider that  $r_s\rightarrow\infty$ and therefore the exciton condensate is tunable exclusively by $d/l_{b}$. The fact that no evidence of an exciton solid has been found in quantum Hall bilayers 
suggests then that either the solid phase is inaccessible for magnetoexcitons, or alternatively that $r_{s}$ is not the relevant experimental parameter. 

In this work, we identify a new dimensionless ratio, $\ell_{e}/\ell_{B}$, where $\ell_{e}$ is the average separation between the excitons, and $\ell_{B}$ defines the effective exciton radius (i.e. the magnetoexciton Bohr radius). 
The inter-particle separation within a layer relates to the carrier density, $n$, and can be rewritten in terms of the LL filling fraction according to $n_{i}=1/\pi\ell_{e}^{2}=\nu_{i}/\pi\ell_{B}^{2}$, where $\nu_{i}$ is filling fraction and $i$ is the layer index. If we define the layer imbalance as $\Delta \nu=\nu_{2}-\nu_{1}$, and consider that the total filling fraction under imbalance remains $\nu_{1}+\nu_{2}=1$, then we can write $\ell_{e}/\ell_{B}=\sqrt{2/(1-|\Delta\nu|)}$. When the two layers are balanced with both at half filling  $\ell_{e}/\ell_{B}=0.7$, i.e the excitons overlap (Fig. 1a).  As layer imbalance increases the excitons become more separated and the $\ell_e/\ell_B$ ratio increases.  We observe a superfluid-to-insulator transition when $\ell_e/\ell_B\sim2$.  We interpret this critical value to represent a transition from a regime at layer balance where the diopoles overlap and repulsion is dominated by intraparticle Coulomb interaction (fluid regime) to the regime when the bilayer excitons become well separated and diople-diople interactions become the dominate repulsive term, stabalizing a solid phase\cite{Vu2023excitonsolid,Hu2024excitonsolid,Joglekar2006excitonsolid,Chen1991excitonsolid,Yang2001excitonsolid,Conti2023supersolid,Boning2011supersolid,Szymanski1994wigner}.

\noindent\textbf{Exciton superfluid}\\
We study transport response in quantum Hall bilayers consisting of two graphene monolayers separated by a multi-layer hBN spacer ~\cite{Li2019pairing,Liu2019interlayer,Liu2022crossover}. Signatures of the condensate are measured using a combination of Hall bar and Corbino geometries (Fig. 1c-f). All devices are made with individual layer contact, allowing us to interrogate the transport response using standard drag, counterflow and parallel flow configurations~\cite{Eisenstein2014review}.  Fig. 1g-i summarizes key characteristics of how these transport signatures evolve as a function of \dlb.  First we consider the layer balanced condition, $\nu_{1}=\nu_{2}=1/2$ (black circles). For large \dlb\ the layers are uncorrelated and form decoupled composite Fermi liquid (CFL) 
. In this regime the Hall drag, $R_{xy}^{drag}$, longitudinal drag, $R_{xx}^{drag}$, and current drag, $I_{drag}/I_{drive}$, are all zero valued.  Below \dlb$\sim$1  there is a marked difference with $R_{xy}^{drag}$ becoming fully quantized to value $h/e^2$ and the current drag becoming unity valued. The longitudinal drag is zero valued both in the large and small \dlb limits, however, a finite value appears at the same critical value as in the hall drag and current drag measurements, providing an additional signature of the transition ~\cite{Liu2022crossover}. The quantized Hall drag and perfect drag current are both signatures of interlayer coherence, while zero longitudinal drag in this coherent flow regime indicates dissipationless transport ~\cite{Eisenstein2014review}.  Taken together the simultaneous observations of quantized Hall drag, perfect drag current, and zero longitudinal resistance therefore establish the existence of an interlayer correlated phase with dissipationless flow, i.e. the exciton superfluid.

A characteristic signature of the magnetoexciton superfluid condensate is that it remains robust under layer imbalance so long as $\nu_{total}=1$ is maintained. Indeed several studies found evidence it may become more robust with layer imbalance though without full understanding of this behavior ~\cite{Champagne.08b,Li.17a, Clarke2005imbalance, Zhang2025exciton,Jog.02}.  Fig. 1i shows that with increasing layer imbalance (blue and red circles) the exciton superfluid persists to larger values of \dlb, consistent with previous measurements.

In Fig.~\ref{fig2}a we plot a full two-dimensional colour plot of the current drag response as a function of \dlb\ and $\Delta\nu$, measured at $T= 0.3$ K. This is, to our knowledge, the first complete map of the interplay between layer separation (\dlb) and layer imbalance ($\Delta\nu$) on the condensate. For \dlb$<1$, a large portion of the phase space is occupied by the condensate phase, evidenced by the perfect current drag (shown as red in the chosen color scale). For \dlb$>1$, the condensate disappears at \dnu$=0$ (diminished current drag - blue color), but reappears under layer imbalance. We note that it is periodically interrupted by a sequence of zero drag regions (blue horizontal stripes) that correspond in filling fraction to the expected single layer FQHE states. Fig.~\ref{fig2}c shows vertical line cuts from Fig. 2a, corresponding to current drag versus $\Delta\nu$ at \dlb=1.18 and 0.57. 
Hall bar measurements produce  a similar response, and allow us to unambiguously identify the FQHE states(see SI Fig.~\ref{SIdragmap}).  The coexistence and competition between the exciton condensate and  FQHE states unlocks a new type of quantum critical point, which merits future study ~\cite{Clarke2005imbalance,Champagne.08b,Zhang2025exciton,Nguyen2024fractionalexciton,Jog.02}. 
A summary of the base-temperature condensate phase diagram versus \dnu and \dlb is showin in Fig. 2b.  The regions labelled as preformed excitons are associated with a current drag value that is non-zero but less than unity, and the BCS to BEC transition reflects the crossover studied in previous literature ~\cite{Liu2022crossover}.

In addition to layer spacing and layer imbalance, the third way to affect the stability of the condensate is by varying the temperature.  Fig.~\ref{fig2}d plots the critical temperature of the condensate transition, $T_{c}$, versus \dnu\ and \dlb, where $T_{c}$ is defined as the temperature where drag ratio deviates from unity. We restrict to  $|\Delta\nu|<0.2$ to avoid the competition with the single-layer FQHE states.
Fig. 2f shows representative horrizontal linecuts corresponding to $T_{c}$ versus \dlb at $\Delta\nu=0$, -0.15 and -0.25.  For a given value of \dlb, $T_c$ is systematically enhanced by increasing $\Delta\nu$.  A similar trend is observed in the superfluid transition width, \DT\ (see SI Fig.~\ref{FigPairing}). 

The observation that $T_c$ and \DT\ vary with $\Delta\nu$, for fixed \dlb , suggests that \dlb\ alone does not fully describe the ground state of the bosonic system.  Fig.~\ref{fig2}e,g replots the same $T_{c}$, but as a function of \dnu\ and \dle, where $\ell_{e}$ is calculated from the exciton density using the definition introduced above. The contours of constant $T_{c}$ now follow vertical trajectories (Fig. 2e) or equivalently the linecuts in Fig. 2f are seen to collapse onto a single curve when plotted as a function of \dle\ (Fig. 2g) .  This confirms that  $\ell_e$, and therefore the exciton density,  plays a critical role in determining the temperature stability of the condensate phase ~\cite{Lozovik2012,Perali2013}. This provides a physical picture that links the magneto-exciton phase diagram with the previously studied zero-field bilayer excitons. 

\begin{figure*}
\includegraphics[width=0.73\linewidth]{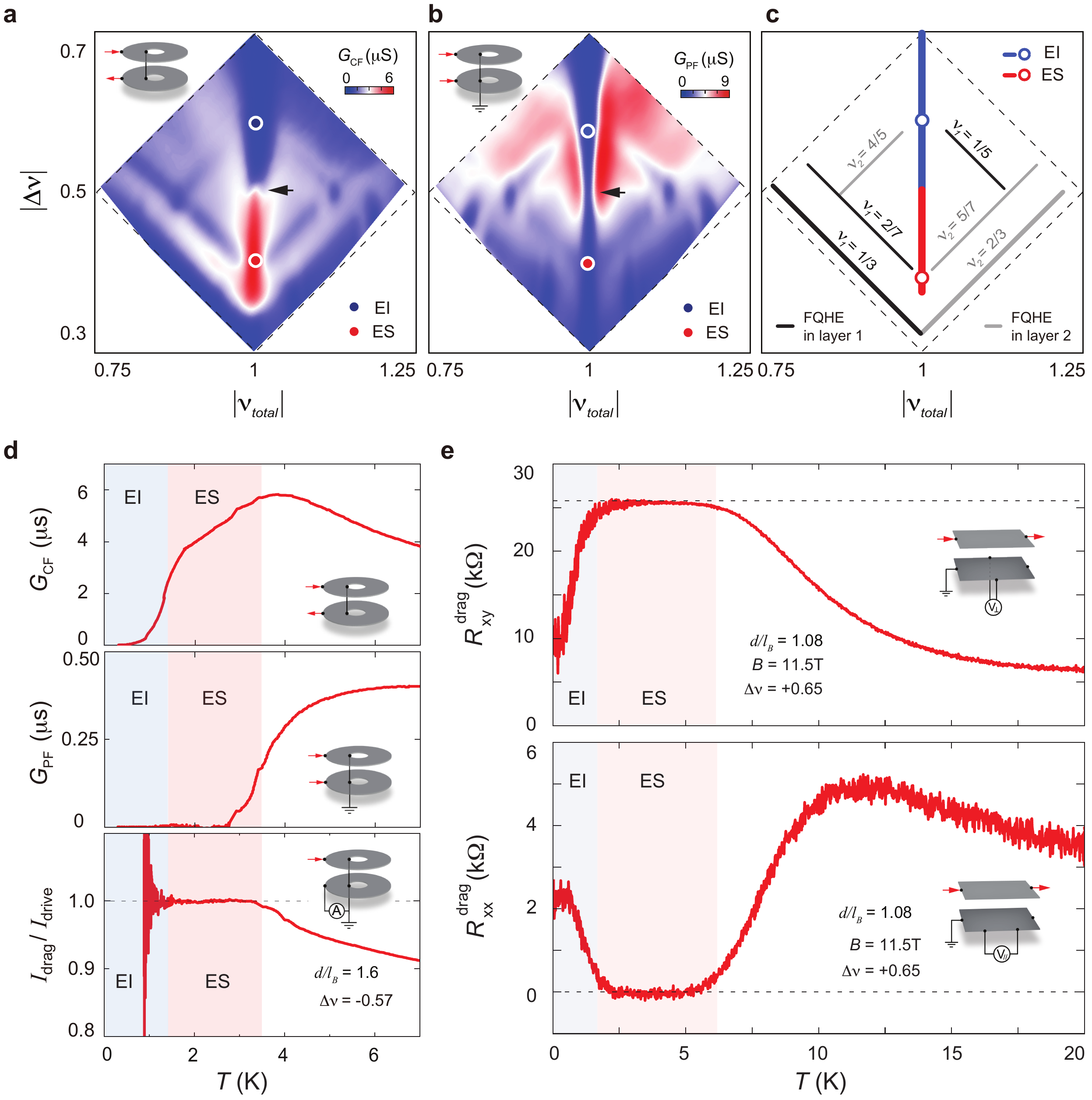}
\caption{\label{fig3}{\bf{An exciton insulator at large $\Delta\nu$.}}  (a) Counterflow conductance $G_{CF}$  and (b) parallel flow conductance $G_{PF}$  as a function of total Landau level filling $\nu_{total}$ and layer imbalance $\Delta\nu$. The measurement is performed near the transition boundary near \dnu $=0.5$ and $\nu_{total} = 1$, at \dlb\ $=1.6$. (c) Schematic diagram labeling the most prominent features in panels (a) and (b). The location of the exciton condensate and insulator in the $\nu_{total}$-\dnu\ map are marked by red and blue vertical lines. 
(d) Counterflow conductance $G_{CF}$ (top panel), parallel flow conductance $G_{PF}$ (middle panel) and current drag ratio $I_{drag}/I_{drive}$ (bottom panel) as a function of $T$ measured at \dnu $=-0.57$ and $\nu_{total}=1$. The measurement is performed at \dlb$=1.6$ in a Corbino-shaped sample. (e) Hall drag (top panel) and longitudinal drag (bottom panel) as a function $T$ measured at \dnu $=+0.65$ and $\nu_{total}=1$. The measurement is performed at \dlb$=1.08$ in a Hall-bar shaped sample. 
}
\end{figure*}

\begin{figure*}
\includegraphics[width=1\linewidth]{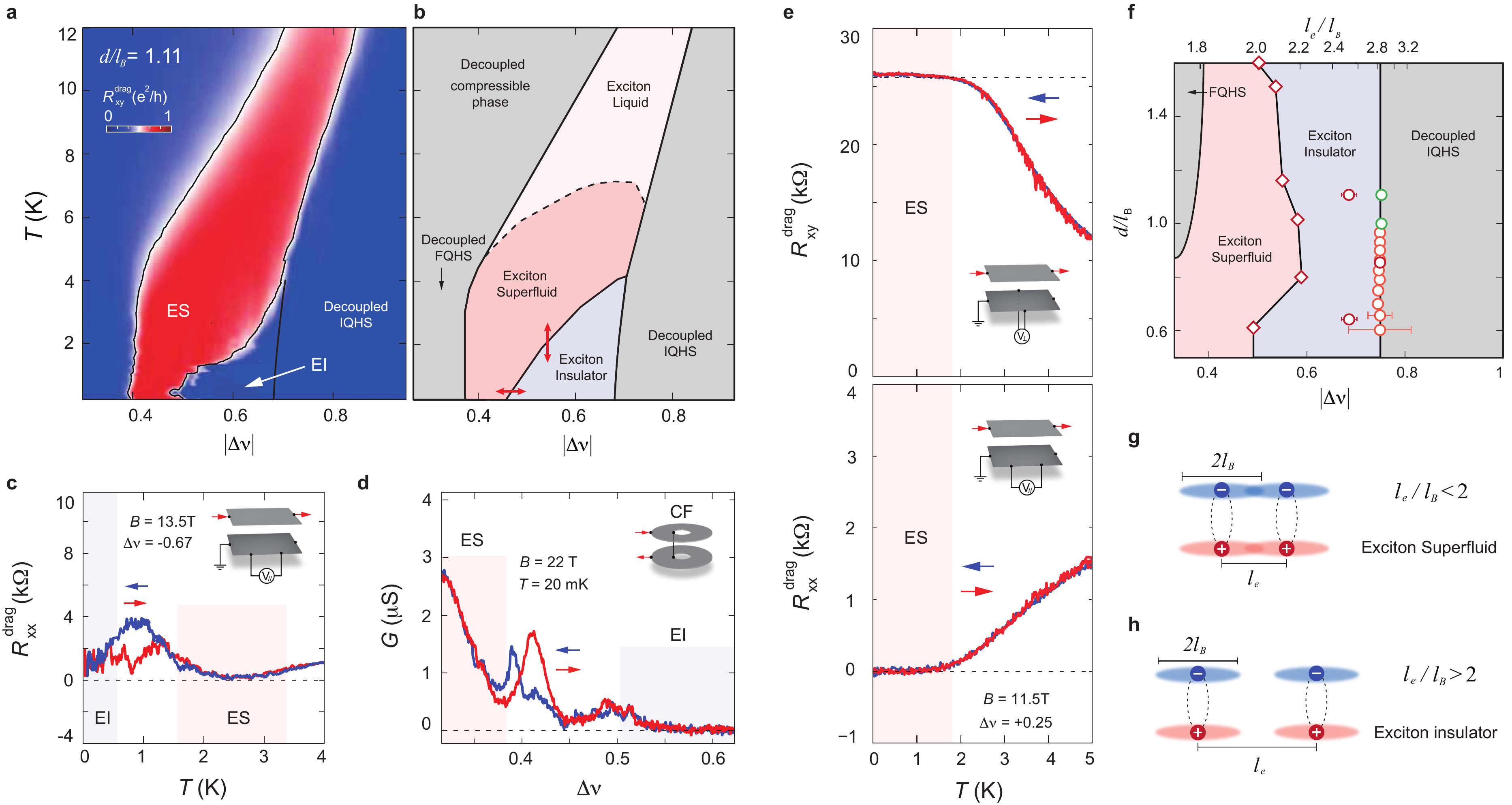}
\caption{\label{fig4}{\bf{A first-order re-entrant transition between the exciton solid and superfluid phases.}}  (a) Hall drag response as a function of temperature $T$ and layer imbalance \dnu\ measured at $\nu_{total} = 1$ and \dlb $=1.11$. (b) Schematic diagram of the the \dnu$-T$ map marking different exciton phases. The regime occupied by the exciton superfluid is defined by quantized Hall drag, concomitant with vanishing longitudinal drag (also see SI Fig.~\ref{dnuT}).  (c) $R_{xx}^{drag}$ measured across the reentrant transition with increasing and decreasing temperature. (d) $G_{CF}$ measured across the low-temperature phase boundary between the superfluid and insulator with increasing and decreasing $\Delta\nu$. (e) Hall drag (top panel) and longitudinal drag (bottom panel) as a function $T$ measured at $\nu_{total}=1$ and \dnu $=-0.2$, where the exciton superfluid is the low-temperature ground state.   (f) Schematic phase diagram of the \dnu$-$\dlb\ map marking the transition boundaries between different phases at $\nu_{total} = 1$.  (g-h) The exciton superfluid-solid transition with the scaling argument. According to the scaling argument, an exciton solid phase is expected to be stabilized at $\ell_e/\ell_B > 2$, whereas a superfluid phase is expected at $\ell_e/\ell_B < 2$.  }
\end{figure*}

\hspace{\linewidth}

\noindent\textbf{Superfluid to insulator transition}\\
Next we investigate the regime of large layer imbalance corresponding to filling fractions beyond the boundary marked by the dashed white line in Fig. 2b.  In this range the interayer exciton enters a low density dilute regime, corresponding to a near total depletion of one layer and near full ocuppancy of the other.  Current drag measurement indicates that in this limit the superfluid phase is no longer observed for any experimental range of \dlb (Fig. 2a,c). 

In Fig.~\ref{fig3}, we examine the transition from the superfluid to this new layer-imbalanced phase at a fixed value of \dlb=1.6.  Figs. 3a and 3b plot the measured parallel flow and counterflow conductance, respectively, versus total filling fraction, $\nu_{total}$ (horizontal axis) and layer imbalance, $\Delta\nu$ (vertical axis). The data was acquired from the same Corbino device as in Fig. 2.  The exciton condensate follows a vertical trajectory through the center of these maps following the total filling fraction $|\nu_{total}|=1$ ~\cite{Eis.04,Eisenstein2014review,Li.17a,Liu.17a,Li2019pairing,Shi2022}. At approximately $\Delta\nu=0.5$ a sharp transition in the transport behaviour is seen.  For $\Delta\nu<0.5$, the device shows high conductance in counterflow (red in Fig. 3a) and low conductance in parallel flow (blue in Fig 3b), consistent with the expected response for superfluid bilayer excitons. For $\Delta\nu>0.5$, the device instead shows a vanishing conductance in both counterflow and parallel flow, indicating a transition to an insulating phase. 
 As summarized in the diagram shown in Fig. 1c, the insulating phase continues to track the $\nu_{total}=1$ trajectory with increasing layer imbalance,  and is not obviously connected to any of the nearby single layer or double layer FQHE states (see also Fig.~\ref{Figlocalization}).  This indicates that, like the exciton sperfluid, the insulating phase associates with interlayer excitons and so we label as an exciton insulator (EI).

The unique nature of the exciton insulator is highlighted by an intriguing reentrant-type transition observed with increasing temperature. Fig.~\ref{fig3}d plots the temperature dependence of the counterflow conductance, parallel flow conductance, and current drag ratio, measured using the same Corbino-shaped sample as in Fig.~\ref{fig3}a-b. The low temperature insulator (labelled EI) is defined by vanishing parallel and counterflow conductance.  The current drag ratio is ill-defined due to the vanishingly small drive current. The insulating phase persists from $T=20~mK$ up to $T \sim 1.5$ K. Above 1.5~K we observe a transition to a phase characterized by non-zero counterflow conductance, zero parallel conductance, and perfect counterflow drag ratio all of which indicate a transition 
to a recovered superfluid phase\cite{Nandi.12}. The superfluid (labelled ES) persists through the intermediate temperature range  $1.5 < T < 4$ K, above which the system behaves as decoupled layers.

Similar measurement on a separate Hall bar device is shown in Fig. 3e. The low temperature region (shaded blue background) exhibits diminishing Hall drag and enhanced longitudinal drag response, consistent with an incipient but not fully formed insulating phase. The  intermediate temperature range, (red-shaded background) exhibits a quantized Hall drag plateau concomitant with zero longitudinal drag, again consistent with a recovered exciton superfluid. 

The appearance of an exciton superfluid at elevated temperature confirms that the layers remain strongly coupled in this regime of large layer imbalance.  This in turn provides further evidence that the low temperature insulator should represent an interlayer excitonic phase since there is no reason to expect exciton pair breaking upon \textit{lowering} the temperature.

\noindent\textbf{Nature of the exciton insulator}\\
Fig.~\ref{fig4}a plots the Hall drag response across the $\Delta \nu-T$ phase space measured at $|\nu_{total}| = 1$. In the chosen color scale, red denotes quantized Hall drag response with a plateau value of $e^{2}/h$, and blue corresponds to  diminishing Hall drag response. The quantized Hall response identifies the exciton superfluid, whereas diminishing Hall drag is observed for the exciton insulator, as well as layer-decoupled phases. The black line that separates the low temperature EI from the decoupled IQHE phases is identified from a peak that is observed in both the longitudinal drag response (Fig.~\ref{dnuT}) and the bulk counterflow conductance (Fig.~\ref{FigHysteresis}).

Fig. 4b shows a schematic representation of the phase diagram measured in Fig. 4a.  The labels are based on combined Hall bar and Corbino measurements of counterflow conductance (Fig.~\ref{fig3}d and Fig.~\ref{SIDT2}) and Hall drag response (Fig.~\ref{fig3}e, Fig.~\ref{fig4}a, and Fig.~\ref{dnuT}).  According to the $\Delta\nu-T$ map, the exciton insulator always warms up into a superfluid phase. This establishes the reentrant behavior as a hallmark signature of the exciton insulator.  
Fig. 4c shows that the temperature driven transition between the insulator and superfluid phases (vetical red arrow in Fig. 4b) is hysteretic, consistent with a first order transition. Notably, the temperature driven transition between the superfluid and the normal fluid, which is expected to be a second order phase transition, shows no such hysteresis (Fig.~\ref{fig4}e). Additionally, hysteretic behavior is observed across the superfluid-to-insulator boundary when varying the layer balance at fixed temperature. This is seen in Fig.~\ref{fig4}d, measured by sweeping the layer imbalance \dnu\ back and forth (horrizontal red arrows in Fig. 4b) at T=20mK.  This further reinforces that the superfluid to insulator transition is first order.

An exciton insulator that is separated from the superfluid with a first-order transition, highlights an unprecedented quantum phase that requires a non-trivial explanation. While our findings do not definitively identify the ground state of the EI, they are consistent with previous predictions that excitons can undergo a superfluid to solid transition within the available phase space ~\cite{Lozovik2004excitonmelting,Mitra2009hexatic,Joglekar2006excitonsolid,Yang2001excitonsolid,Conti2023supersolid,Boning2011supersolid,Astrakharchik2007dipole,Hu2024excitonsolid}. 
A solid-to-superfluid melting transition requires latent heat, which naturally accounts for the observed hysteresis in the temperature-driven transition~\cite{Zhou2021bilayerWigner,Zarenia2017excitonsolid,Zeng2023excitonDW,Yoshioka1990excitonsolid,Vu2023excitonsolid}. Taken together, our experimental signatures point to the exciton insulator being a solid phase. While quantum Monte Carlo calculations have suggested that an exciton superfluid could transition directly into an exciton solid upon varying density ~\cite{Boning2011supersolid,Astrakharchik2007dipole,Hu2024excitonsolid}, the unusual temperature dependence we observe—where the presumed solid melts into a superfluid upon heating—challenges the picture of a conventional solid ground state.

An alternate possibility for the origin of the insulating phase is disorder driven localization since in the large layer imbalance regime only the tails of the LLs are occupied.  We emphasize that the insulating phase is restricted to $\nu_{total}=1$ and does not persist when fixing one layer filling fraction and varying the other (moving horrizontally in Fig. 3a). Consequently, if disorder is responsible, it must be a unique form that localizes excitons only, while leaving individual-layer charge carriers unaffected ~\cite{Abergel2013exciton}.  At $\nu_{total}=1$, screening may be suppressed by the emergence of charge-neutral excitons, which could enhance the influence of disorder ~\cite{Lozovik2012} and drive localization.  While this could explain the density driven superfluid-insulator transition, we believe the temperature dependence argues against this since the superfluid phase above the reentrant transition temperature exhibits off-diagonal long-range order that appears unaffected by disorder localization.  There are no physical examples we are aware of in which a superfluid phase undergoes disorder-driven localization upon lowering the temperature. 

Finally, we show that the low-temperature phase boundary between the exciton insulator and superfluid follows a simple scaling argument. Fig.~\ref{fig4}f plots a region of the base temperature phase diagram, summarizing the various phases we access by tuning \dlb and $\Delta\nu$.  We plot $\Delta\nu$ along the bottom axis, and identify the associated value of $\ell_{e}/\ell_{B}$ along the top axis. The superfluid-to-insulator transition occurs around $\ell_{e}/\ell_{B}\sim2$  throughout the experimentally accessible range of \dlb. In  quantum Hall bilayers, $\ell_{B}$ defines the Bohr radius for magneto-excitons. This indicates that the relative stability between the exciton superfluid and insulator is defined by the relation between the exciton-exciton spacing ($\ell_{e}$) and the exciton size ($\ell_{B}$) (Fig. 4g-h), with the insulator phase stabilized when the exciton spacing exceeds the Bohr radius.  
The scaling argument here bears a striking resemblance of the Wigner crystal in the QHE regime of a single quantum wall ~\cite{Andrei1988Wigner,Jiang1990Wigner,Ma2020Wigner,Gervais2004Wigner,Goldman1990Wigner,tsui_wignercrystal_stm_2024}, which forms in the dilute limit where the electron-electron spacing exceeds two times the magnetic length ~\cite{Goerbig2003scaling,Shi2020}. This provides another support for a crystalline order underlying the exciton insulator. 

\hspace{\linewidth}

\noindent\textbf{Conclusion}\\
In summary we observe a superfluid to insulator transition in a dilute bilayer exciton system with the phase transition driven by varying either exciton density or temperature. Hystertic behaviour across the phase boundary is observed in both cases.  In the context of prior thoeretical work we interpret the first order phase transition as a signature that the insulator is an exciton solid.  The peculiar observation that, upon raising the temperature, the insulator gives way to a reentrant superfluid phase provides compelling evidence that the low temperature solid phase could be a correlated bosonic state. This raises intriguiging possibilities about the nature of the solid. For example,  numerous theoretical works have identified that an exotic solid phase, termed an exciton supersolid, could be realizable in a bilayer exciton system in a regime of phase space (dilute excitons with strong dipolar interactions) similar to where we observe the insulating behaviour\cite{Joglekar2006excitonsolid,Conti2023supersolid,De2002excitonsolid,Zarenia2017excitonsolid,Chui2020excitonsolid,Yoshioka1990excitonsolid,Chen1991excitonsolid,Yang2001excitonsolid,Chester1970supersolid}. The supersolid is conjectured to exhibit properties of both a solid, such as broken translational symmetry, and a superfluid, such as zero viscosity\cite{Chester1970supersolid,Penrose1956supersolid,Andreev1969supersolid,Leggett1970supersolid,Meisel1992supersolid}. However, until recently this state remained largely a theoretical concept\cite{Kim2004He4,Day2007He4,Hunt2009He4}.  Signatures of the supersolid phase have been observed in dipolar quantum gasses, in the context of cold atoms~\cite{Tanzi2019AMO,Chomaz2019AMO,Bottcher2019AMO,Norcia2021AMO}, but transport properties of this state have not been accessible in experiment.  That the exciton solid we observe manifests insulating behaviour could result from defect or boundary pinning, similar to 2D electronic wigner crystals. Establishing a solid state platform to study the flow dynamics of a potential supersolid opens many new opportunities to probe this phase.  Moreover, we anticipate our results will guide efforts to access similar states in zero-field bilayer heterostructures ~\cite{Wang2019excitonB0,Ma2021excitonB0,Fogler2014exciton,Conti2023supersolid}.

\section{Data availability}
Source data are available for this paper. All other data that support the findings of this study are available from the corresponding authors upon reasonable request.

\section{acknowledgments}
J.I.A.L. and C.R.D. wish to thank Sankar Das Sarma and Kun Yang for stimulating discussions. This research was primarily supported by the US Department of Energy, Office of Science, Basic Energy Sciences, under award no. DE-SC0019481 (transport measurements). Heterostructure and device fabrication was supported by the NSF MRSEC program at Columbia through the Center for Precision-Assembled Quantum Materials (DMR-2011738). Data analysis (Q.S.) was partially supported by Department of Energy (DE-SC0016703). 
J.I.A.L. acknowledges support from the Sloan research fellowship and NSF DMR-2143384. N.J.Z. acknowledge support from the Air Force Office of Scientific Research.  R.Q.N. acknowledges  support from the National Science Foundation EPSCoR Program under NSF Award OIA-2327206. K.W. and T.T. acknowledge support from the JSPS KAKENHI (Grant Numbers 20H00354, 21H05233 and 23H02052) and World Premier International Research Center Initiative (WPI), MEXT, Japan. A portion of this work was performed at the National High Magnetic Field Laboratory, which is supported by National Science Foundation Cooperative Agreement No. DMR-1157490 and the State of Florida.

\section*{Competing financial interests}
The authors declare no competing financial interests.

\bibliography{Li_ref}%

\begin{thebibliography}{75}%
\makeatletter
\providecommand \@ifxundefined [1]{%
 \@ifx{#1\undefined}
}%
\providecommand \@ifnum [1]{%
 \ifnum #1\expandafter \@firstoftwo
 \else \expandafter \@secondoftwo
 \fi
}%
\providecommand \@ifx [1]{%
 \ifx #1\expandafter \@firstoftwo
 \else \expandafter \@secondoftwo
 \fi
}%
\providecommand \natexlab [1]{#1}%
\providecommand \enquote  [1]{``#1''}%
\providecommand \bibnamefont  [1]{#1}%
\providecommand \bibfnamefont [1]{#1}%
\providecommand \citenamefont [1]{#1}%
\providecommand \href@noop [0]{\@secondoftwo}%
\providecommand \href [0]{\begingroup \@sanitize@url \@href}%
\providecommand \@href[1]{\@@startlink{#1}\@@href}%
\providecommand \@@href[1]{\endgroup#1\@@endlink}%
\providecommand \@sanitize@url [0]{\catcode `\\12\catcode `\$12\catcode
  `\&12\catcode `\#12\catcode `\^12\catcode `\_12\catcode `\%12\relax}%
\providecommand \@@startlink[1]{}%
\providecommand \@@endlink[0]{}%
\providecommand \url  [0]{\begingroup\@sanitize@url \@url }%
\providecommand \@url [1]{\endgroup\@href {#1}{\urlprefix }}%
\providecommand \urlprefix  [0]{URL }%
\providecommand \Eprint [0]{\href }%
\providecommand \doibase [0]{http://dx.doi.org/}%
\providecommand \selectlanguage [0]{\@gobble}%
\providecommand \bibinfo  [0]{\@secondoftwo}%
\providecommand \bibfield  [0]{\@secondoftwo}%
\providecommand \translation [1]{[#1]}%
\providecommand \BibitemOpen [0]{}%
\providecommand \bibitemStop [0]{}%
\providecommand \bibitemNoStop [0]{.\EOS\space}%
\providecommand \EOS [0]{\spacefactor3000\relax}%
\providecommand \BibitemShut  [1]{\csname bibitem#1\endcsname}%
\let\auto@bib@innerbib\@empty
\bibitem [{\citenamefont {Allen}\ and\ \citenamefont
  {Misener}(1938)}]{Allen1938He4}%
  \BibitemOpen
  \bibfield  {author} {\bibinfo {author} {\bibfnamefont {J.~F.}\ \bibnamefont
  {Allen}}\ and\ \bibinfo {author} {\bibfnamefont {A.}~\bibnamefont
  {Misener}},\ }\href@noop {} {\bibfield  {journal} {\bibinfo  {journal}
  {Nature}\ }\textbf {\bibinfo {volume} {142}},\ \bibinfo {pages} {643}
  (\bibinfo {year} {1938})}\BibitemShut {NoStop}%
\bibitem [{\citenamefont {Kapitza}(1938)}]{Kapitza1938He4}%
  \BibitemOpen
  \bibfield  {author} {\bibinfo {author} {\bibfnamefont {P.}~\bibnamefont
  {Kapitza}},\ }\href@noop {} {\bibfield  {journal} {\bibinfo  {journal}
  {Nature}\ }\textbf {\bibinfo {volume} {141}},\ \bibinfo {pages} {74}
  (\bibinfo {year} {1938})}\BibitemShut {NoStop}%
\bibitem [{\citenamefont {Anderson}\ \emph {et~al.}(1995)\citenamefont
  {Anderson}, \citenamefont {Ensher}, \citenamefont {Matthews}, \citenamefont
  {Wieman},\ and\ \citenamefont {Cornell}}]{Anderson1995BEC}%
  \BibitemOpen
  \bibfield  {author} {\bibinfo {author} {\bibfnamefont {M.~H.}\ \bibnamefont
  {Anderson}}, \bibinfo {author} {\bibfnamefont {J.~R.}\ \bibnamefont
  {Ensher}}, \bibinfo {author} {\bibfnamefont {M.~R.}\ \bibnamefont
  {Matthews}}, \bibinfo {author} {\bibfnamefont {C.~E.}\ \bibnamefont
  {Wieman}}, \ and\ \bibinfo {author} {\bibfnamefont {E.~A.}\ \bibnamefont
  {Cornell}},\ }\href {\doibase 10.1126/science.269.5221.198} {\bibfield
  {journal} {\bibinfo  {journal} {Science}\ }\textbf {\bibinfo {volume}
  {269}},\ \bibinfo {pages} {198} (\bibinfo {year} {1995})}\BibitemShut
  {NoStop}%
\bibitem [{\citenamefont {Davis}\ \emph {et~al.}(1995)\citenamefont {Davis},
  \citenamefont {Mewes}, \citenamefont {Andrews}, \citenamefont {van Druten},
  \citenamefont {Durfee}, \citenamefont {Kurn},\ and\ \citenamefont
  {Ketterle}}]{Davis1995BEC}%
  \BibitemOpen
  \bibfield  {author} {\bibinfo {author} {\bibfnamefont {K.~B.}\ \bibnamefont
  {Davis}}, \bibinfo {author} {\bibfnamefont {M.-O.}\ \bibnamefont {Mewes}},
  \bibinfo {author} {\bibfnamefont {M.~R.}\ \bibnamefont {Andrews}}, \bibinfo
  {author} {\bibfnamefont {N.~J.}\ \bibnamefont {van Druten}}, \bibinfo
  {author} {\bibfnamefont {D.~S.}\ \bibnamefont {Durfee}}, \bibinfo {author}
  {\bibfnamefont {D.~M.}\ \bibnamefont {Kurn}}, \ and\ \bibinfo {author}
  {\bibfnamefont {W.}~\bibnamefont {Ketterle}},\ }\href {\doibase
  10.1103/PhysRevLett.75.3969} {\bibfield  {journal} {\bibinfo  {journal}
  {Phys. Rev. Lett.}\ }\textbf {\bibinfo {volume} {75}},\ \bibinfo {pages}
  {3969} (\bibinfo {year} {1995})}\BibitemShut {NoStop}%
\bibitem [{\citenamefont {Kellogg}\ \emph {et~al.}(2004)\citenamefont
  {Kellogg}, \citenamefont {Eisenstein}, \citenamefont {Pfeiffer},\ and\
  \citenamefont {West}}]{Kel.04}%
  \BibitemOpen
  \bibfield  {author} {\bibinfo {author} {\bibfnamefont {M.}~\bibnamefont
  {Kellogg}}, \bibinfo {author} {\bibfnamefont {J.~P.}\ \bibnamefont
  {Eisenstein}}, \bibinfo {author} {\bibfnamefont {L.~N.}\ \bibnamefont
  {Pfeiffer}}, \ and\ \bibinfo {author} {\bibfnamefont {K.~W.}\ \bibnamefont
  {West}},\ }\href {\doibase 10.1103/PhysRevLett.93.036801} {\bibfield
  {journal} {\bibinfo  {journal} {Phys. Rev. Lett.}\ }\textbf {\bibinfo
  {volume} {93}},\ \bibinfo {pages} {036801} (\bibinfo {year}
  {2004})}\BibitemShut {NoStop}%
\bibitem [{\citenamefont {Kellogg}(2005)}]{Kel.05}%
  \BibitemOpen
  \bibfield  {author} {\bibinfo {author} {\bibfnamefont {M.~J.}\ \bibnamefont
  {Kellogg}},\ }\emph {\bibinfo {title} {Evidence for Excitonic Superfluidity
  in a Bilayer Two-Dimensional Electron System}},\ \href@noop {} {Ph.D.
  thesis},\ \bibinfo  {school} {California Institute of Technology} (\bibinfo
  {year} {2005})\BibitemShut {NoStop}%
\bibitem [{\citenamefont {Tutuc}\ \emph {et~al.}(2004)\citenamefont {Tutuc},
  \citenamefont {Shayegan},\ and\ \citenamefont {Huse}}]{Tutuc.04}%
  \BibitemOpen
  \bibfield  {author} {\bibinfo {author} {\bibfnamefont {E.}~\bibnamefont
  {Tutuc}}, \bibinfo {author} {\bibfnamefont {M.}~\bibnamefont {Shayegan}}, \
  and\ \bibinfo {author} {\bibfnamefont {D.~A.}\ \bibnamefont {Huse}},\ }\href
  {\doibase 10.1103/PhysRevLett.93.036802} {\bibfield  {journal} {\bibinfo
  {journal} {Phys. Rev. Lett.}\ }\textbf {\bibinfo {volume} {93}},\ \bibinfo
  {pages} {036802} (\bibinfo {year} {2004})}\BibitemShut {NoStop}%
\bibitem [{\citenamefont {Wiersma}\ \emph {et~al.}(2004)\citenamefont
  {Wiersma}, \citenamefont {Lok}, \citenamefont {Kraus}, \citenamefont
  {Dietsche}, \citenamefont {von Klitzing}, \citenamefont {Schuh},
  \citenamefont {Bichler}, \citenamefont {Tranitz},\ and\ \citenamefont
  {Wegscheider}}]{Wiersma.04}%
  \BibitemOpen
  \bibfield  {author} {\bibinfo {author} {\bibfnamefont {R.~D.}\ \bibnamefont
  {Wiersma}}, \bibinfo {author} {\bibfnamefont {J.~G.~S.}\ \bibnamefont {Lok}},
  \bibinfo {author} {\bibfnamefont {S.}~\bibnamefont {Kraus}}, \bibinfo
  {author} {\bibfnamefont {W.}~\bibnamefont {Dietsche}}, \bibinfo {author}
  {\bibfnamefont {K.}~\bibnamefont {von Klitzing}}, \bibinfo {author}
  {\bibfnamefont {D.}~\bibnamefont {Schuh}}, \bibinfo {author} {\bibfnamefont
  {M.}~\bibnamefont {Bichler}}, \bibinfo {author} {\bibfnamefont {H.-P.}\
  \bibnamefont {Tranitz}}, \ and\ \bibinfo {author} {\bibfnamefont
  {W.}~\bibnamefont {Wegscheider}},\ }\href {\doibase
  10.1103/PhysRevLett.93.266805} {\bibfield  {journal} {\bibinfo  {journal}
  {Phys. Rev. Lett.}\ }\textbf {\bibinfo {volume} {93}},\ \bibinfo {pages}
  {266805} (\bibinfo {year} {2004})}\BibitemShut {NoStop}%
\bibitem [{\citenamefont {Nandi}\ \emph {et~al.}(2012)\citenamefont {Nandi},
  \citenamefont {Finck}, \citenamefont {Eisenstein}, \citenamefont {Pfeiffer},\
  and\ \citenamefont {West}}]{Nandi.12}%
  \BibitemOpen
  \bibfield  {author} {\bibinfo {author} {\bibfnamefont {D.}~\bibnamefont
  {Nandi}}, \bibinfo {author} {\bibfnamefont {A.~D.~K.}\ \bibnamefont {Finck}},
  \bibinfo {author} {\bibfnamefont {J.~P.}\ \bibnamefont {Eisenstein}},
  \bibinfo {author} {\bibfnamefont {L.~N.}\ \bibnamefont {Pfeiffer}}, \ and\
  \bibinfo {author} {\bibfnamefont {K.~W.}\ \bibnamefont {West}},\ }\href
  {\doibase 10.1038/nature11302} {\bibfield  {journal} {\bibinfo  {journal}
  {Nature}\ }\textbf {\bibinfo {volume} {488}},\ \bibinfo {pages} {481}
  (\bibinfo {year} {2012})}\BibitemShut {NoStop}%
\bibitem [{\citenamefont {Eisenstein}(2014)}]{Eisenstein2014review}%
  \BibitemOpen
  \bibfield  {author} {\bibinfo {author} {\bibfnamefont {J.~P.}\ \bibnamefont
  {Eisenstein}},\ }\href {\doibase 10.1146/annurev-conmatphys-031113-133832}
  {\bibfield  {journal} {\bibinfo  {journal} {Annu. Rev. of Condens. Matter
  Phys.}\ }\textbf {\bibinfo {volume} {5}},\ \bibinfo {pages} {159} (\bibinfo
  {year} {2014})}\BibitemShut {NoStop}%
\bibitem [{\citenamefont {Li}\ \emph {et~al.}(2017)\citenamefont {Li},
  \citenamefont {Taniguchi}, \citenamefont {Watanabe}, \citenamefont {Hone},\
  and\ \citenamefont {Dean}}]{Li.17a}%
  \BibitemOpen
  \bibfield  {author} {\bibinfo {author} {\bibfnamefont {J.~I.~A.}\
  \bibnamefont {Li}}, \bibinfo {author} {\bibfnamefont {T.}~\bibnamefont
  {Taniguchi}}, \bibinfo {author} {\bibfnamefont {K.}~\bibnamefont {Watanabe}},
  \bibinfo {author} {\bibfnamefont {J.}~\bibnamefont {Hone}}, \ and\ \bibinfo
  {author} {\bibfnamefont {C.~R.}\ \bibnamefont {Dean}},\ }\href {\doibase
  10.1038/nphys4140} {\bibfield  {journal} {\bibinfo  {journal} {Nat. Phys.}\
  }\textbf {\bibinfo {volume} {13}},\ \bibinfo {pages} {751} (\bibinfo {year}
  {2017})}\BibitemShut {NoStop}%
\bibitem [{\citenamefont {Liu}\ \emph {et~al.}(2017)\citenamefont {Liu},
  \citenamefont {Taniguchi}, \citenamefont {Watanabe}, \citenamefont
  {Halperin},\ and\ \citenamefont {Kim}}]{Liu.17a}%
  \BibitemOpen
  \bibfield  {author} {\bibinfo {author} {\bibfnamefont {X.}~\bibnamefont
  {Liu}}, \bibinfo {author} {\bibfnamefont {T.}~\bibnamefont {Taniguchi}},
  \bibinfo {author} {\bibfnamefont {K.}~\bibnamefont {Watanabe}}, \bibinfo
  {author} {\bibfnamefont {B.}~\bibnamefont {Halperin}}, \ and\ \bibinfo
  {author} {\bibfnamefont {P.}~\bibnamefont {Kim}},\ }\href {\doibase
  10.1038/nphys4116} {\bibfield  {journal} {\bibinfo  {journal} {Nat. Phys.}\
  }\textbf {\bibinfo {volume} {13}},\ \bibinfo {pages} {746} (\bibinfo {year}
  {2017})}\BibitemShut {NoStop}%
\bibitem [{\citenamefont {Liu}\ \emph {et~al.}(2022)\citenamefont {Liu},
  \citenamefont {Li}, \citenamefont {Watanabe}, \citenamefont {Taniguchi},
  \citenamefont {Hone}, \citenamefont {Halperin}, \citenamefont {Kim},\ and\
  \citenamefont {Dean}}]{Liu2022crossover}%
  \BibitemOpen
  \bibfield  {author} {\bibinfo {author} {\bibfnamefont {X.}~\bibnamefont
  {Liu}}, \bibinfo {author} {\bibfnamefont {J.}~\bibnamefont {Li}}, \bibinfo
  {author} {\bibfnamefont {K.}~\bibnamefont {Watanabe}}, \bibinfo {author}
  {\bibfnamefont {T.}~\bibnamefont {Taniguchi}}, \bibinfo {author}
  {\bibfnamefont {J.}~\bibnamefont {Hone}}, \bibinfo {author} {\bibfnamefont
  {B.~I.}\ \bibnamefont {Halperin}}, \bibinfo {author} {\bibfnamefont
  {P.}~\bibnamefont {Kim}}, \ and\ \bibinfo {author} {\bibfnamefont {C.~R.}\
  \bibnamefont {Dean}},\ }\href@noop {} {\bibfield  {journal} {\bibinfo
  {journal} {Science}\ }\textbf {\bibinfo {volume} {375}},\ \bibinfo {pages}
  {205} (\bibinfo {year} {2022})}\BibitemShut {NoStop}%
\bibitem [{\citenamefont {Burg}\ \emph {et~al.}(2018)\citenamefont {Burg},
  \citenamefont {Prasad}, \citenamefont {Kim}, \citenamefont {Taniguchi},
  \citenamefont {Watanabe}, \citenamefont {MacDonald}, \citenamefont
  {Register},\ and\ \citenamefont {Tutuc}}]{Burg2018strongly}%
  \BibitemOpen
  \bibfield  {author} {\bibinfo {author} {\bibfnamefont {G.~W.}\ \bibnamefont
  {Burg}}, \bibinfo {author} {\bibfnamefont {N.}~\bibnamefont {Prasad}},
  \bibinfo {author} {\bibfnamefont {K.}~\bibnamefont {Kim}}, \bibinfo {author}
  {\bibfnamefont {T.}~\bibnamefont {Taniguchi}}, \bibinfo {author}
  {\bibfnamefont {K.}~\bibnamefont {Watanabe}}, \bibinfo {author}
  {\bibfnamefont {A.~H.}\ \bibnamefont {MacDonald}}, \bibinfo {author}
  {\bibfnamefont {L.~F.}\ \bibnamefont {Register}}, \ and\ \bibinfo {author}
  {\bibfnamefont {E.}~\bibnamefont {Tutuc}},\ }\href {\doibase
  10.1103/PhysRevLett.120.177702} {\bibfield  {journal} {\bibinfo  {journal}
  {Phys. Rev. Lett.}\ }\textbf {\bibinfo {volume} {120}},\ \bibinfo {pages}
  {177702} (\bibinfo {year} {2018})}\BibitemShut {NoStop}%
\bibitem [{\citenamefont {Shi}\ \emph {et~al.}(2020)\citenamefont {Shi},
  \citenamefont {Shih}, \citenamefont {Gustafsson}, \citenamefont {Rhodes},
  \citenamefont {Kim}, \citenamefont {Watanabe}, \citenamefont {Taniguchi},
  \citenamefont {Papić}, \citenamefont {Hone},\ and\ \citenamefont
  {Dean}}]{Shi2020}%
  \BibitemOpen
  \bibfield  {author} {\bibinfo {author} {\bibfnamefont {Q.}~\bibnamefont
  {Shi}}, \bibinfo {author} {\bibfnamefont {E.-M.}\ \bibnamefont {Shih}},
  \bibinfo {author} {\bibfnamefont {M.~V.}\ \bibnamefont {Gustafsson}},
  \bibinfo {author} {\bibfnamefont {D.}~\bibnamefont {Rhodes}}, \bibinfo
  {author} {\bibfnamefont {B.}~\bibnamefont {Kim}}, \bibinfo {author}
  {\bibfnamefont {K.}~\bibnamefont {Watanabe}}, \bibinfo {author}
  {\bibfnamefont {T.}~\bibnamefont {Taniguchi}}, \bibinfo {author}
  {\bibfnamefont {Z.}~\bibnamefont {Papić}}, \bibinfo {author} {\bibfnamefont
  {J.}~\bibnamefont {Hone}}, \ and\ \bibinfo {author} {\bibfnamefont {C.~R.}\
  \bibnamefont {Dean}},\ }\href {\doibase 10.1038/s41565-020-0685-6} {\bibfield
   {journal} {\bibinfo  {journal} {Nature Nanotechnology}\ }\textbf {\bibinfo
  {volume} {15}},\ \bibinfo {pages} {569} (\bibinfo {year} {2020})}\BibitemShut
  {NoStop}%
\bibitem [{\citenamefont {Shi}\ \emph {et~al.}(2022)\citenamefont {Shi},
  \citenamefont {Shih}, \citenamefont {Rhodes}, \citenamefont {Kim},
  \citenamefont {Barmak}, \citenamefont {Watanabe}, \citenamefont {Taniguchi},
  \citenamefont {Papić}, \citenamefont {Abanin}, \citenamefont {Hone},\ and\
  \citenamefont {Dean}}]{Shi2022}%
  \BibitemOpen
  \bibfield  {author} {\bibinfo {author} {\bibfnamefont {Q.}~\bibnamefont
  {Shi}}, \bibinfo {author} {\bibfnamefont {E.-M.}\ \bibnamefont {Shih}},
  \bibinfo {author} {\bibfnamefont {D.}~\bibnamefont {Rhodes}}, \bibinfo
  {author} {\bibfnamefont {B.}~\bibnamefont {Kim}}, \bibinfo {author}
  {\bibfnamefont {K.}~\bibnamefont {Barmak}}, \bibinfo {author} {\bibfnamefont
  {K.}~\bibnamefont {Watanabe}}, \bibinfo {author} {\bibfnamefont
  {T.}~\bibnamefont {Taniguchi}}, \bibinfo {author} {\bibfnamefont
  {Z.}~\bibnamefont {Papić}}, \bibinfo {author} {\bibfnamefont {D.~A.}\
  \bibnamefont {Abanin}}, \bibinfo {author} {\bibfnamefont {J.}~\bibnamefont
  {Hone}}, \ and\ \bibinfo {author} {\bibfnamefont {C.~R.}\ \bibnamefont
  {Dean}},\ }\href {\doibase 10.1038/s41565-022-01104-5} {\bibfield  {journal}
  {\bibinfo  {journal} {Nature Nanotechnology}\ }\textbf {\bibinfo {volume}
  {17}},\ \bibinfo {pages} {577} (\bibinfo {year} {2022})}\BibitemShut
  {NoStop}%
\bibitem [{\citenamefont {Fisher}\ \emph {et~al.}(1989)\citenamefont {Fisher},
  \citenamefont {Weichman}, \citenamefont {Grinstein},\ and\ \citenamefont
  {Fisher}}]{Fisher1989BoseMott}%
  \BibitemOpen
  \bibfield  {author} {\bibinfo {author} {\bibfnamefont {M.~P.~A.}\
  \bibnamefont {Fisher}}, \bibinfo {author} {\bibfnamefont {P.~B.}\
  \bibnamefont {Weichman}}, \bibinfo {author} {\bibfnamefont {G.}~\bibnamefont
  {Grinstein}}, \ and\ \bibinfo {author} {\bibfnamefont {D.~S.}\ \bibnamefont
  {Fisher}},\ }\href {\doibase 10.1103/PhysRevB.40.546} {\bibfield  {journal}
  {\bibinfo  {journal} {Phys. Rev. B}\ }\textbf {\bibinfo {volume} {40}},\
  \bibinfo {pages} {546} (\bibinfo {year} {1989})}\BibitemShut {NoStop}%
\bibitem [{\citenamefont {Penrose}\ and\ \citenamefont
  {Onsager}(1956)}]{Penrose1956supersolid}%
  \BibitemOpen
  \bibfield  {author} {\bibinfo {author} {\bibfnamefont {O.}~\bibnamefont
  {Penrose}}\ and\ \bibinfo {author} {\bibfnamefont {L.}~\bibnamefont
  {Onsager}},\ }\href {\doibase 10.1103/PhysRev.104.576} {\bibfield  {journal}
  {\bibinfo  {journal} {Phys. Rev.}\ }\textbf {\bibinfo {volume} {104}},\
  \bibinfo {pages} {576} (\bibinfo {year} {1956})}\BibitemShut {NoStop}%
\bibitem [{\citenamefont {Andreev}\ and\ \citenamefont
  {Lifshits}(1969)}]{Andreev1969supersolid}%
  \BibitemOpen
  \bibfield  {author} {\bibinfo {author} {\bibfnamefont {A.}~\bibnamefont
  {Andreev}}\ and\ \bibinfo {author} {\bibfnamefont {I.}~\bibnamefont
  {Lifshits}},\ }\href@noop {} {\bibfield  {journal} {\bibinfo  {journal} {Zhur
  Eksper Teoret Fiziki}\ }\textbf {\bibinfo {volume} {56}},\ \bibinfo {pages}
  {2057} (\bibinfo {year} {1969})}\BibitemShut {NoStop}%
\bibitem [{\citenamefont {Leggett}(1970)}]{Leggett1970supersolid}%
  \BibitemOpen
  \bibfield  {author} {\bibinfo {author} {\bibfnamefont {A.~J.}\ \bibnamefont
  {Leggett}},\ }\href {\doibase 10.1103/PhysRevLett.25.1543} {\bibfield
  {journal} {\bibinfo  {journal} {Phys. Rev. Lett.}\ }\textbf {\bibinfo
  {volume} {25}},\ \bibinfo {pages} {1543} (\bibinfo {year}
  {1970})}\BibitemShut {NoStop}%
\bibitem [{\citenamefont {Meisel}(1992)}]{Meisel1992supersolid}%
  \BibitemOpen
  \bibfield  {author} {\bibinfo {author} {\bibfnamefont {M.~W.}\ \bibnamefont
  {Meisel}},\ }\href@noop {} {\bibfield  {journal} {\bibinfo  {journal}
  {Physica B: Condensed Matter}\ }\textbf {\bibinfo {volume} {178}},\ \bibinfo
  {pages} {121} (\bibinfo {year} {1992})}\BibitemShut {NoStop}%
\bibitem [{\citenamefont {Tanzi}\ \emph {et~al.}(2019)\citenamefont {Tanzi},
  \citenamefont {Lucioni}, \citenamefont {Fam{\`a}}, \citenamefont {Catani},
  \citenamefont {Fioretti}, \citenamefont {Gabbanini}, \citenamefont {Bisset},
  \citenamefont {Santos},\ and\ \citenamefont {Modugno}}]{Tanzi2019AMO}%
  \BibitemOpen
  \bibfield  {author} {\bibinfo {author} {\bibfnamefont {L.}~\bibnamefont
  {Tanzi}}, \bibinfo {author} {\bibfnamefont {E.}~\bibnamefont {Lucioni}},
  \bibinfo {author} {\bibfnamefont {F.}~\bibnamefont {Fam{\`a}}}, \bibinfo
  {author} {\bibfnamefont {J.}~\bibnamefont {Catani}}, \bibinfo {author}
  {\bibfnamefont {A.}~\bibnamefont {Fioretti}}, \bibinfo {author}
  {\bibfnamefont {C.}~\bibnamefont {Gabbanini}}, \bibinfo {author}
  {\bibfnamefont {R.~N.}\ \bibnamefont {Bisset}}, \bibinfo {author}
  {\bibfnamefont {L.}~\bibnamefont {Santos}}, \ and\ \bibinfo {author}
  {\bibfnamefont {G.}~\bibnamefont {Modugno}},\ }\href@noop {} {\bibfield
  {journal} {\bibinfo  {journal} {Physical review letters}\ }\textbf {\bibinfo
  {volume} {122}},\ \bibinfo {pages} {130405} (\bibinfo {year}
  {2019})}\BibitemShut {NoStop}%
\bibitem [{\citenamefont {Chomaz}\ \emph {et~al.}(2019)\citenamefont {Chomaz},
  \citenamefont {Petter}, \citenamefont {Ilzh{\"o}fer}, \citenamefont {Natale},
  \citenamefont {Trautmann}, \citenamefont {Politi}, \citenamefont
  {Durastante}, \citenamefont {Van~Bijnen}, \citenamefont {Patscheider},
  \citenamefont {Sohmen} \emph {et~al.}}]{Chomaz2019AMO}%
  \BibitemOpen
  \bibfield  {author} {\bibinfo {author} {\bibfnamefont {L.}~\bibnamefont
  {Chomaz}}, \bibinfo {author} {\bibfnamefont {D.}~\bibnamefont {Petter}},
  \bibinfo {author} {\bibfnamefont {P.}~\bibnamefont {Ilzh{\"o}fer}}, \bibinfo
  {author} {\bibfnamefont {G.}~\bibnamefont {Natale}}, \bibinfo {author}
  {\bibfnamefont {A.}~\bibnamefont {Trautmann}}, \bibinfo {author}
  {\bibfnamefont {C.}~\bibnamefont {Politi}}, \bibinfo {author} {\bibfnamefont
  {G.}~\bibnamefont {Durastante}}, \bibinfo {author} {\bibfnamefont
  {R.}~\bibnamefont {Van~Bijnen}}, \bibinfo {author} {\bibfnamefont
  {A.}~\bibnamefont {Patscheider}}, \bibinfo {author} {\bibfnamefont
  {M.}~\bibnamefont {Sohmen}},  \emph {et~al.},\ }\href@noop {} {\bibfield
  {journal} {\bibinfo  {journal} {Physical Review X}\ }\textbf {\bibinfo
  {volume} {9}},\ \bibinfo {pages} {021012} (\bibinfo {year}
  {2019})}\BibitemShut {NoStop}%
\bibitem [{\citenamefont {B{\"o}ttcher}\ \emph {et~al.}(2019)\citenamefont
  {B{\"o}ttcher}, \citenamefont {Schmidt}, \citenamefont {Wenzel},
  \citenamefont {Hertkorn}, \citenamefont {Guo}, \citenamefont {Langen},\ and\
  \citenamefont {Pfau}}]{Bottcher2019AMO}%
  \BibitemOpen
  \bibfield  {author} {\bibinfo {author} {\bibfnamefont {F.}~\bibnamefont
  {B{\"o}ttcher}}, \bibinfo {author} {\bibfnamefont {J.-N.}\ \bibnamefont
  {Schmidt}}, \bibinfo {author} {\bibfnamefont {M.}~\bibnamefont {Wenzel}},
  \bibinfo {author} {\bibfnamefont {J.}~\bibnamefont {Hertkorn}}, \bibinfo
  {author} {\bibfnamefont {M.}~\bibnamefont {Guo}}, \bibinfo {author}
  {\bibfnamefont {T.}~\bibnamefont {Langen}}, \ and\ \bibinfo {author}
  {\bibfnamefont {T.}~\bibnamefont {Pfau}},\ }\href@noop {} {\bibfield
  {journal} {\bibinfo  {journal} {Physical Review X}\ }\textbf {\bibinfo
  {volume} {9}},\ \bibinfo {pages} {011051} (\bibinfo {year}
  {2019})}\BibitemShut {NoStop}%
\bibitem [{\citenamefont {Norcia}\ \emph {et~al.}(2021)\citenamefont {Norcia},
  \citenamefont {Politi}, \citenamefont {Klaus}, \citenamefont {Poli},
  \citenamefont {Sohmen}, \citenamefont {Mark}, \citenamefont {Bisset},
  \citenamefont {Santos},\ and\ \citenamefont {Ferlaino}}]{Norcia2021AMO}%
  \BibitemOpen
  \bibfield  {author} {\bibinfo {author} {\bibfnamefont {M.~A.}\ \bibnamefont
  {Norcia}}, \bibinfo {author} {\bibfnamefont {C.}~\bibnamefont {Politi}},
  \bibinfo {author} {\bibfnamefont {L.}~\bibnamefont {Klaus}}, \bibinfo
  {author} {\bibfnamefont {E.}~\bibnamefont {Poli}}, \bibinfo {author}
  {\bibfnamefont {M.}~\bibnamefont {Sohmen}}, \bibinfo {author} {\bibfnamefont
  {M.~J.}\ \bibnamefont {Mark}}, \bibinfo {author} {\bibfnamefont {R.~N.}\
  \bibnamefont {Bisset}}, \bibinfo {author} {\bibfnamefont {L.}~\bibnamefont
  {Santos}}, \ and\ \bibinfo {author} {\bibfnamefont {F.}~\bibnamefont
  {Ferlaino}},\ }\href@noop {} {\bibfield  {journal} {\bibinfo  {journal}
  {Nature}\ }\textbf {\bibinfo {volume} {596}},\ \bibinfo {pages} {357}
  (\bibinfo {year} {2021})}\BibitemShut {NoStop}%
\bibitem [{\citenamefont {Conti}\ \emph {et~al.}(2023)\citenamefont {Conti},
  \citenamefont {Perali}, \citenamefont {Hamilton}, \citenamefont {Milo\ifmmode
  \check{s}\else \v{s}\fi{}evi\ifmmode~\acute{c}\else \'{c}\fi{}},
  \citenamefont {Peeters},\ and\ \citenamefont
  {Neilson}}]{Conti2023supersolid}%
  \BibitemOpen
  \bibfield  {author} {\bibinfo {author} {\bibfnamefont {S.}~\bibnamefont
  {Conti}}, \bibinfo {author} {\bibfnamefont {A.}~\bibnamefont {Perali}},
  \bibinfo {author} {\bibfnamefont {A.~R.}\ \bibnamefont {Hamilton}}, \bibinfo
  {author} {\bibfnamefont {M.~V.}\ \bibnamefont {Milo\ifmmode \check{s}\else
  \v{s}\fi{}evi\ifmmode~\acute{c}\else \'{c}\fi{}}}, \bibinfo {author}
  {\bibfnamefont {F.~m. c.~M.}\ \bibnamefont {Peeters}}, \ and\ \bibinfo
  {author} {\bibfnamefont {D.}~\bibnamefont {Neilson}},\ }\href {\doibase
  10.1103/PhysRevLett.130.057001} {\bibfield  {journal} {\bibinfo  {journal}
  {Phys. Rev. Lett.}\ }\textbf {\bibinfo {volume} {130}},\ \bibinfo {pages}
  {057001} (\bibinfo {year} {2023})}\BibitemShut {NoStop}%
\bibitem [{\citenamefont {Lozovik}\ and\ \citenamefont
  {Yudson}(1975)}]{Lozovik1975exciton}%
  \BibitemOpen
  \bibfield  {author} {\bibinfo {author} {\bibfnamefont {Y.~E.}\ \bibnamefont
  {Lozovik}}\ and\ \bibinfo {author} {\bibfnamefont {V.}~\bibnamefont
  {Yudson}},\ }\href@noop {} {\bibfield  {journal} {\bibinfo  {journal} {JETP
  Lett.}\ }\textbf {\bibinfo {volume} {22}},\ \bibinfo {pages} {274} (\bibinfo
  {year} {1975})}\BibitemShut {NoStop}%
\bibitem [{\citenamefont {Pogrebinsky}(1977)}]{Pogrebinsky1977exciton}%
  \BibitemOpen
  \bibfield  {author} {\bibinfo {author} {\bibfnamefont {M.~B.}\ \bibnamefont
  {Pogrebinsky}},\ }\href@noop {} {\bibfield  {journal} {\bibinfo  {journal}
  {Sov. Phys. Semicond.}\ }\textbf {\bibinfo {volume} {11}},\ \bibinfo {pages}
  {372} (\bibinfo {year} {1977})}\BibitemShut {NoStop}%
\bibitem [{\citenamefont {Blatt}\ \emph {et~al.}(1962)\citenamefont {Blatt},
  \citenamefont {B\"oer},\ and\ \citenamefont {Brandt}}]{Blatt1962exciton}%
  \BibitemOpen
  \bibfield  {author} {\bibinfo {author} {\bibfnamefont {J.~M.}\ \bibnamefont
  {Blatt}}, \bibinfo {author} {\bibfnamefont {K.~W.}\ \bibnamefont {B\"oer}}, \
  and\ \bibinfo {author} {\bibfnamefont {W.}~\bibnamefont {Brandt}},\ }\href
  {\doibase 10.1103/PhysRev.126.1691} {\bibfield  {journal} {\bibinfo
  {journal} {Phys. Rev.}\ }\textbf {\bibinfo {volume} {126}},\ \bibinfo {pages}
  {1691} (\bibinfo {year} {1962})}\BibitemShut {NoStop}%
\bibitem [{\citenamefont {Li}\ \emph {et~al.}(2019)\citenamefont {Li},
  \citenamefont {Shi}, \citenamefont {Zeng}, \citenamefont {Watanabe},
  \citenamefont {Taniguchi}, \citenamefont {Hone},\ and\ \citenamefont
  {Dean}}]{Li2019pairing}%
  \BibitemOpen
  \bibfield  {author} {\bibinfo {author} {\bibfnamefont {J.}~\bibnamefont
  {Li}}, \bibinfo {author} {\bibfnamefont {Q.}~\bibnamefont {Shi}}, \bibinfo
  {author} {\bibfnamefont {Y.}~\bibnamefont {Zeng}}, \bibinfo {author}
  {\bibfnamefont {K.}~\bibnamefont {Watanabe}}, \bibinfo {author}
  {\bibfnamefont {T.}~\bibnamefont {Taniguchi}}, \bibinfo {author}
  {\bibfnamefont {J.}~\bibnamefont {Hone}}, \ and\ \bibinfo {author}
  {\bibfnamefont {C.}~\bibnamefont {Dean}},\ }\href@noop {} {\bibfield
  {journal} {\bibinfo  {journal} {Nature Physics}\ }\textbf {\bibinfo {volume}
  {15}},\ \bibinfo {pages} {898} (\bibinfo {year} {2019})}\BibitemShut
  {NoStop}%
\bibitem [{\citenamefont {Liu}\ \emph {et~al.}(2019)\citenamefont {Liu},
  \citenamefont {Hao}, \citenamefont {Watanabe}, \citenamefont {Taniguchi},
  \citenamefont {Halperin},\ and\ \citenamefont {Kim}}]{Liu2019interlayer}%
  \BibitemOpen
  \bibfield  {author} {\bibinfo {author} {\bibfnamefont {X.}~\bibnamefont
  {Liu}}, \bibinfo {author} {\bibfnamefont {Z.}~\bibnamefont {Hao}}, \bibinfo
  {author} {\bibfnamefont {K.}~\bibnamefont {Watanabe}}, \bibinfo {author}
  {\bibfnamefont {T.}~\bibnamefont {Taniguchi}}, \bibinfo {author}
  {\bibfnamefont {B.~I.}\ \bibnamefont {Halperin}}, \ and\ \bibinfo {author}
  {\bibfnamefont {P.}~\bibnamefont {Kim}},\ }\href@noop {} {\bibfield
  {journal} {\bibinfo  {journal} {Nature Physics}\ }\textbf {\bibinfo {volume}
  {15}},\ \bibinfo {pages} {893} (\bibinfo {year} {2019})}\BibitemShut
  {NoStop}%
\bibitem [{\citenamefont {Zhang}\ \emph {et~al.}(2025)\citenamefont {Zhang},
  \citenamefont {Nguyen}, \citenamefont {Batra}, \citenamefont {Liu},
  \citenamefont {Watanabe}, \citenamefont {Taniguchi}, \citenamefont
  {Feldman},\ and\ \citenamefont {Li}}]{Zhang2025exciton}%
  \BibitemOpen
  \bibfield  {author} {\bibinfo {author} {\bibfnamefont {N.~J.}\ \bibnamefont
  {Zhang}}, \bibinfo {author} {\bibfnamefont {R.~Q.}\ \bibnamefont {Nguyen}},
  \bibinfo {author} {\bibfnamefont {N.}~\bibnamefont {Batra}}, \bibinfo
  {author} {\bibfnamefont {X.}~\bibnamefont {Liu}}, \bibinfo {author}
  {\bibfnamefont {K.}~\bibnamefont {Watanabe}}, \bibinfo {author}
  {\bibfnamefont {T.}~\bibnamefont {Taniguchi}}, \bibinfo {author}
  {\bibfnamefont {D.}~\bibnamefont {Feldman}}, \ and\ \bibinfo {author}
  {\bibfnamefont {J.}~\bibnamefont {Li}},\ }\href@noop {} {\bibfield  {journal}
  {\bibinfo  {journal} {Nature}\ }\textbf {\bibinfo {volume} {637}},\ \bibinfo
  {pages} {327} (\bibinfo {year} {2025})}\BibitemShut {NoStop}%
\bibitem [{\citenamefont {Nguyen}\ \emph {et~al.}(2025)\citenamefont {Nguyen},
  \citenamefont {Ma}, \citenamefont {Chaturvedi}, \citenamefont {Watanabe},
  \citenamefont {Taniguchi}, \citenamefont {Shan},\ and\ \citenamefont
  {Mak}}]{nguyen_perfectdrag2025}%
  \BibitemOpen
  \bibfield  {author} {\bibinfo {author} {\bibfnamefont {P.~X.}\ \bibnamefont
  {Nguyen}}, \bibinfo {author} {\bibfnamefont {L.}~\bibnamefont {Ma}}, \bibinfo
  {author} {\bibfnamefont {R.}~\bibnamefont {Chaturvedi}}, \bibinfo {author}
  {\bibfnamefont {K.}~\bibnamefont {Watanabe}}, \bibinfo {author}
  {\bibfnamefont {T.}~\bibnamefont {Taniguchi}}, \bibinfo {author}
  {\bibfnamefont {J.}~\bibnamefont {Shan}}, \ and\ \bibinfo {author}
  {\bibfnamefont {K.~F.}\ \bibnamefont {Mak}},\ }\href {\doibase
  10.1126/science.adl1829} {\bibfield  {journal} {\bibinfo  {journal}
  {Science}\ }\textbf {\bibinfo {volume} {388}},\ \bibinfo {pages} {274}
  (\bibinfo {year} {2025})}\BibitemShut {NoStop}%
\bibitem [{\citenamefont {Qi}\ \emph {et~al.}(2025)\citenamefont {Qi},
  \citenamefont {Joe}, \citenamefont {Zhang}, \citenamefont {Xie},
  \citenamefont {Feng}, \citenamefont {Lu}, \citenamefont {Wang}, \citenamefont
  {Taniguchi}, \citenamefont {Watanabe}, \citenamefont {Tongay},\ and\
  \citenamefont {Wang}}]{Qi_perfectdrag2025}%
  \BibitemOpen
  \bibfield  {author} {\bibinfo {author} {\bibfnamefont {R.}~\bibnamefont
  {Qi}}, \bibinfo {author} {\bibfnamefont {A.~Y.}\ \bibnamefont {Joe}},
  \bibinfo {author} {\bibfnamefont {Z.}~\bibnamefont {Zhang}}, \bibinfo
  {author} {\bibfnamefont {J.}~\bibnamefont {Xie}}, \bibinfo {author}
  {\bibfnamefont {Q.}~\bibnamefont {Feng}}, \bibinfo {author} {\bibfnamefont
  {Z.}~\bibnamefont {Lu}}, \bibinfo {author} {\bibfnamefont {Z.}~\bibnamefont
  {Wang}}, \bibinfo {author} {\bibfnamefont {T.}~\bibnamefont {Taniguchi}},
  \bibinfo {author} {\bibfnamefont {K.}~\bibnamefont {Watanabe}}, \bibinfo
  {author} {\bibfnamefont {S.}~\bibnamefont {Tongay}}, \ and\ \bibinfo {author}
  {\bibfnamefont {F.}~\bibnamefont {Wang}},\ }\href {\doibase
  10.1126/science.adl1839} {\bibfield  {journal} {\bibinfo  {journal}
  {Science}\ }\textbf {\bibinfo {volume} {388}},\ \bibinfo {pages} {278}
  (\bibinfo {year} {2025})}\BibitemShut {NoStop}%
\bibitem [{\citenamefont {Chester}(1970)}]{Chester1970supersolid}%
  \BibitemOpen
  \bibfield  {author} {\bibinfo {author} {\bibfnamefont {G.~V.}\ \bibnamefont
  {Chester}},\ }\href {\doibase 10.1103/PhysRevA.2.256} {\bibfield  {journal}
  {\bibinfo  {journal} {Phys. Rev. A}\ }\textbf {\bibinfo {volume} {2}},\
  \bibinfo {pages} {256} (\bibinfo {year} {1970})}\BibitemShut {NoStop}%
\bibitem [{\citenamefont {Vu}\ and\ \citenamefont
  {Das~Sarma}(2023)}]{Vu2023excitonsolid}%
  \BibitemOpen
  \bibfield  {author} {\bibinfo {author} {\bibfnamefont {D.}~\bibnamefont
  {Vu}}\ and\ \bibinfo {author} {\bibfnamefont {S.}~\bibnamefont {Das~Sarma}},\
  }\href {\doibase 10.1103/PhysRevB.108.235158} {\bibfield  {journal} {\bibinfo
   {journal} {Phys. Rev. B}\ }\textbf {\bibinfo {volume} {108}},\ \bibinfo
  {pages} {235158} (\bibinfo {year} {2023})}\BibitemShut {NoStop}%
\bibitem [{\citenamefont {Hu}\ and\ \citenamefont
  {Yang}(2024)}]{Hu2024excitonsolid}%
  \BibitemOpen
  \bibfield  {author} {\bibinfo {author} {\bibfnamefont {Z.}~\bibnamefont
  {Hu}}\ and\ \bibinfo {author} {\bibfnamefont {K.}~\bibnamefont {Yang}},\
  }\href {\doibase 10.1103/PhysRevB.110.195307} {\bibfield  {journal} {\bibinfo
   {journal} {Phys. Rev. B}\ }\textbf {\bibinfo {volume} {110}},\ \bibinfo
  {pages} {195307} (\bibinfo {year} {2024})}\BibitemShut {NoStop}%
\bibitem [{\citenamefont {Chui}\ \emph {et~al.}(2020)\citenamefont {Chui},
  \citenamefont {Wang},\ and\ \citenamefont {Wan}}]{Chui2020excitonsolid}%
  \BibitemOpen
  \bibfield  {author} {\bibinfo {author} {\bibfnamefont {S.}~\bibnamefont
  {Chui}}, \bibinfo {author} {\bibfnamefont {N.}~\bibnamefont {Wang}}, \ and\
  \bibinfo {author} {\bibfnamefont {C.~Y.}\ \bibnamefont {Wan}},\ }\href@noop
  {} {\bibfield  {journal} {\bibinfo  {journal} {Physical Review B}\ }\textbf
  {\bibinfo {volume} {102}},\ \bibinfo {pages} {125420} (\bibinfo {year}
  {2020})}\BibitemShut {NoStop}%
\bibitem [{\citenamefont {Yoshioka}\ and\ \citenamefont
  {MacDonald}(1990)}]{Yoshioka1990excitonsolid}%
  \BibitemOpen
  \bibfield  {author} {\bibinfo {author} {\bibfnamefont {D.}~\bibnamefont
  {Yoshioka}}\ and\ \bibinfo {author} {\bibfnamefont {A.~H.}\ \bibnamefont
  {MacDonald}},\ }\href@noop {} {\bibfield  {journal} {\bibinfo  {journal}
  {Journal of the Physical Society of Japan}\ }\textbf {\bibinfo {volume}
  {59}},\ \bibinfo {pages} {4211} (\bibinfo {year} {1990})}\BibitemShut
  {NoStop}%
\bibitem [{\citenamefont {Joglekar}\ \emph {et~al.}(2006)\citenamefont
  {Joglekar}, \citenamefont {Balatsky},\ and\ \citenamefont
  {Sarma}}]{Joglekar2006excitonsolid}%
  \BibitemOpen
  \bibfield  {author} {\bibinfo {author} {\bibfnamefont {Y.~N.}\ \bibnamefont
  {Joglekar}}, \bibinfo {author} {\bibfnamefont {A.~V.}\ \bibnamefont
  {Balatsky}}, \ and\ \bibinfo {author} {\bibfnamefont {S.~D.}\ \bibnamefont
  {Sarma}},\ }\href@noop {} {\bibfield  {journal} {\bibinfo  {journal}
  {Physical Review B}\ }\textbf {\bibinfo {volume} {74}},\ \bibinfo {pages}
  {233302} (\bibinfo {year} {2006})}\BibitemShut {NoStop}%
\bibitem [{\citenamefont {Zarenia}\ \emph {et~al.}(2017)\citenamefont
  {Zarenia}, \citenamefont {Neilson},\ and\ \citenamefont
  {Peeters}}]{Zarenia2017excitonsolid}%
  \BibitemOpen
  \bibfield  {author} {\bibinfo {author} {\bibfnamefont {M.}~\bibnamefont
  {Zarenia}}, \bibinfo {author} {\bibfnamefont {D.}~\bibnamefont {Neilson}}, \
  and\ \bibinfo {author} {\bibfnamefont {F.}~\bibnamefont {Peeters}},\
  }\href@noop {} {\bibfield  {journal} {\bibinfo  {journal} {Scientific
  reports}\ }\textbf {\bibinfo {volume} {7}},\ \bibinfo {pages} {1} (\bibinfo
  {year} {2017})}\BibitemShut {NoStop}%
\bibitem [{\citenamefont {De~Palo}\ \emph {et~al.}(2002)\citenamefont
  {De~Palo}, \citenamefont {Rapisarda},\ and\ \citenamefont
  {Senatore}}]{De2002excitonsolid}%
  \BibitemOpen
  \bibfield  {author} {\bibinfo {author} {\bibfnamefont {S.}~\bibnamefont
  {De~Palo}}, \bibinfo {author} {\bibfnamefont {F.}~\bibnamefont {Rapisarda}},
  \ and\ \bibinfo {author} {\bibfnamefont {G.}~\bibnamefont {Senatore}},\
  }\href@noop {} {\bibfield  {journal} {\bibinfo  {journal} {Physical review
  letters}\ }\textbf {\bibinfo {volume} {88}},\ \bibinfo {pages} {206401}
  (\bibinfo {year} {2002})}\BibitemShut {NoStop}%
\bibitem [{\citenamefont {Chen}\ and\ \citenamefont
  {Quinn}(1991)}]{Chen1991excitonsolid}%
  \BibitemOpen
  \bibfield  {author} {\bibinfo {author} {\bibfnamefont {X.}~\bibnamefont
  {Chen}}\ and\ \bibinfo {author} {\bibfnamefont {J.}~\bibnamefont {Quinn}},\
  }\href@noop {} {\bibfield  {journal} {\bibinfo  {journal} {Physical review
  letters}\ }\textbf {\bibinfo {volume} {67}},\ \bibinfo {pages} {895}
  (\bibinfo {year} {1991})}\BibitemShut {NoStop}%
\bibitem [{\citenamefont {Yang}(2001)}]{Yang2001excitonsolid}%
  \BibitemOpen
  \bibfield  {author} {\bibinfo {author} {\bibfnamefont {K.}~\bibnamefont
  {Yang}},\ }\href@noop {} {\bibfield  {journal} {\bibinfo  {journal} {Physical
  Review Letters}\ }\textbf {\bibinfo {volume} {87}},\ \bibinfo {pages}
  {056802} (\bibinfo {year} {2001})}\BibitemShut {NoStop}%
\bibitem [{\citenamefont {B\"oning}\ \emph {et~al.}(2011)\citenamefont
  {B\"oning}, \citenamefont {Filinov},\ and\ \citenamefont
  {Bonitz}}]{Boning2011supersolid}%
  \BibitemOpen
  \bibfield  {author} {\bibinfo {author} {\bibfnamefont {J.}~\bibnamefont
  {B\"oning}}, \bibinfo {author} {\bibfnamefont {A.}~\bibnamefont {Filinov}}, \
  and\ \bibinfo {author} {\bibfnamefont {M.}~\bibnamefont {Bonitz}},\ }\href
  {\doibase 10.1103/PhysRevB.84.075130} {\bibfield  {journal} {\bibinfo
  {journal} {Phys. Rev. B}\ }\textbf {\bibinfo {volume} {84}},\ \bibinfo
  {pages} {075130} (\bibinfo {year} {2011})}\BibitemShut {NoStop}%
\bibitem [{\citenamefont {Szyma\ifmmode~\acute{n}\else \'{n}\fi{}ski}\ \emph
  {et~al.}(1994)\citenamefont {Szyma\ifmmode~\acute{n}\else \'{n}\fi{}ski},
  \citenamefont {\ifmmode~\acute{S}\else \'{S}\fi{}wierkowski},\ and\
  \citenamefont {Neilson}}]{Szymanski1994wigner}%
  \BibitemOpen
  \bibfield  {author} {\bibinfo {author} {\bibfnamefont {J.}~\bibnamefont
  {Szyma\ifmmode~\acute{n}\else \'{n}\fi{}ski}}, \bibinfo {author}
  {\bibfnamefont {L.}~\bibnamefont {\ifmmode~\acute{S}\else
  \'{S}\fi{}wierkowski}}, \ and\ \bibinfo {author} {\bibfnamefont
  {D.}~\bibnamefont {Neilson}},\ }\href {\doibase 10.1103/PhysRevB.50.11002}
  {\bibfield  {journal} {\bibinfo  {journal} {Phys. Rev. B}\ }\textbf {\bibinfo
  {volume} {50}},\ \bibinfo {pages} {11002} (\bibinfo {year}
  {1994})}\BibitemShut {NoStop}%
\bibitem [{\citenamefont {Astrakharchik}\ \emph {et~al.}(2007)\citenamefont
  {Astrakharchik}, \citenamefont {Boronat}, \citenamefont {Kurbakov},\ and\
  \citenamefont {Lozovik}}]{Astrakharchik2007dipole}%
  \BibitemOpen
  \bibfield  {author} {\bibinfo {author} {\bibfnamefont {G.~E.}\ \bibnamefont
  {Astrakharchik}}, \bibinfo {author} {\bibfnamefont {J.}~\bibnamefont
  {Boronat}}, \bibinfo {author} {\bibfnamefont {I.~L.}\ \bibnamefont
  {Kurbakov}}, \ and\ \bibinfo {author} {\bibfnamefont {Y.~E.}\ \bibnamefont
  {Lozovik}},\ }\href {\doibase 10.1103/PhysRevLett.98.060405} {\bibfield
  {journal} {\bibinfo  {journal} {Phys. Rev. Lett.}\ }\textbf {\bibinfo
  {volume} {98}},\ \bibinfo {pages} {060405} (\bibinfo {year}
  {2007})}\BibitemShut {NoStop}%
\bibitem [{\citenamefont {Champagne}\ \emph {et~al.}(2008)\citenamefont
  {Champagne}, \citenamefont {Finck}, \citenamefont {Eisenstein}, \citenamefont
  {Pfeiffer},\ and\ \citenamefont {West}}]{Champagne.08b}%
  \BibitemOpen
  \bibfield  {author} {\bibinfo {author} {\bibfnamefont {A.~R.}\ \bibnamefont
  {Champagne}}, \bibinfo {author} {\bibfnamefont {A.~D.~K.}\ \bibnamefont
  {Finck}}, \bibinfo {author} {\bibfnamefont {J.~P.}\ \bibnamefont
  {Eisenstein}}, \bibinfo {author} {\bibfnamefont {L.~N.}\ \bibnamefont
  {Pfeiffer}}, \ and\ \bibinfo {author} {\bibfnamefont {K.~W.}\ \bibnamefont
  {West}},\ }\href {\doibase 10.1103/PhysRevB.78.205310} {\bibfield  {journal}
  {\bibinfo  {journal} {Phys. Rev. B}\ }\textbf {\bibinfo {volume} {78}},\
  \bibinfo {pages} {205310} (\bibinfo {year} {2008})}\BibitemShut {NoStop}%
\bibitem [{\citenamefont {Clarke}\ \emph {et~al.}(2005)\citenamefont {Clarke},
  \citenamefont {Micolich}, \citenamefont {Hamilton}, \citenamefont {Simmons},
  \citenamefont {Hanna}, \citenamefont {Rodriguez}, \citenamefont {Pepper},\
  and\ \citenamefont {Ritchie}}]{Clarke2005imbalance}%
  \BibitemOpen
  \bibfield  {author} {\bibinfo {author} {\bibfnamefont {W.~R.}\ \bibnamefont
  {Clarke}}, \bibinfo {author} {\bibfnamefont {A.~P.}\ \bibnamefont
  {Micolich}}, \bibinfo {author} {\bibfnamefont {A.~R.}\ \bibnamefont
  {Hamilton}}, \bibinfo {author} {\bibfnamefont {M.~Y.}\ \bibnamefont
  {Simmons}}, \bibinfo {author} {\bibfnamefont {C.~B.}\ \bibnamefont {Hanna}},
  \bibinfo {author} {\bibfnamefont {J.~R.}\ \bibnamefont {Rodriguez}}, \bibinfo
  {author} {\bibfnamefont {M.}~\bibnamefont {Pepper}}, \ and\ \bibinfo {author}
  {\bibfnamefont {D.~A.}\ \bibnamefont {Ritchie}},\ }\href {\doibase
  10.1103/PhysRevB.71.081304} {\bibfield  {journal} {\bibinfo  {journal} {Phys.
  Rev. B}\ }\textbf {\bibinfo {volume} {71}},\ \bibinfo {pages} {081304}
  (\bibinfo {year} {2005})}\BibitemShut {NoStop}%
\bibitem [{\citenamefont {Joglekar}\ and\ \citenamefont
  {MacDonald}(2002)}]{Jog.02}%
  \BibitemOpen
  \bibfield  {author} {\bibinfo {author} {\bibfnamefont {Y.~N.}\ \bibnamefont
  {Joglekar}}\ and\ \bibinfo {author} {\bibfnamefont {A.~H.}\ \bibnamefont
  {MacDonald}},\ }\href {\doibase 10.1103/PhysRevB.65.235319} {\bibfield
  {journal} {\bibinfo  {journal} {Phys. Rev. B}\ }\textbf {\bibinfo {volume}
  {65}},\ \bibinfo {pages} {235319} (\bibinfo {year} {2002})}\BibitemShut
  {NoStop}%
\bibitem [{\citenamefont {Nguyen}\ \emph {et~al.}(2024)\citenamefont {Nguyen},
  \citenamefont {Zhang}, \citenamefont {Batra}, \citenamefont {Liu},
  \citenamefont {Watanabe}, \citenamefont {Taniguchi}, \citenamefont
  {Feldman},\ and\ \citenamefont {Li}}]{Nguyen2024fractionalexciton}%
  \BibitemOpen
  \bibfield  {author} {\bibinfo {author} {\bibfnamefont {R.~Q.}\ \bibnamefont
  {Nguyen}}, \bibinfo {author} {\bibfnamefont {N.~J.}\ \bibnamefont {Zhang}},
  \bibinfo {author} {\bibfnamefont {N.}~\bibnamefont {Batra}}, \bibinfo
  {author} {\bibfnamefont {X.}~\bibnamefont {Liu}}, \bibinfo {author}
  {\bibfnamefont {K.}~\bibnamefont {Watanabe}}, \bibinfo {author}
  {\bibfnamefont {T.}~\bibnamefont {Taniguchi}}, \bibinfo {author}
  {\bibfnamefont {D.}~\bibnamefont {Feldman}}, \ and\ \bibinfo {author}
  {\bibfnamefont {J.}~\bibnamefont {Li}},\ }\href@noop {} {\bibfield  {journal}
  {\bibinfo  {journal} {arXiv preprint arXiv:2410.24208}\ } (\bibinfo {year}
  {2024})}\BibitemShut {NoStop}%
\bibitem [{\citenamefont {Lozovik}\ \emph {et~al.}(2012)\citenamefont
  {Lozovik}, \citenamefont {Ogarkov},\ and\ \citenamefont
  {Sokolik}}]{Lozovik2012}%
  \BibitemOpen
  \bibfield  {author} {\bibinfo {author} {\bibfnamefont {Y.~E.}\ \bibnamefont
  {Lozovik}}, \bibinfo {author} {\bibfnamefont {S.~L.}\ \bibnamefont
  {Ogarkov}}, \ and\ \bibinfo {author} {\bibfnamefont {A.~A.}\ \bibnamefont
  {Sokolik}},\ }\href {\doibase 10.1103/PhysRevB.86.045429} {\bibfield
  {journal} {\bibinfo  {journal} {Phys. Rev. B}\ }\textbf {\bibinfo {volume}
  {86}},\ \bibinfo {pages} {045429} (\bibinfo {year} {2012})}\BibitemShut
  {NoStop}%
\bibitem [{\citenamefont {Perali}\ \emph {et~al.}(2013)\citenamefont {Perali},
  \citenamefont {Neilson},\ and\ \citenamefont {Hamilton}}]{Perali2013}%
  \BibitemOpen
  \bibfield  {author} {\bibinfo {author} {\bibfnamefont {A.}~\bibnamefont
  {Perali}}, \bibinfo {author} {\bibfnamefont {D.}~\bibnamefont {Neilson}}, \
  and\ \bibinfo {author} {\bibfnamefont {A.~R.}\ \bibnamefont {Hamilton}},\
  }\href {\doibase 10.1103/PhysRevLett.110.146803} {\bibfield  {journal}
  {\bibinfo  {journal} {Phys. Rev. Lett.}\ }\textbf {\bibinfo {volume} {110}},\
  \bibinfo {pages} {146803} (\bibinfo {year} {2013})}\BibitemShut {NoStop}%
\bibitem [{\citenamefont {Eisenstein}\ and\ \citenamefont
  {MacDonald}(2004)}]{Eis.04}%
  \BibitemOpen
  \bibfield  {author} {\bibinfo {author} {\bibfnamefont {J.~P.}\ \bibnamefont
  {Eisenstein}}\ and\ \bibinfo {author} {\bibfnamefont {A.~H.}\ \bibnamefont
  {MacDonald}},\ }\href {\doibase doi:10.1038/nature03081} {\bibfield
  {journal} {\bibinfo  {journal} {Nature}\ }\textbf {\bibinfo {volume} {432}},\
  \bibinfo {pages} {691} (\bibinfo {year} {2004})}\BibitemShut {NoStop}%
\bibitem [{\citenamefont {Lozovik}\ \emph {et~al.}(2004)\citenamefont
  {Lozovik}, \citenamefont {Volkov},\ and\ \citenamefont
  {Willander}}]{Lozovik2004excitonmelting}%
  \BibitemOpen
  \bibfield  {author} {\bibinfo {author} {\bibfnamefont {Y.~E.}\ \bibnamefont
  {Lozovik}}, \bibinfo {author} {\bibfnamefont {S.~Y.}\ \bibnamefont {Volkov}},
  \ and\ \bibinfo {author} {\bibfnamefont {M.}~\bibnamefont {Willander}},\
  }\href@noop {} {\bibfield  {journal} {\bibinfo  {journal} {Journal of
  Experimental and Theoretical Physics Letters}\ }\textbf {\bibinfo {volume}
  {79}},\ \bibinfo {pages} {473} (\bibinfo {year} {2004})}\BibitemShut
  {NoStop}%
\bibitem [{\citenamefont {Mitra}\ \emph {et~al.}(2009)\citenamefont {Mitra},
  \citenamefont {Williams},\ and\ \citenamefont {de~Melo}}]{Mitra2009hexatic}%
  \BibitemOpen
  \bibfield  {author} {\bibinfo {author} {\bibfnamefont {K.}~\bibnamefont
  {Mitra}}, \bibinfo {author} {\bibfnamefont {C.}~\bibnamefont {Williams}}, \
  and\ \bibinfo {author} {\bibfnamefont {C.}~\bibnamefont {de~Melo}},\
  }\href@noop {} {\bibfield  {journal} {\bibinfo  {journal} {arXiv preprint
  arXiv:0903.4655}\ } (\bibinfo {year} {2009})}\BibitemShut {NoStop}%
\bibitem [{\citenamefont {Zhou}\ \emph {et~al.}(2021)\citenamefont {Zhou},
  \citenamefont {Sung}, \citenamefont {Brutschea}, \citenamefont {Esterlis},
  \citenamefont {Wang}, \citenamefont {Scuri}, \citenamefont {Gelly},
  \citenamefont {Heo}, \citenamefont {Taniguchi}, \citenamefont {Watanabe}
  \emph {et~al.}}]{Zhou2021bilayerWigner}%
  \BibitemOpen
  \bibfield  {author} {\bibinfo {author} {\bibfnamefont {Y.}~\bibnamefont
  {Zhou}}, \bibinfo {author} {\bibfnamefont {J.}~\bibnamefont {Sung}}, \bibinfo
  {author} {\bibfnamefont {E.}~\bibnamefont {Brutschea}}, \bibinfo {author}
  {\bibfnamefont {I.}~\bibnamefont {Esterlis}}, \bibinfo {author}
  {\bibfnamefont {Y.}~\bibnamefont {Wang}}, \bibinfo {author} {\bibfnamefont
  {G.}~\bibnamefont {Scuri}}, \bibinfo {author} {\bibfnamefont {R.~J.}\
  \bibnamefont {Gelly}}, \bibinfo {author} {\bibfnamefont {H.}~\bibnamefont
  {Heo}}, \bibinfo {author} {\bibfnamefont {T.}~\bibnamefont {Taniguchi}},
  \bibinfo {author} {\bibfnamefont {K.}~\bibnamefont {Watanabe}},  \emph
  {et~al.},\ }\href@noop {} {\bibfield  {journal} {\bibinfo  {journal}
  {Nature}\ }\textbf {\bibinfo {volume} {595}},\ \bibinfo {pages} {48}
  (\bibinfo {year} {2021})}\BibitemShut {NoStop}%
\bibitem [{\citenamefont {Zeng}\ \emph {et~al.}(2023)\citenamefont {Zeng},
  \citenamefont {Xia}, \citenamefont {Dery}, \citenamefont {Watanabe},
  \citenamefont {Taniguchi}, \citenamefont {Shan},\ and\ \citenamefont
  {Mak}}]{Zeng2023excitonDW}%
  \BibitemOpen
  \bibfield  {author} {\bibinfo {author} {\bibfnamefont {Y.}~\bibnamefont
  {Zeng}}, \bibinfo {author} {\bibfnamefont {Z.}~\bibnamefont {Xia}}, \bibinfo
  {author} {\bibfnamefont {R.}~\bibnamefont {Dery}}, \bibinfo {author}
  {\bibfnamefont {K.}~\bibnamefont {Watanabe}}, \bibinfo {author}
  {\bibfnamefont {T.}~\bibnamefont {Taniguchi}}, \bibinfo {author}
  {\bibfnamefont {J.}~\bibnamefont {Shan}}, \ and\ \bibinfo {author}
  {\bibfnamefont {K.~F.}\ \bibnamefont {Mak}},\ }\href@noop {} {\bibfield
  {journal} {\bibinfo  {journal} {Nature Materials}\ ,\ \bibinfo {pages} {1}}
  (\bibinfo {year} {2023})}\BibitemShut {NoStop}%
\bibitem [{\citenamefont {Abergel}\ \emph {et~al.}(2013)\citenamefont
  {Abergel}, \citenamefont {Rodriguez-Vega}, \citenamefont {Rossi},\ and\
  \citenamefont {Das~Sarma}}]{Abergel2013exciton}%
  \BibitemOpen
  \bibfield  {author} {\bibinfo {author} {\bibfnamefont {D.~S.~L.}\
  \bibnamefont {Abergel}}, \bibinfo {author} {\bibfnamefont {M.}~\bibnamefont
  {Rodriguez-Vega}}, \bibinfo {author} {\bibfnamefont {E.}~\bibnamefont
  {Rossi}}, \ and\ \bibinfo {author} {\bibfnamefont {S.}~\bibnamefont
  {Das~Sarma}},\ }\href {\doibase 10.1103/PhysRevB.88.235402} {\bibfield
  {journal} {\bibinfo  {journal} {Phys. Rev. B}\ }\textbf {\bibinfo {volume}
  {88}},\ \bibinfo {pages} {235402} (\bibinfo {year} {2013})}\BibitemShut
  {NoStop}%
\bibitem [{\citenamefont {Andrei}\ \emph {et~al.}(1988)\citenamefont {Andrei},
  \citenamefont {Deville}, \citenamefont {Glattli}, \citenamefont {Williams},
  \citenamefont {Paris},\ and\ \citenamefont {Etienne}}]{Andrei1988Wigner}%
  \BibitemOpen
  \bibfield  {author} {\bibinfo {author} {\bibfnamefont {E.}~\bibnamefont
  {Andrei}}, \bibinfo {author} {\bibfnamefont {G.}~\bibnamefont {Deville}},
  \bibinfo {author} {\bibfnamefont {D.}~\bibnamefont {Glattli}}, \bibinfo
  {author} {\bibfnamefont {F.}~\bibnamefont {Williams}}, \bibinfo {author}
  {\bibfnamefont {E.}~\bibnamefont {Paris}}, \ and\ \bibinfo {author}
  {\bibfnamefont {B.}~\bibnamefont {Etienne}},\ }\href@noop {} {\bibfield
  {journal} {\bibinfo  {journal} {Physical review letters}\ }\textbf {\bibinfo
  {volume} {60}},\ \bibinfo {pages} {2765} (\bibinfo {year}
  {1988})}\BibitemShut {NoStop}%
\bibitem [{\citenamefont {Jiang}\ \emph {et~al.}(1990)\citenamefont {Jiang},
  \citenamefont {Willett}, \citenamefont {Stormer}, \citenamefont {Tsui},
  \citenamefont {Pfeiffer},\ and\ \citenamefont {West}}]{Jiang1990Wigner}%
  \BibitemOpen
  \bibfield  {author} {\bibinfo {author} {\bibfnamefont {H.~W.}\ \bibnamefont
  {Jiang}}, \bibinfo {author} {\bibfnamefont {R.~L.}\ \bibnamefont {Willett}},
  \bibinfo {author} {\bibfnamefont {H.~L.}\ \bibnamefont {Stormer}}, \bibinfo
  {author} {\bibfnamefont {D.~C.}\ \bibnamefont {Tsui}}, \bibinfo {author}
  {\bibfnamefont {L.~N.}\ \bibnamefont {Pfeiffer}}, \ and\ \bibinfo {author}
  {\bibfnamefont {K.~W.}\ \bibnamefont {West}},\ }\href {\doibase
  10.1103/PhysRevLett.65.633} {\bibfield  {journal} {\bibinfo  {journal} {Phys.
  Rev. Lett.}\ }\textbf {\bibinfo {volume} {65}},\ \bibinfo {pages} {633}
  (\bibinfo {year} {1990})}\BibitemShut {NoStop}%
\bibitem [{\citenamefont {Ma}\ \emph {et~al.}(2020)\citenamefont {Ma},
  \citenamefont {Rosales}, \citenamefont {Deng}, \citenamefont {Chung},
  \citenamefont {Pfeiffer}, \citenamefont {West}, \citenamefont {Baldwin},
  \citenamefont {Winkler},\ and\ \citenamefont {Shayegan}}]{Ma2020Wigner}%
  \BibitemOpen
  \bibfield  {author} {\bibinfo {author} {\bibfnamefont {M.~K.}\ \bibnamefont
  {Ma}}, \bibinfo {author} {\bibfnamefont {K.~V.}\ \bibnamefont {Rosales}},
  \bibinfo {author} {\bibfnamefont {H.}~\bibnamefont {Deng}}, \bibinfo {author}
  {\bibfnamefont {Y.}~\bibnamefont {Chung}}, \bibinfo {author} {\bibfnamefont
  {L.}~\bibnamefont {Pfeiffer}}, \bibinfo {author} {\bibfnamefont
  {K.}~\bibnamefont {West}}, \bibinfo {author} {\bibfnamefont {K.}~\bibnamefont
  {Baldwin}}, \bibinfo {author} {\bibfnamefont {R.}~\bibnamefont {Winkler}}, \
  and\ \bibinfo {author} {\bibfnamefont {M.}~\bibnamefont {Shayegan}},\
  }\href@noop {} {\bibfield  {journal} {\bibinfo  {journal} {Physical review
  letters}\ }\textbf {\bibinfo {volume} {125}},\ \bibinfo {pages} {036601}
  (\bibinfo {year} {2020})}\BibitemShut {NoStop}%
\bibitem [{\citenamefont {Gervais}\ \emph {et~al.}(2004)\citenamefont
  {Gervais}, \citenamefont {Engel}, \citenamefont {Stormer}, \citenamefont
  {Tsui}, \citenamefont {Baldwin}, \citenamefont {West},\ and\ \citenamefont
  {Pfeiffer}}]{Gervais2004Wigner}%
  \BibitemOpen
  \bibfield  {author} {\bibinfo {author} {\bibfnamefont {G.}~\bibnamefont
  {Gervais}}, \bibinfo {author} {\bibfnamefont {L.~W.}\ \bibnamefont {Engel}},
  \bibinfo {author} {\bibfnamefont {H.~L.}\ \bibnamefont {Stormer}}, \bibinfo
  {author} {\bibfnamefont {D.~C.}\ \bibnamefont {Tsui}}, \bibinfo {author}
  {\bibfnamefont {K.~W.}\ \bibnamefont {Baldwin}}, \bibinfo {author}
  {\bibfnamefont {K.~W.}\ \bibnamefont {West}}, \ and\ \bibinfo {author}
  {\bibfnamefont {L.~N.}\ \bibnamefont {Pfeiffer}},\ }\href {\doibase
  10.1103/PhysRevLett.93.266804} {\bibfield  {journal} {\bibinfo  {journal}
  {Phys. Rev. Lett.}\ }\textbf {\bibinfo {volume} {93}},\ \bibinfo {pages}
  {266804} (\bibinfo {year} {2004})}\BibitemShut {NoStop}%
\bibitem [{\citenamefont {Goldman}\ \emph {et~al.}(1990)\citenamefont
  {Goldman}, \citenamefont {Santos}, \citenamefont {Shayegan},\ and\
  \citenamefont {Cunningham}}]{Goldman1990Wigner}%
  \BibitemOpen
  \bibfield  {author} {\bibinfo {author} {\bibfnamefont {V.~J.}\ \bibnamefont
  {Goldman}}, \bibinfo {author} {\bibfnamefont {M.}~\bibnamefont {Santos}},
  \bibinfo {author} {\bibfnamefont {M.}~\bibnamefont {Shayegan}}, \ and\
  \bibinfo {author} {\bibfnamefont {J.~E.}\ \bibnamefont {Cunningham}},\ }\href
  {\doibase 10.1103/PhysRevLett.65.2189} {\bibfield  {journal} {\bibinfo
  {journal} {Phys. Rev. Lett.}\ }\textbf {\bibinfo {volume} {65}},\ \bibinfo
  {pages} {2189} (\bibinfo {year} {1990})}\BibitemShut {NoStop}%
\bibitem [{\citenamefont {Tsui}\ \emph {et~al.}(2024)\citenamefont {Tsui},
  \citenamefont {He}, \citenamefont {Hu}, \citenamefont {Lake}, \citenamefont
  {Wang}, \citenamefont {Watanabe}, \citenamefont {Taniguchi}, \citenamefont
  {Zaletel},\ and\ \citenamefont {Yazdani}}]{tsui_wignercrystal_stm_2024}%
  \BibitemOpen
  \bibfield  {author} {\bibinfo {author} {\bibfnamefont {Y.-C.}\ \bibnamefont
  {Tsui}}, \bibinfo {author} {\bibfnamefont {M.}~\bibnamefont {He}}, \bibinfo
  {author} {\bibfnamefont {Y.}~\bibnamefont {Hu}}, \bibinfo {author}
  {\bibfnamefont {E.}~\bibnamefont {Lake}}, \bibinfo {author} {\bibfnamefont
  {T.}~\bibnamefont {Wang}}, \bibinfo {author} {\bibfnamefont {K.}~\bibnamefont
  {Watanabe}}, \bibinfo {author} {\bibfnamefont {T.}~\bibnamefont {Taniguchi}},
  \bibinfo {author} {\bibfnamefont {M.~P.}\ \bibnamefont {Zaletel}}, \ and\
  \bibinfo {author} {\bibfnamefont {A.}~\bibnamefont {Yazdani}},\ }\href
  {\doibase 10.1038/s41586-024-07212-7} {\bibfield  {journal} {\bibinfo
  {journal} {Nature}\ }\textbf {\bibinfo {volume} {628}},\ \bibinfo {pages}
  {287} (\bibinfo {year} {2024})}\BibitemShut {NoStop}%
\bibitem [{\citenamefont {Goerbig}\ and\ \citenamefont
  {Smith}(2003)}]{Goerbig2003scaling}%
  \BibitemOpen
  \bibfield  {author} {\bibinfo {author} {\bibfnamefont {M.}~\bibnamefont
  {Goerbig}}\ and\ \bibinfo {author} {\bibfnamefont {C.~M.}\ \bibnamefont
  {Smith}},\ }\href@noop {} {\bibfield  {journal} {\bibinfo  {journal} {EPL
  (Europhysics Letters)}\ }\textbf {\bibinfo {volume} {63}},\ \bibinfo {pages}
  {736} (\bibinfo {year} {2003})}\BibitemShut {NoStop}%
\bibitem [{\citenamefont {Kim}\ and\ \citenamefont {Chan}(2004)}]{Kim2004He4}%
  \BibitemOpen
  \bibfield  {author} {\bibinfo {author} {\bibfnamefont {E.}~\bibnamefont
  {Kim}}\ and\ \bibinfo {author} {\bibfnamefont {M.~H.}\ \bibnamefont {Chan}},\
  }\href@noop {} {\bibfield  {journal} {\bibinfo  {journal} {Science}\ }\textbf
  {\bibinfo {volume} {305}},\ \bibinfo {pages} {1941} (\bibinfo {year}
  {2004})}\BibitemShut {NoStop}%
\bibitem [{\citenamefont {Day}\ and\ \citenamefont
  {Beamish}(2007)}]{Day2007He4}%
  \BibitemOpen
  \bibfield  {author} {\bibinfo {author} {\bibfnamefont {J.}~\bibnamefont
  {Day}}\ and\ \bibinfo {author} {\bibfnamefont {J.}~\bibnamefont {Beamish}},\
  }\href@noop {} {\bibfield  {journal} {\bibinfo  {journal} {Nature}\ }\textbf
  {\bibinfo {volume} {450}},\ \bibinfo {pages} {853} (\bibinfo {year}
  {2007})}\BibitemShut {NoStop}%
\bibitem [{\citenamefont {Hunt}\ \emph {et~al.}(2009)\citenamefont {Hunt},
  \citenamefont {Pratt}, \citenamefont {Gadagkar}, \citenamefont {Yamashita},
  \citenamefont {Balatsky},\ and\ \citenamefont {Davis}}]{Hunt2009He4}%
  \BibitemOpen
  \bibfield  {author} {\bibinfo {author} {\bibfnamefont {B.}~\bibnamefont
  {Hunt}}, \bibinfo {author} {\bibfnamefont {E.}~\bibnamefont {Pratt}},
  \bibinfo {author} {\bibfnamefont {V.}~\bibnamefont {Gadagkar}}, \bibinfo
  {author} {\bibfnamefont {M.}~\bibnamefont {Yamashita}}, \bibinfo {author}
  {\bibfnamefont {A.~V.}\ \bibnamefont {Balatsky}}, \ and\ \bibinfo {author}
  {\bibfnamefont {J.}~\bibnamefont {Davis}},\ }\href@noop {} {\bibfield
  {journal} {\bibinfo  {journal} {Science}\ }\textbf {\bibinfo {volume}
  {324}},\ \bibinfo {pages} {632} (\bibinfo {year} {2009})}\BibitemShut
  {NoStop}%
\bibitem [{\citenamefont {Wang}\ \emph {et~al.}(2019)\citenamefont {Wang},
  \citenamefont {Rhodes}, \citenamefont {Watanabe}, \citenamefont {Taniguchi},
  \citenamefont {Hone}, \citenamefont {Shan},\ and\ \citenamefont
  {Mak}}]{Wang2019excitonB0}%
  \BibitemOpen
  \bibfield  {author} {\bibinfo {author} {\bibfnamefont {Z.}~\bibnamefont
  {Wang}}, \bibinfo {author} {\bibfnamefont {D.~A.}\ \bibnamefont {Rhodes}},
  \bibinfo {author} {\bibfnamefont {K.}~\bibnamefont {Watanabe}}, \bibinfo
  {author} {\bibfnamefont {T.}~\bibnamefont {Taniguchi}}, \bibinfo {author}
  {\bibfnamefont {J.~C.}\ \bibnamefont {Hone}}, \bibinfo {author}
  {\bibfnamefont {J.}~\bibnamefont {Shan}}, \ and\ \bibinfo {author}
  {\bibfnamefont {K.~F.}\ \bibnamefont {Mak}},\ }\href@noop {} {\bibfield
  {journal} {\bibinfo  {journal} {Nature}\ }\textbf {\bibinfo {volume} {574}},\
  \bibinfo {pages} {76} (\bibinfo {year} {2019})}\BibitemShut {NoStop}%
\bibitem [{\citenamefont {Ma}\ \emph {et~al.}(2021)\citenamefont {Ma},
  \citenamefont {Nguyen}, \citenamefont {Wang}, \citenamefont {Zeng},
  \citenamefont {Watanabe}, \citenamefont {Taniguchi}, \citenamefont
  {MacDonald}, \citenamefont {Mak},\ and\ \citenamefont
  {Shan}}]{Ma2021excitonB0}%
  \BibitemOpen
  \bibfield  {author} {\bibinfo {author} {\bibfnamefont {L.}~\bibnamefont
  {Ma}}, \bibinfo {author} {\bibfnamefont {P.~X.}\ \bibnamefont {Nguyen}},
  \bibinfo {author} {\bibfnamefont {Z.}~\bibnamefont {Wang}}, \bibinfo {author}
  {\bibfnamefont {Y.}~\bibnamefont {Zeng}}, \bibinfo {author} {\bibfnamefont
  {K.}~\bibnamefont {Watanabe}}, \bibinfo {author} {\bibfnamefont
  {T.}~\bibnamefont {Taniguchi}}, \bibinfo {author} {\bibfnamefont {A.~H.}\
  \bibnamefont {MacDonald}}, \bibinfo {author} {\bibfnamefont {K.~F.}\
  \bibnamefont {Mak}}, \ and\ \bibinfo {author} {\bibfnamefont
  {J.}~\bibnamefont {Shan}},\ }\href@noop {} {\bibfield  {journal} {\bibinfo
  {journal} {Nature}\ }\textbf {\bibinfo {volume} {598}},\ \bibinfo {pages}
  {585} (\bibinfo {year} {2021})}\BibitemShut {NoStop}%
\bibitem [{\citenamefont {Fogler}\ \emph {et~al.}(2014)\citenamefont {Fogler},
  \citenamefont {Butov},\ and\ \citenamefont {Novoselov}}]{Fogler2014exciton}%
  \BibitemOpen
  \bibfield  {author} {\bibinfo {author} {\bibfnamefont {M.}~\bibnamefont
  {Fogler}}, \bibinfo {author} {\bibfnamefont {L.}~\bibnamefont {Butov}}, \
  and\ \bibinfo {author} {\bibfnamefont {K.}~\bibnamefont {Novoselov}},\
  }\href@noop {} {\bibfield  {journal} {\bibinfo  {journal} {Nature
  communications}\ }\textbf {\bibinfo {volume} {5}},\ \bibinfo {pages} {4555}
  (\bibinfo {year} {2014})}\BibitemShut {NoStop}%
\bibitem [{\citenamefont {Hatke}\ \emph {et~al.}(2019)\citenamefont {Hatke},
  \citenamefont {Deng}, \citenamefont {Liu}, \citenamefont {Engel},
  \citenamefont {Pfeiffer}, \citenamefont {West}, \citenamefont {Baldwin},\
  and\ \citenamefont {Shayegan}}]{Hatke2019wigner}%
  \BibitemOpen
  \bibfield  {author} {\bibinfo {author} {\bibfnamefont {A.}~\bibnamefont
  {Hatke}}, \bibinfo {author} {\bibfnamefont {H.}~\bibnamefont {Deng}},
  \bibinfo {author} {\bibfnamefont {Y.}~\bibnamefont {Liu}}, \bibinfo {author}
  {\bibfnamefont {L.}~\bibnamefont {Engel}}, \bibinfo {author} {\bibfnamefont
  {L.}~\bibnamefont {Pfeiffer}}, \bibinfo {author} {\bibfnamefont
  {K.}~\bibnamefont {West}}, \bibinfo {author} {\bibfnamefont {K.}~\bibnamefont
  {Baldwin}}, \ and\ \bibinfo {author} {\bibfnamefont {M.}~\bibnamefont
  {Shayegan}},\ }\href@noop {} {\bibfield  {journal} {\bibinfo  {journal}
  {Science Advances}\ }\textbf {\bibinfo {volume} {5}},\ \bibinfo {pages}
  {eaao2848} (\bibinfo {year} {2019})}\BibitemShut {NoStop}%
\bibitem [{\citenamefont {Ahn}\ and\ \citenamefont
  {Das~Sarma}(2023)}]{Ahn2023Wigner}%
  \BibitemOpen
  \bibfield  {author} {\bibinfo {author} {\bibfnamefont {S.}~\bibnamefont
  {Ahn}}\ and\ \bibinfo {author} {\bibfnamefont {S.}~\bibnamefont
  {Das~Sarma}},\ }\href {\doibase 10.1103/PhysRevB.107.195435} {\bibfield
  {journal} {\bibinfo  {journal} {Phys. Rev. B}\ }\textbf {\bibinfo {volume}
  {107}},\ \bibinfo {pages} {195435} (\bibinfo {year} {2023})}\BibitemShut
  {NoStop}%
\bibitem [{\citenamefont {Randeria}\ and\ \citenamefont
  {Taylor}(2014)}]{Randeria2014}%
  \BibitemOpen
  \bibfield  {author} {\bibinfo {author} {\bibfnamefont {M.}~\bibnamefont
  {Randeria}}\ and\ \bibinfo {author} {\bibfnamefont {E.}~\bibnamefont
  {Taylor}},\ }\href {\doibase 10.1146/annurev-conmatphys-031113-133829}
  {\bibfield  {journal} {\bibinfo  {journal} {Annual Review of Condensed Matter
  Physics}\ }\textbf {\bibinfo {volume} {5}},\ \bibinfo {pages} {209} (\bibinfo
  {year} {2014})}\BibitemShut {NoStop}%
\end{thebibliography}%

\newpage

\newpage
\clearpage

\pagebreak
\begin{widetext}
\section{Supplementary Materials}

\begin{center}
\textbf{\large Evidence for a Superfluid-to-insulator Transition of Bilayer Excitons }\\
\vspace{10pt}
Yihang Zeng, Dihao Sun, Q. Shi,  A. Okounkova, K. Watanabe, T. Taniguchi, James Hone, C.R. Dean$^{\dag}$, J.I.A. Li$^{\dag}$\\ 
\vspace{10pt}
$^{\dag}$ Corresponding author. Email: cdean@phys.columbia.edu, jia$\_$li@brown.edu
\end{center}

\noindent\textbf{This PDF file includes:}

\noindent{Supplementary Text}

\noindent{Materials and Methods}




\renewcommand{\thefigure}{S\arabic{figure}}
\setcounter{figure}{0}
\setcounter{equation}{0}

\vspace{10pt}

\begin{table}[h!]
\centering
\renewcommand{\arraystretch}{1.4}
\begin{tabular}{|| c | c | c ||} 
 \hline 
\hspace{2pt} Sample \hspace{2pt} & \hspace{2pt} interlayer separation $d$ (nm) \hspace{2pt} & \hspace{2pt} Geometry \hspace{2pt} \\  
 \hline\hline
 C1 & 7.4 & Corbino \\ 
 \hline C2 & 6.4 & Corbino  \\ 
 \hline C3 & 3.9 & Corbino  \\ 
 \hline C4 & 6.6 & Corbino  \\ 
 \hline H1 & 7.4 & hall bar  \\ 
\hline H2 & 8.0 & hall bar \\ 
\hline C5 & 4.5 & Corbino \\ 
 \hline
\end{tabular}
\caption{{\bf{List of devices}} An extensive list of devices are measured in this work. This table provides important parameters for these devices, including the interlayer separation $d$ as well as the sample geometry. Furthermore, we summarize the sample used for each data set in the following. Fig.~\ref{fig1}i, Fig.~\ref{fig2}a-c, Fig.~\ref{fig3}a-d, Fig.~\ref{FigPF}a-e, and Fig.~\ref{SIhysteresis} are taken from sample C1. Fig.~\ref{fig1}g,h Fig.~\ref{fig2}e-g,Fig.~\ref{fig4}a, Fig.~\ref{SIDT1}, Fig.~\ref{Tc}, Fig.~\ref{FigPairing}, Fig.~\ref{dnuTdrag}, and Fig.~\ref{SIdragmap} are taken from sample H1, Fig.~\ref{fig3}e, Fig.~\ref{fig4}c,e, Fig.~\ref{dnuT}, and Fig.~\ref{FigHysteresis} are taken from sample H2. Fig.~\ref{fig4}d and Fig.~\ref{IVsolid} are taken from sample C5, which is the same sample used in Ref.~\cite{Nguyen2024fractionalexciton} and ~\cite{Zhang2025exciton}.}
\label{table:1}
\end{table}

\section{CURRENT-INDUCED RE-ENTRANT TRANSITION}

\begin{figure*}[!t]
\includegraphics[width=0.98\linewidth]{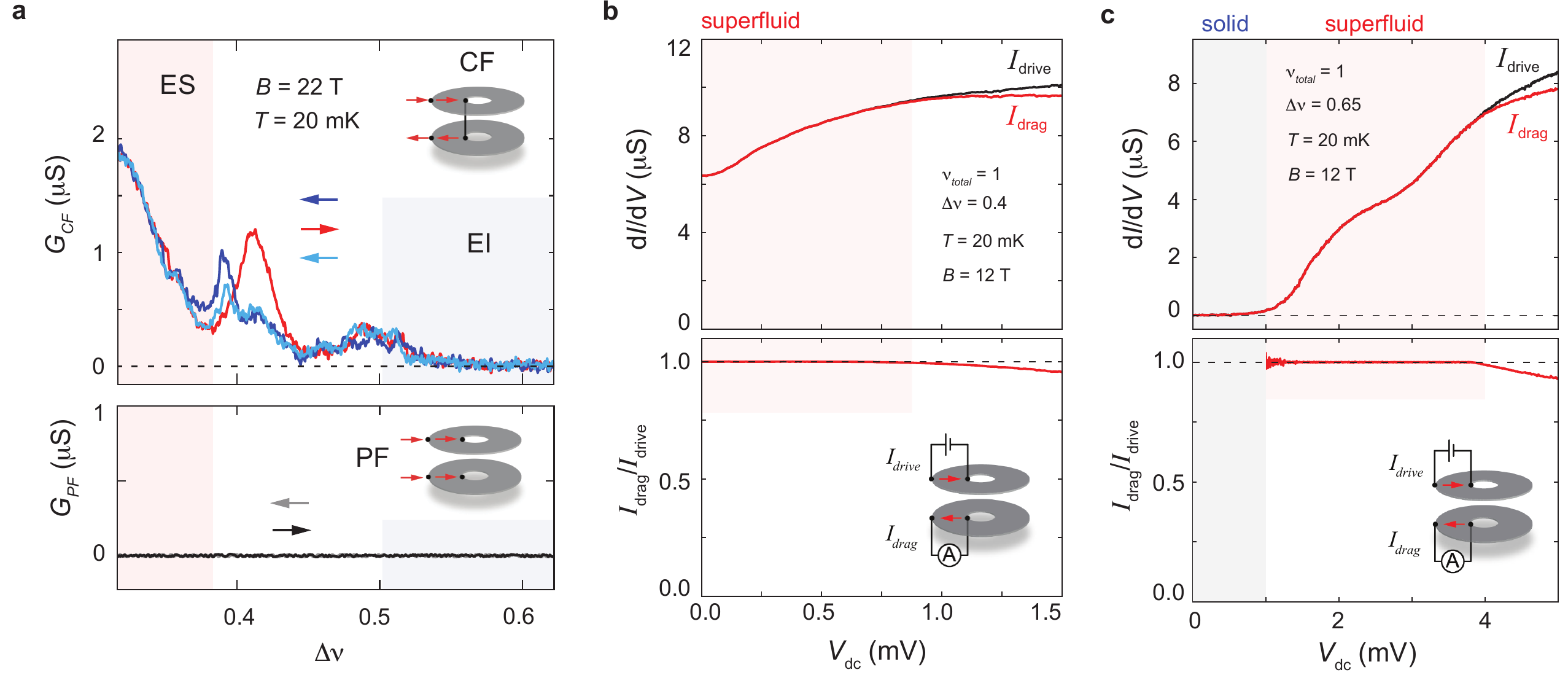}
\caption{\label{IVsolid} {\bf{Current-voltage (IV) characteristics of the exciton phases. }} (a) Counterflow conductance $G_{CF}$ (top panel) and parallel flow conductance $G_{PF}$ (bottom panel) as a function of \dnu\ measured at \nutotal $=1$. (b) Differential conductance $dI/dV$ (top panel) and drag ratio $I_{\text{drag}}/I_{\text{drive}}$ (bottom panel) as a function of d.c. voltage bias $V_{dc}$ measured from an exciton superfluid at \dnu $=0.4$ and \nutotal $=1$. (c) Differential conductance $dI/dV$ (top panel) and drag ratio $I_{\text{drag}}/I_{\text{drive}}$ (bottom panel) as a function of d.c. voltage bias $V_{dc}$ measured from an exciton insulator at \dnu $=0.75$ and \nutotal $=1$. All measurements are performed at $B=12$ T and $T = 20$ mK in sample C5. 
}
\end{figure*} 

Fig.~\ref{IVsolid}a compares the CF and PF responses, $G_{CF}$ and $G_{PF}$,  across the low-temperature phase boundary between the exciton superfluid and insulating phases. On the low-\dnu\ side of the boundary, the superfluid phase exhibits high conductance in the CF configuration but remains insulating in the PF configuration. This behavior is consistent with the hallmark transport signature of exciton superfluid. 

In contrast, the exciton insulator displays vanishing conductance in both $G_{CF}$ and $G_{PF}$. Across the transition boundary, $G_{PF}$ remains zero, further corroborating the robustness of exciton pairing in both the superfluid and insulating phases. 

Moreover, the behavior of $G_{CF}$ near the transition boundary is sensitively dependent on the direction of sweeping \dnu. Notably, $G_{CF}$ is swept back and forth three times in Fig.~\ref{IVsolid}a, highlighting the robustness of the hysteretic transition. This hysteresis underscores the first-order nature of the phase boundary between the exciton superfluid and insulating phases, consistent with an excitonic crystalline order underlying the insulator.

Fig.~\ref{IVsolid}b-c plots the IV characteristics of the exciton superfluid and insulating phases, measured in the drag geometry. By dividing the graphene layers into two separate electrical circuits, drag geometry produces one of the most defining transport responses of the exciton superfluid. As a drive current $I_{\text{drive}}$ is sent through the drive layer, charges of a specific sign move in a given direction. Due to exciton pairing, carriers in the drag layer are dragged in the same direction as those in the drive layer, generating a drag current $I_{\text{drag}}$  ~\cite{Nandi.12,Zhang2025exciton}. 

Since excitons consist of oppositely charged constituents, $I_{\text{drive}}$ and $I_{\text{drag}}$ flow in opposite directions but share the same amplitude. As a result, the exciton condensate acts like a perfect transformer, converting $100\%$ of the current in the drive layer into the drag circuit. This is captured by the perfect drag response in Fig.~\ref{IVsolid}b below a critical voltage bias.
As the d.c. voltage bias increases, the drag ratio deviates from unity, with the the drive current $I_{\text{drive}}$ becoming larger than the drag current $I_{\text{drag}}$. This behavior indicates the existence of a critical threshold of d.c. voltage bias and temperature, beyond which the generation of unpaired charge carriers lead to the breakdown of the perfect drag condition.

The exciton insulator exhibits a distinct IV curve characterized by two critical voltage biases. At small voltage bias below the first threshold, both $I_{\text{drive}}$ and $I_{\text{drag}}$ remains zero, consistent with observations in Fig.~\ref{fig3}d. As the voltage bias increases above the first critical value, the I-V curve shows a re-entrant transition resembling the temperature-driven transition in Fig.~\ref{fig3}d. On the high bias side of this transition, the sample recovers the perfect drag behavior, $I_{\text{drag}}/I_{\text{drive}}=1$, characteristic of a superfluid phase. 

As the d.c. voltage bias increases above a second threshold around $V_{dc} = 4$ mV, $I_{\text{drive}}$ exceeds$I_{\text{drag}}$, indicating that a large exciton current induces pair breaking.  The I-V characteristics of the exciton insulator thus mirror the temperature-driven reentrant transition, providing further evidence for an exciton solid phase in the dilute limit of bilayer excitons.


\section{EXCITON PHASES IN PARALLEL FLOW AND COUNTERFLOW MEASUREMENTS}

\begin{figure*}
\includegraphics[width=1\linewidth]{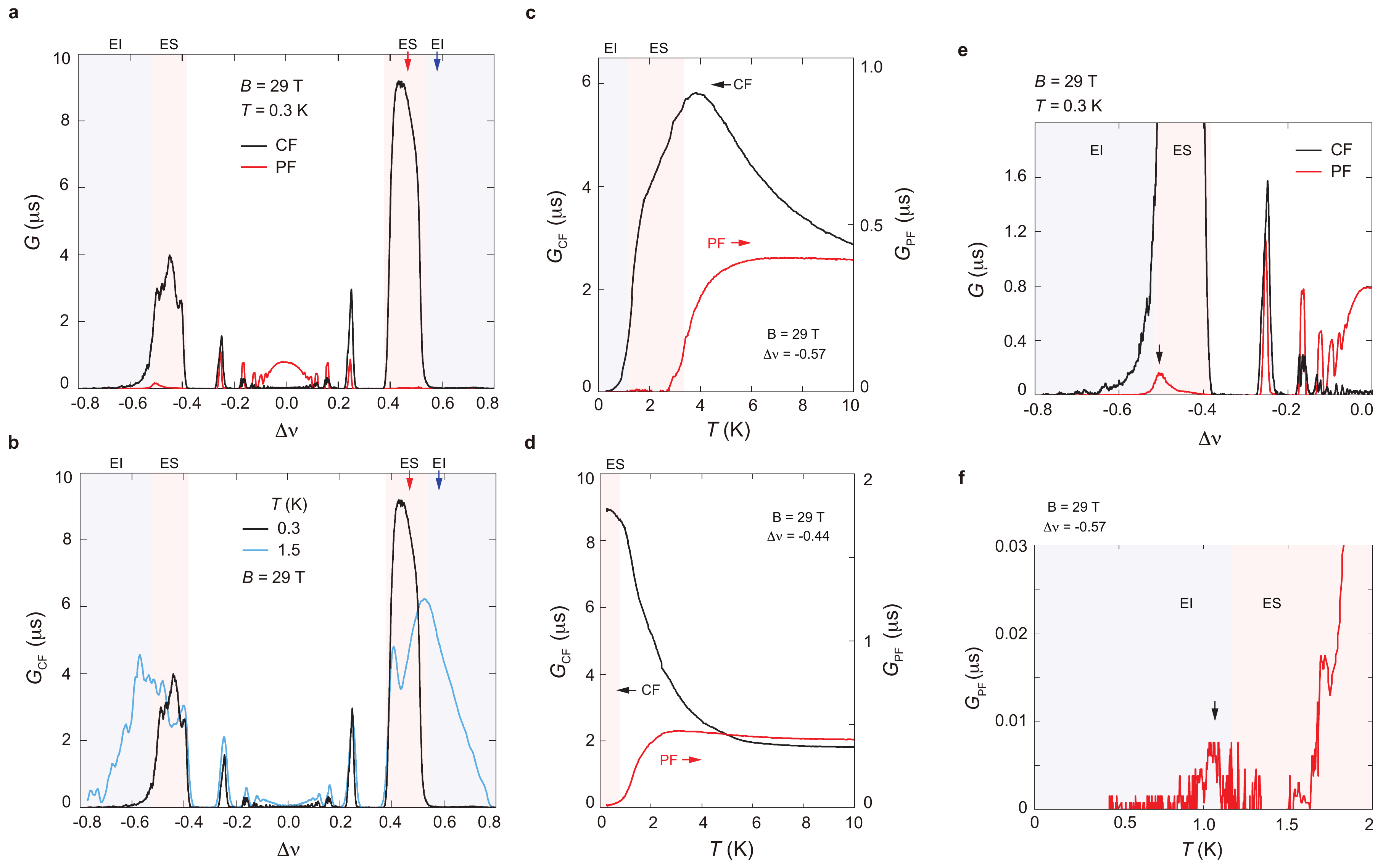}
\caption{\label{FigPF} {\bf{Parallel flow and counterflow measurement.}} (a) Counterflow conductance ($G_{CF}$, black trace) and parallel flow conductance ($G_{PF}$, red trace)  as a function of \dnu measured at $T = 0.3$.  (b) Counterflow conductance as a function of \dnu measured at $T = 0.3$ K and $1.5$ K. The $\Delta \nu$-regime of the exciton superfluid (ES) and exciton insulator (EI) are marked by red and blue shaded background, respectively. The superfluid corresponds to a large counterflow conductance, whereas it is insulating in parallel flow measurement. On the other hand, the exciton insulator displays zero conductance in both parallel flow and counterflow measurements. However, the insulator undergoes a reentrant transition into a superfluid phase with increasing temperature. (c) Counterflow and parallel flow conductance, $G_{CF}$ and $G_{PF}$ (black and red traces), as a function of temperature measured at \dnu $=-0.57$, where the low temperature ground state is an exciton insulator.  (d) Counterflow and parallel flow conductance, $G_{CF}$ and $G_{PF}$ (black and red traces), as a function of temperature measured at \dnu $=-0.44$, where the low temperature ground state is an exciton superfluid. All measurements are performed in a Corbino-shaped sample (C1) at $|\nu_{total}| = 1$ and $B = 29$ T. (e) $G_{CF}$ and $G_{PF}$ as a function of \dnu near the low-temperature transition induced by \dnu. A peak in $G_{PF}$, marked by the vertical arrow, is observed near the transition. (f)  $G_{PF}$ as a function of $T$ measured at \dnu $=-0.57$. Near the EI-to-ES transition with increasing temperature, $G_{PF}$ exhibits a small peak, which is marked by the black vertical arrow.
}
\end{figure*} 

In the Corbino-shaped samples, both the exciton superfluid and exciton insulator exhibit zero conductance in the parallel flow measurement (Fig.~\ref{fig3}b and Fig.~\ref{FigPF}a). This is consistent with the fact that excitons are overall charge neutral. The superfluid and insulator can be distinguished based on their counterflow conductance. As shown in Fig.~\ref{fig3}a and Fig.~\ref{FigPF}a, the exciton superfluid (ES), marked with the red shaded background, exhibits large counterflow conductance, whereas the exciton insulator (EI) is insulating under the counterflow measurement. 

With increasing temperature, the EI undergoes a reentrant-type transition, where the counterflow conductance increases with temperature and becomes non-zero above the transition temperature. The reentrant phase behaves like an exciton superfluid (Fig.~\ref{fig3}d,e).  The parallel flow conductance remains zero in the reentrant phase (Fig.~\ref{FigPF}c), which is consistent with an exciton superfluid phase. 

Interestingly, the phase boundary between the ES and EI phases is accompanied by a small peak in the parallel flow conductance. As shown in Fig.~\ref{FigPF}e, a peak in $G_{PF}$ is observed near the low-temperature boundary between the ES and EI phases, which is marked by the black vertical arrow. Similarly, a small peak in $G_{PF}$ emerges near the reentrant-type transition with varying temperature. it is worth noting that counterflow drag response in the hall bar shaped samples display a similar behavior. As shown in Fig.~\ref{fig4}c and d, the boundary between the ES and EI phases coincides with a peak in the longitudinal drag response. In the counterflow drag measurement, a non-zero longitudinal response could arise from dissipative exciton flow. However, the origin of the conductance peak in the Corbino-shaped sample remains an open question. 

When both layers are tuned to half filling, interlayer correlation is suppressed and each graphene layer is occupied by a layer-decoupled composite fermi liquid (CFL) phase. The CFL exhibits a large parallel flow conductance with suppressed counterflow conductance (Fig.~\ref{FigPF}a and e). This is in stark contrast with the exciton superfluid, which shows enhanced counterflow conductance and vanishing parallel flow conductance. It is worth noting that the counterflow conductance of the CFL shows little dependence on temperature. As shown in Fig.~\ref{FigPF}b, the counterflow conductance of the CFL remains low at $T = 1.5$ K. In comparison, the exciton insulator exhibits enhanced $G_{CF}$ at $T = 1.5$ K. This suggests that the enhanced counterflow conductance in the Corbino-shaped sample is directly linked to the exciton superfluid.

Away from the balanced condition, $G_{PF}$ shows a series of oscillations with increasing $\Delta\nu$, which are in excellent agreement with the Jain sequence of layer-decoupled fractional quantum Hall effect states. If interlayer correlation is negligible, the bilayer system is expected to recover the 4-flux sequence of FQHE states in the regime of large layer imbalance. Instead, the behavior of the parallel flow conductance points towards a single incompressible state that occupy a large range of $\Delta \nu$. This is indicative of an exciton insulator, which is inconsistent with the layer-decoupled FQHE state or a single-layer Wigner solid ~\cite{Hatke2019wigner}.

\section{$\Delta \nu-T$ MAP AT $\nu_{total} = 1$ AND HYSTERETIC TRANSITIONS}

\begin{figure*} [!t]
\includegraphics[width=1\linewidth]{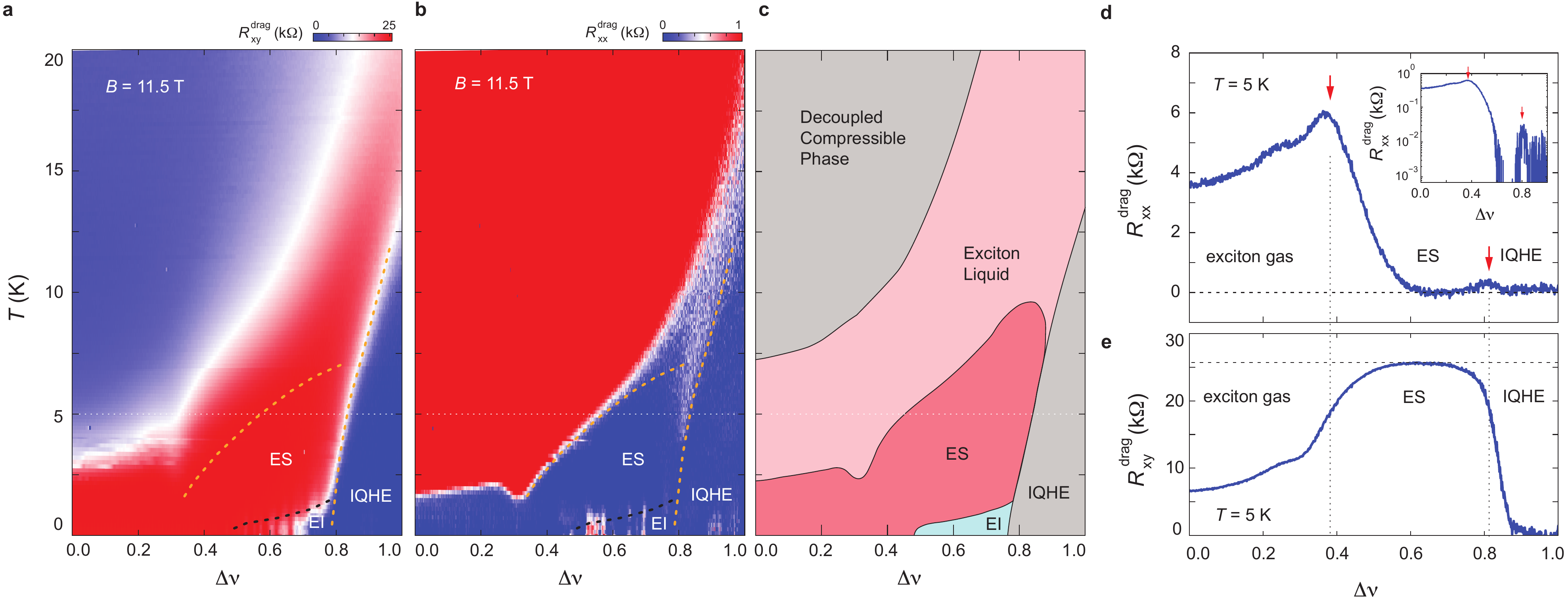}
\caption{\label{dnuT} {\bf{The phase boundaries between exciton superfluid, exciton insualtor and the IQHE state.}} (a) Hall drag and (b) longitudinal drag responses as a function of \dnu\ and $T$, measured from sample H2. The phase boundaries of the superfluid phase is marked with the orange dashed lines, whereas the superfluid to insulator transition boundary is marked by the black dashed line. (c) Schematic diagram of main phases in panel (a-b).  (d) Longitudinal drag and (e) Hall drag as a function of \dnu measured at $T = 5$ K. The exciton superfluid and IQHE are separated by a peak in longitudinal drag response. The transition also coincides with a dramatic change in the Hall drag response. The superfluid phase corresponds to quantized Hall drag at $R_{xy}^{drag} = h/e^2$, whereas Hall drag diminishes in the IQHE state.} 
\end{figure*} 

Fig.~\ref{dnuT} plots the counterflow drag response as a function of \dnu and $T$, measured from a hallbar-shaped sample with interlayer separation of $d = 8.0$ nm. In the limit of large layer imbalance, the observed phase diagram resembles the data shown in Fig.~\ref{fig4}a-b, which is measured from a sample with $d = 7.4$ nm. 

In the temperature range directly above the exciton insulator (EI), Fig.~\ref{dnuT}a-c shows hallmark signatures of an exciton superfluid. This includes quantized Hall drag, which is shown as color red in Fig.~\ref{dnuT}a, and vanishing longitudinal drag, denoted by color blue in Fig.~\ref{dnuT}b.  

The presence of the reentrant superfluid phase is further supported by the line trace at $T = 5$ K taken along the horizontal white dashed line from Fig.~\ref{dnuT}a-b. 
As shown in Fig.~\ref{dnuT}c-d, the range of \dnu marked by ``ES'', indicating the exciton superfluid phase, exhibits quantized Hall drag and vanishing longitudinal drag. The exciton superfluid is separated from the normal fluid phase of exciton, which occupies the lower \dnu part of the phase diagram, by a peak in the longitudinal drag response, which is marked by the red arrow on the left in Fig.~\ref{dnuT}c-d. The exciton normal fluid phase displays nonzero drag in both longitudinal and Hall channels. On the other hand, the exciton superfluid transitions into the integer quantum Hall effect (IQHE) state with increasing $\Delta \nu$. The IQHE is marked by vanishing drag response in both longitudinal and Hall channels. Interestingly, the transition boundary between the superfluid and the IQHE phases is marked by a peak in the longitudinal drag (red arrow in Fig.~\ref{dnuT} on the right).

\begin{figure*} 
\includegraphics[width=0.7\linewidth]{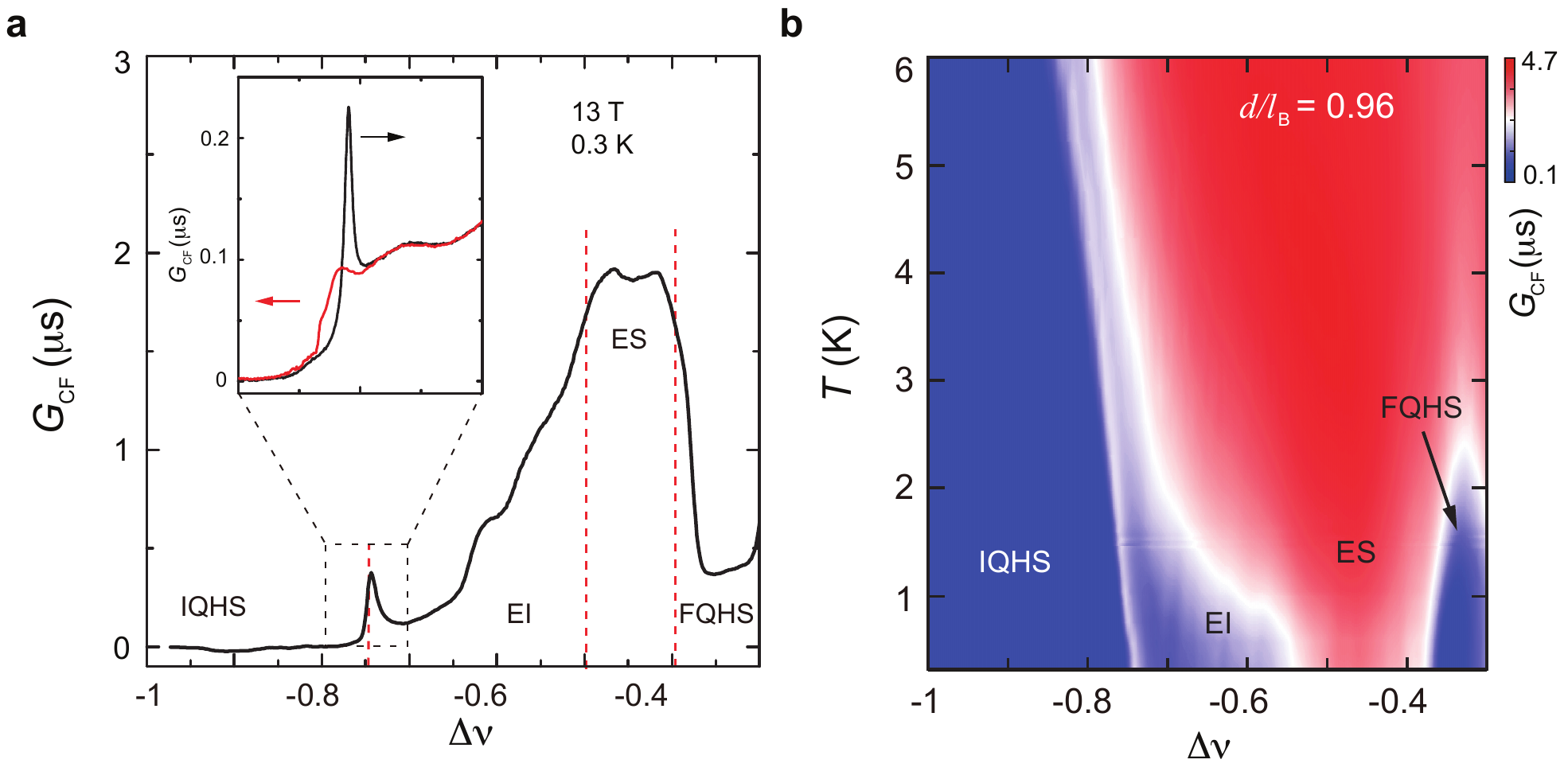}
\caption{\label{SIhysteresis} {\bf{$\Delta\nu$-induced transition.}} (a) Counterflow conductance $G_{CF}$ as a function of \dnu\ measured from sample C1. The exciton solid occupies the range of $-0.75 < \Delta \nu < -0.5$, which is separated from the layer-decoupled IQHE state by a sharp peak in $G_{CF}$. The inset: $G_{CF}$ is measured while \dnu\ is swept back and forth. Depending on the direction of the \dnu\ sweep, the transition boundary between the solid and IQHE states exhibits hysteresis. (b) Temperature-layer-imbalance ($T-\Delta \nu$) map of $G_{CF}$. The boundary between the exciton solid and IQHE states are marked by a peak in $G_{CF}$, which is shown as white in the chosen color scale. This phase boundary is consistent with the white dashed lines in Fig.~\ref{fig3}f-h. )}
\end{figure*}


\begin{figure*} [!t]
\includegraphics[width=0.65\linewidth]{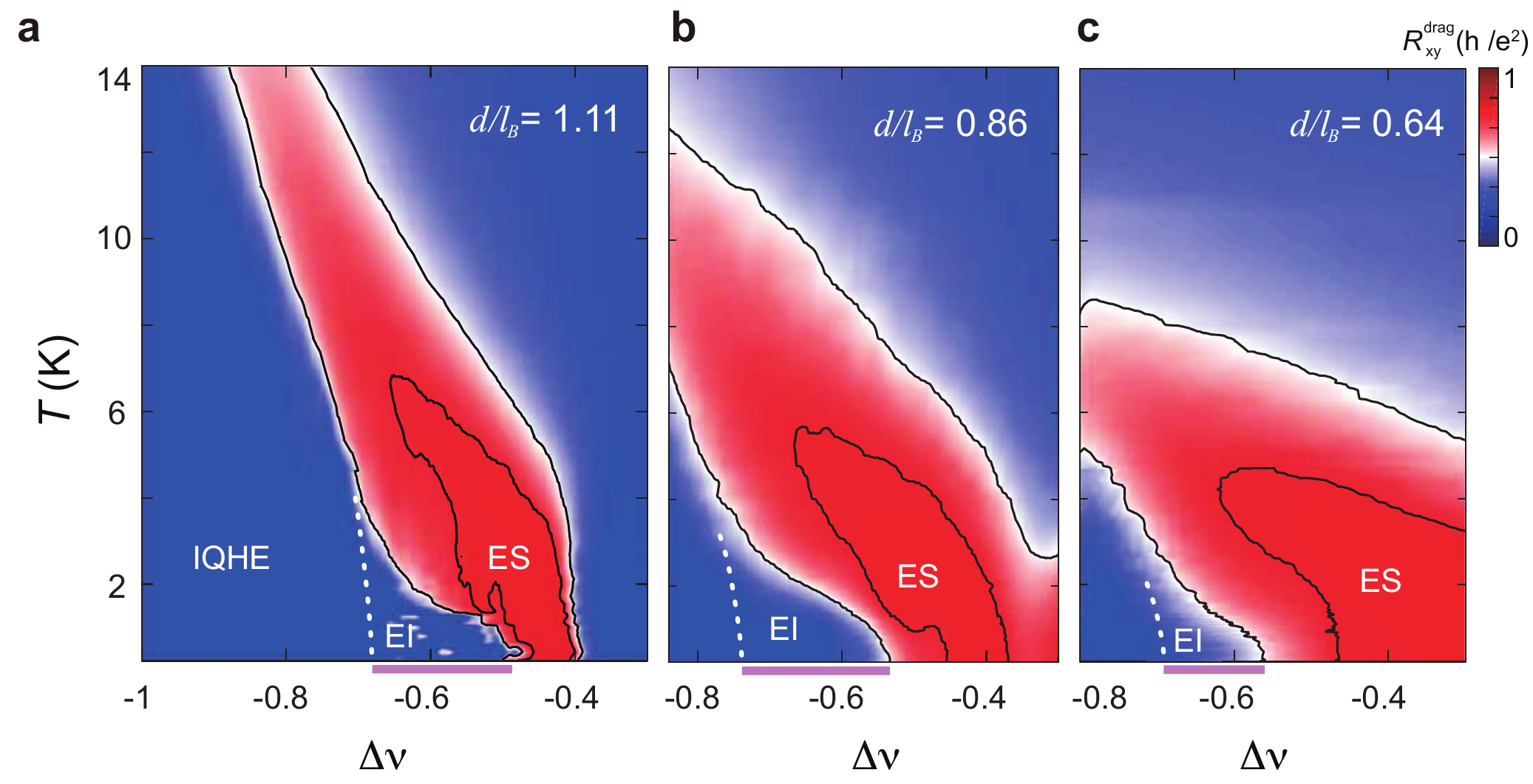}
\caption{\label{dnuTdrag} {\bf{The phase boundaries between exciton superfluid, exciton insualtor and the IQHE state.}} Hall drag as a function of \dnu\ and $T$ measured from sample H1 at (a) \dlb $= 1.11$, (b) $ 0.86$, and (c) $0.64$. The phase boundaries of the exciton superfluid (ES) phase is marked with the black solid lines, whereas the exciton insulator (EI) to IQHE transition boundary is marked by the white dashed lines. }
\end{figure*} 



A transition boundary between the layer-decoupled IQHE state and exciton phases is also observed in a Corbino-shaped sample, as shown in Fig.~\ref{SIhysteresis}. In this sample, the transition is marked by a peak in the counterflow conductance. Most remarkably, this transition appears to be the first order between the exciton insulator and the IQHE state at low temperature. As shown in the inset of Fig.~\ref{SIhysteresis}a, the peak in the counterflow conductance shows hysteretic behavior as \dnu is swept back and forth across the low-temperature transition. With increasing $T$, the exciton insulator disappears and the peak in the counterflow conductance marks the superfluid-to-solid transition (Fig.~\ref{SIhysteresis}b), which is in excellent agreement with the phase diagram in Fig.~\ref{dnuT}a-c. 



\section{COUNTERFLOW DRAG RESPONSE AT EXTREME LAYER IMBALANCE}

\begin{figure*}[!t]
\includegraphics[width=0.9\linewidth]{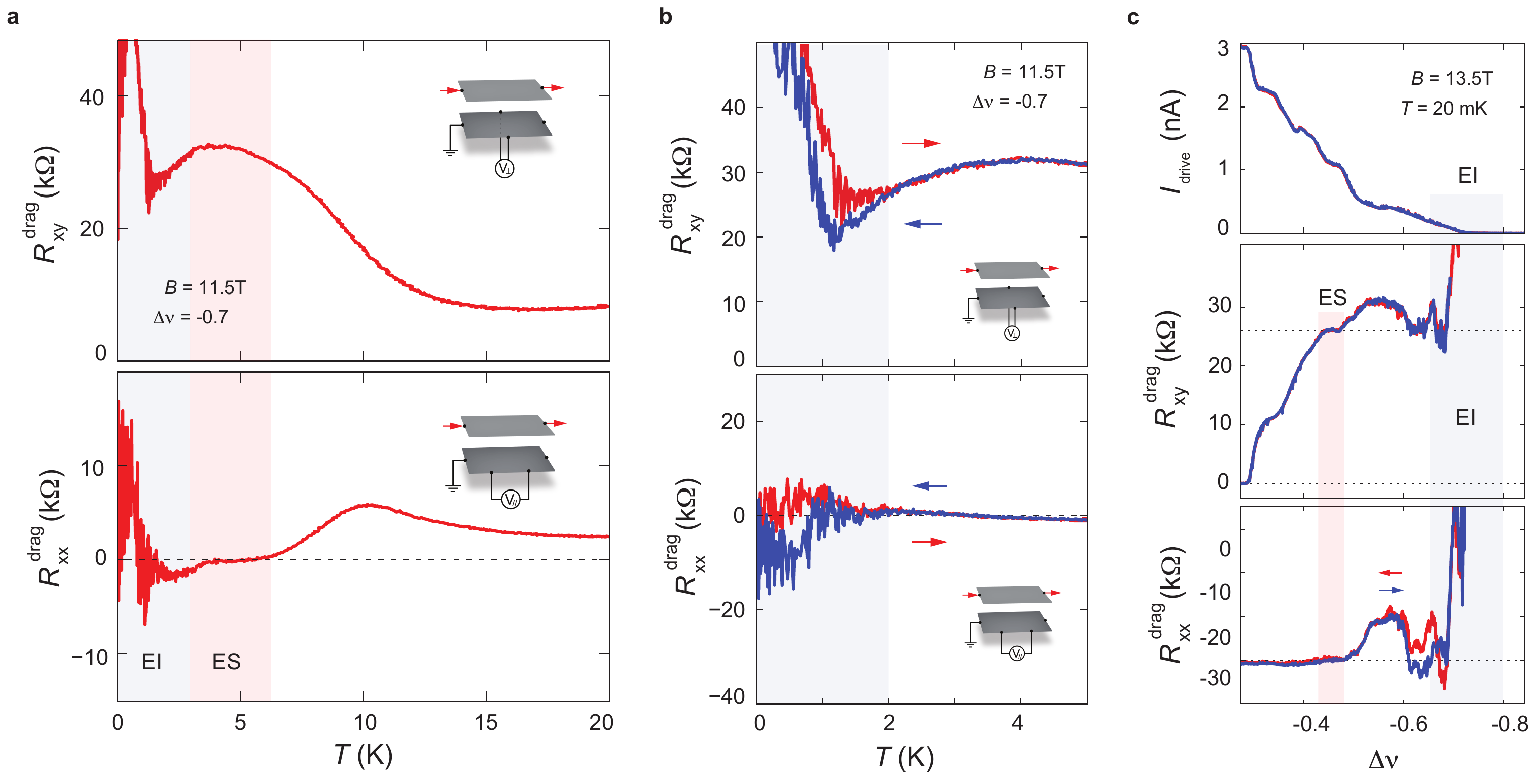}
\caption{\label{FigHysteresis} {\bf{Hysteretic transition between the exciton solid and fluid phases.}} (a) Temperature dependence of Hall drag $R_{xy}^{drag}$ (top panel) and longitudinal drag $R_{xx}^{drag}$ (bottom panel) measured at $\nu_{total} = 1$ \dnu $=-0.7$. (b) Temperature-induced hysteresis in drag responses across the phase boundary between the low-temperature exciton insulator and the high-temperature fluid phase. The measurement is performed at $\nu_{total} = 1$ and \dnu $=-0.7$. (c) Counterflow drag responses measured at $\nu_{total} = 1$ as \dnu\ is swept back and forth. The measurement is performed at a constant temperature of $T = 20$ mK. The drive layer becomes insulating in the limit of large positive \dnu, where drive current $I_{drive}$ diminishes (top panel).  Since the counterflow flow drag response is defined as $R^{drag} = V^{drag}/I_{drive}$, both longitudinal and Hall drag responses diverges in the presence of an exciton insulator. )
}
\end{figure*} 

With increasing layer imbalance, the evolution from the exciton superfluid to layer-decoupled IQHE state represents an intriguing portion of the phase space that has not been investigated by past experimental efforts. 

At moderate layer imbalance, the ground state of the quantum Hall bilayer is an exciton superfluid at $\nu_{total} = 1$. In this scenario, the exciton counterflow in the bulk of the sample gives rise to the ``which layer'' uncertainty and a coherent edge mode. As a result, both drive and drag layer exhibits quantized Hall response, whereas longitudinal responses vanish. It is important to point out that the signature behavior of the exciton superfluid is unaffected by layer imbalance. As long as the excitons flow remains dissipationless, quantized Hall drag and vanishing longitudinal drag responses are expected.

In the limit of extreme layer imbalance, \dnu $=\pm 1$, the quantum Hall bilayer is occupied by two layer-decoupled IQHE states. At \dnu $=+1$, \emph{i.e.} $\nu_1 = 1$ and $\nu_2 = 0$, layer 1 features an integer quantum Hall effect state, whereas layer 2 has no conductive edge channels. The situation is reversed with the edge mode occupying layer 2 at \dnu $=-1$. As a result of the edge mode, the transport properties in this regime is strongly layer dependent. The layer with the edge mode exhibits vanishing longitudinal resistance with quantized Hall resistance, whereas the layer without an edge mode behaves like an insulator.

An exciton solid phase represents an interesting scenario. If the solid phase does not support exciton flow, the sample bulk behaves like an insulator. This should give rise to a layer-dependent transport response. In a counterflow drag measurement, we drive current through layer 1 and measure Hall and longitudinal drag responses from layer 2. At large positive \dnu, current in the drive layer flows through the integer edge mode on layer 1. As exciton flow is suppressed by the emerging solid phase, voltage response on the drag layer vanishes, which is reflected by vanishing drag responses vanish in both longitudinal and transverse channels. This is consistent with our observation in Fig.~\ref{fig4}a.  At large negative \dnu, current flow on the drive layer is suppressed by the lack of edge channel (combined with the lack of exciton flow in sample bulk). Since the counterflow drag response is defined as the ratio between the voltage response on the drag layer and the drive layer current, $R_{xx}^{drag}$ and $R_{xy}^{drag}$ diverge as the exciton solid phase emerges. 

Fig.~\ref{FigHysteresis}a shows the diverging drag response on the low-temperature side of the superfluid-to-solid transition at large negative $\Delta\nu$. Notably, the superfluid-to-solid transition in the regime of large negative \dnu\ exhibits hysteretic behavior, as shown in Fig.~\ref{FigHysteresis}b. This is consistent with the expected behavior of a first-order melting transition.

Fig.~\ref{FigHysteresis}c plots the \dnu-dependence of drive layer current $I_{drive}$, longitudinal and Hall drag responses measured at low temperature. The superfluid-to-solid transition is shown to be accompanied by the suppression of drive layer current in the top panel. On the low-\dnu\ side of transition, the exciton superfluid is characterized by quantized Hall drag and vanishing longitudinal drag. On the high-\dnu\ side of transition, both Hall and longitudinal drag diverge. A hysteretic behavior is observed in the longitudinal drag measurement across the transition boundary.

It is worth noting that the nature of the superfluid-to-solid transition is independent of the sign of \dnu\ in a Corbino-shaped sample. In the Corbino geometry, transport response is solely determined by exciton flow is in sample bulk. As such, the transport response across the transition boundary is consistent with Fig.~\ref{fig3}a-b for both positive and negative $\Delta\nu$. 


\begin{figure*}
\includegraphics[width=0.9\linewidth]{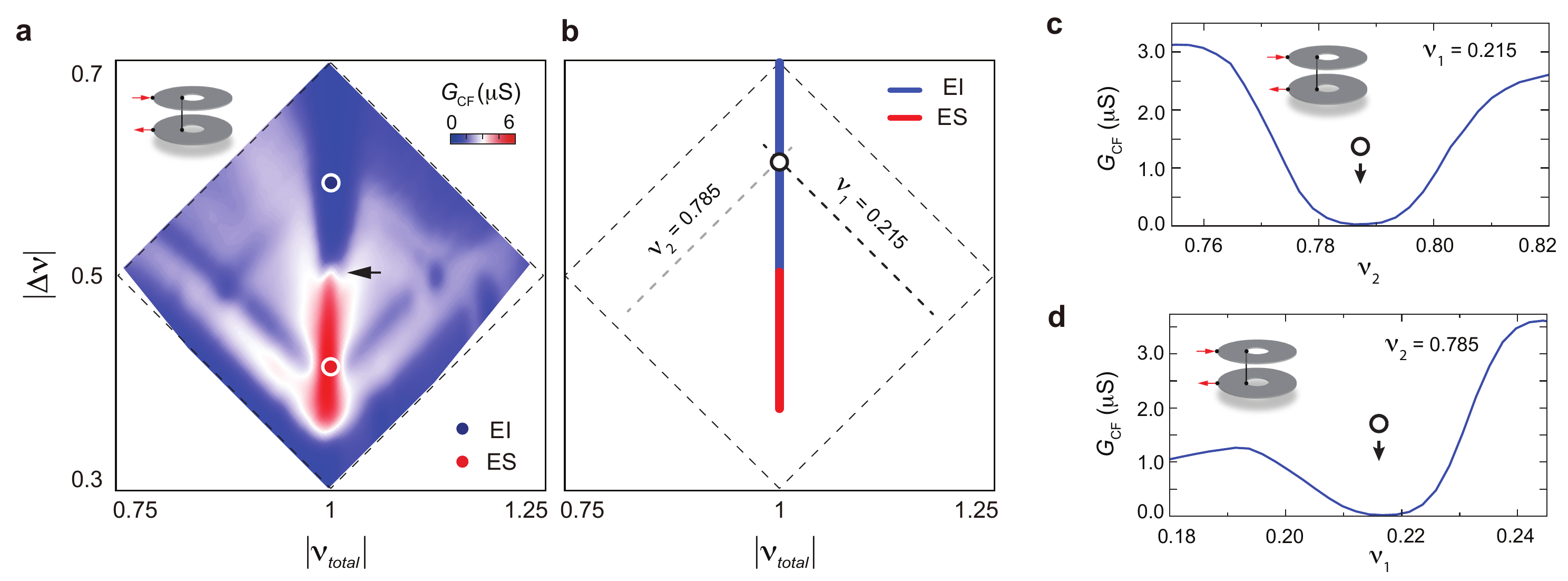}
\caption{\label{Figlocalization} {\bf{The influence of disorder localization.}}
Since the exciton insulator occurs when both graphene layers are tuned to be near the Landau level edge, the influence of the disorder localization should be considered. (a) Counterflow conductance as a function of \dnu\ and $\nu_{total}$ near the transition boundary between the exciton insulator and exciton superfluid. (b) Schematic diagram of the $\Delta \nu-\nu_{total}$ map.  Two trajectories along constant $\nu_1$ and $\nu_2$ are highlighted as black and gray dashed lines. (c) Counterflow conductance $G_{CF}$ as a function of $\nu_2$, while $\nu_1$ is held constant at $\nu_1 = 0.215$. (d) Counterflow conductance $G_{CF}$ as a function of $\nu_1$, while $\nu_2$ is held constant at $\nu_2 = 0.785$.  (c)  and (d) are measured along trajectories marked by gray and black dashed lines in panel (b). 
As shown in (c) and (d), an insulating phase with vanishing $G_{CF}$ is observed at $\nu_{total}=1$. If disorder plays the dominating role, the sample is expected to remain insulating in the regimes defined by $\nu_1 < 0.215$ in panel (c) and $\nu_2 > 0.785$ in panel (d). In these regimes, one of the layers is tuned closer to the Landau level edge, which enhances the influence of disorder localization ~\cite{Ahn2023Wigner}. Since two graphene layers form a series circuit in the counterflow measurement, disorder localization in one layer will induce a vanishing counterflow conductance. Instead, we observed a highly conductive counterflow behavior as $\nu_2 > 0.785$ and $\nu_1 < 0.215$. Our findings suggest that the influence of disorder localization is negligible.
}
\end{figure*}

\section{The Influence of Layer Imbalance}

\begin{figure*}
\includegraphics[width=1\linewidth]{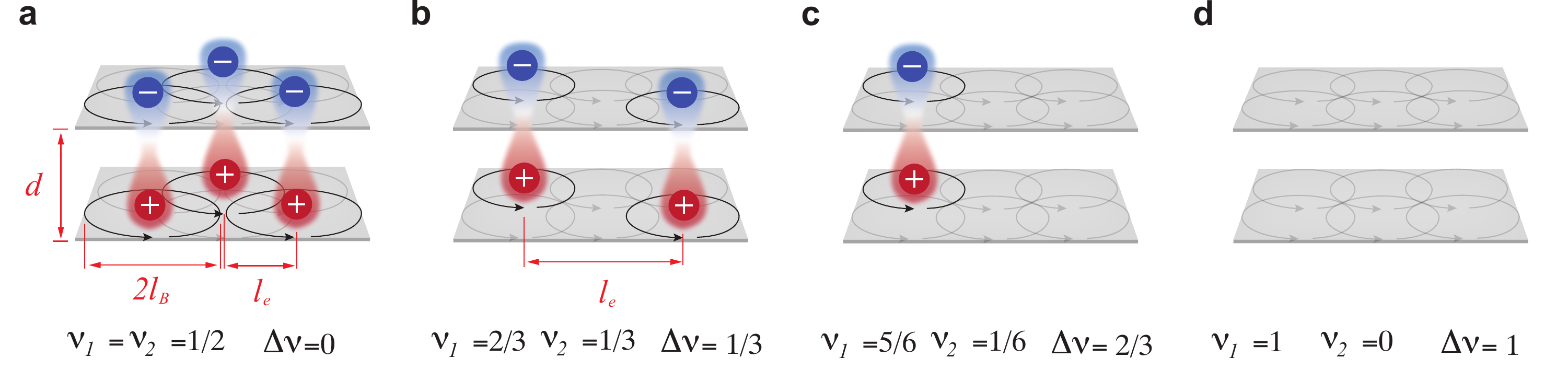}
\caption{\label{Figdnu} {\bf{The excitonic density as a function of layer imbalance $\Delta\nu$.}} Schematic diagram of interlayer excitons in a quantum Hall bilayer at $\nu_{total}=1$. Black solid (gray) circles indicates occupied (unoccupied) Landau orbits. (a) Under layer-balanced condition, \dnu $=\nu_1-\nu_2 =0$, interlayer excitons occupy half the available Landau orbital. As such, $\ell_e = \sqrt{2}\ell_B$. In the presence of non-zero layer imbalance, the exciton density is reduced. For instance, excitons occupy $1/3$ of Landau orbitals at \dnu $=1/3$ (panel b), whereas only $1/6$ of the orbitals host excitons at \dnu $=2/3$ (panel c). In the limit of extreme layer imbalance, layer $1$ and $2$ are tuned to integer LL filling of $\nu_1 = 1$ and $\nu_2$ =0. In this scenario, both layer are occupied by integer quantum Hall effect and the exciton density diminishes to zero. Taken together, the inter-exciton spacing can be defined based on layer imbalance \dnu\ as $\ell_e= \ell_B\sqrt{2/(1-|{\Delta \nu}|)}$.
}
\end{figure*}

Fig.~\ref{Figdnu} plots the schematic diagram of exciton density at different values of layer imbalance. Exciton density is maximized under the layer-balanced condition with \dnu $=0$ (Fig.~\ref{Figdnu}a), giving rise to the smallest inter-exciton spacing $\ell_e = \sqrt{2}\ell_B$. At extreme layer imbalance at $\nu_1=1$ and $\nu_2=0$ (\dnu $=1$), exciton density is zero. At an intermediate value of \dnu, inter-exciton spacing is determined by $\ell_e= \ell_B\sqrt{2/(1-|{\Delta \nu}|)}$.

Fig.~\ref{SIDT2} plots the counterflow conductance, measured from a Corbino-shaped graphene bilayer (Fig.~\ref{fig1}d), as a function of \dnu\ and $T$. While \dnu\ is varied from $-0.8$ to $+0.8$, $\nu_{total}$ is fixed at $-1$. In the counterflow measurement, the presence of exciton pairing enhances the counterflow conductance $G_{CF}$, which is shown as red in the chosen color scale. Fig.~\ref{SIDT2}a and b compares the \dnu$-T$ map measured at two different \dlb-values.  At \dlb $=0.63$ (Fig.~\ref{SIDT2}a), a robust condensate phase occupy the layer imbalance range of $-0.6 < \Delta \nu < 0.6$. With increasing temperature, exciton pairing, evidenced by large counterflow conductance, persists to temperature much higher than the superfluid transition temperature $T_c$. This is consistent with the expected behavior in the strong-coupling limit of interlayer excitons ~\cite{Liu2022crossover}.  At \dlb $=1.09$ (Fig.~\ref{SIDT2}b), both the exciton condensate and excitonic pairing are limited to $T < 2$ K near the layer-balanced condition, which is in excellent agreement with the expected behavior in the weak-coupling regime ~\cite{Liu2022crossover}. However, strong exciton pairing is recovered at \dlb $=1.09$ by tuning the sample to the layer-imbalanced regime. In the range of $0.4 < |\Delta \nu| < 0.6$, both exciton condensate and exciton pairing persist to $T > 2$ K. The influence of \dnu\ on the temperature dependence of counterflow conductance is consistent with the observation in Fig.~\ref{fig2}.

At \dlb $=1.09$, the single-layer FQHE states competes against exciton condensate near specific values of $\Delta\nu$. For instance, \dnu $=-1/3$ corresponds to $\nu_1=1/3$ and $\nu_2 = 2/3$, where layer $1$ and $2$ are both occupied by the Laughlin state. The single-layer FQHE states is insulating in the counterflow measurements, which are shown as blue in the chosen color scale.  On the other hand, the influence of this competition is negligible in the strong-coupling limit at \dlb $=0.63$.



\section{Exciton Pairing Strength}

\begin{figure*}
\includegraphics[width=0.95\linewidth]{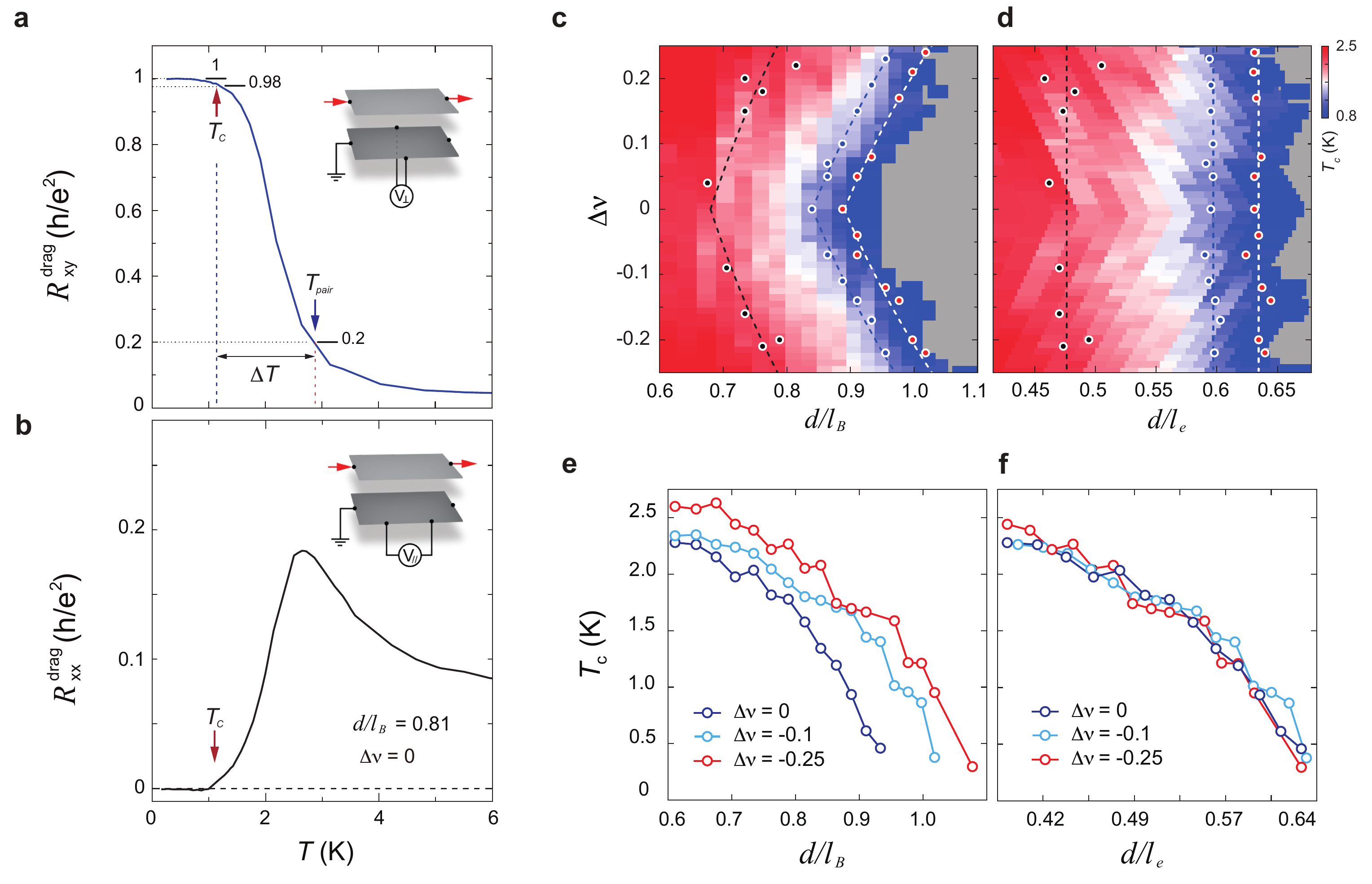}
\caption{\label{Tc} {\bf{$T_c$ defined using longitudinal drag and Hall drag.}} (a) Hall drag \dragxy\ and (b) longitudinal drag \dragxx\ as a function of temperature $T$ measured at \dlb $=0.81$ and \dnu $=0$. Inset shows the schematic diagram for Hall drag and longitudinal drag measurements. As shown in panel (a), $T_c$ can be defined as the temperature at which Hall drag response equals $98\%$ of the plateau value, \dragxy $=0.98h/e^2$. Whereas $T_{pair}$ is defined as the temperature at which Hall drag response drops to $20\%$ of the plateau value, \dragxy $=0.2h/e^2$. The temperature window \DT\ denotes the difference between $T_{pair}$ and $T_c$, \DT $= T_{pair} - T_c$. Similarly, $T_c$ can be defined as the onset of longitudinal drag \dragxx\ with increasing temperature (panel b). Both definitions yields the same value for $T_c$. Panel (c-f) plots the dependence of $T_c$ on \dlb\ and \dle.  $T_c$ as a function of (d) \dnu\ and \dlb, (f) \dnu\ and \dle.  Constant values of $T_c$ are marked by black, blue and red circles.  $T_c$ as a function of (e) \dlb\ and (f) \dle\ measured at different interlayer density imbalance $\Delta\nu$. Here, $T_{c}$ is defined as the temperature at which longitudinal drag \dragxx\ onsets from the noise floor with increasing temperature (marked by vertical red arrow in panel b). The dependence on \dlb\ and \dle\ is in excellent agreement with the results shown in Fig.2, where $T_c$ is determined based on the Hall drag response \dragxy.
}
\end{figure*}

\begin{figure*}
\includegraphics[width=0.65\linewidth]{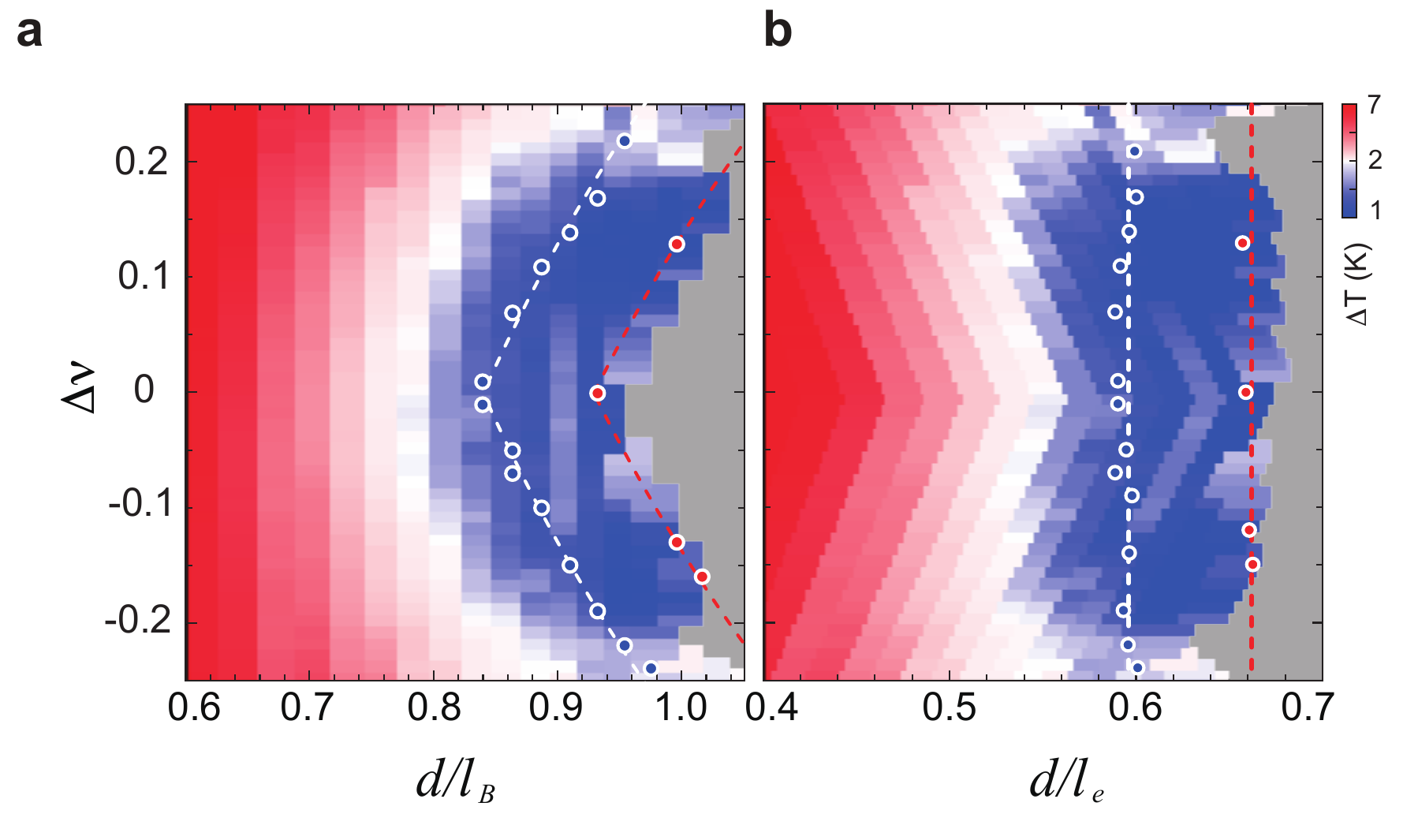}
\caption{\label{FigPairing} {\bf{\DT\ as an indicator of excitonic pairing strength.}} \DT\ is defined according to Fig.~\ref{Tc}a. (a-b) \DT\ as a function of \dnu\ and (a) \dlb, (b) \dle. Blue and red circles mark constant values of \DT. For a constant \dlb, \DT\ increases with increasing $\Delta\nu$. Whereas \DT\ is independent of \dnu\ for a constant \dle. This indicates that the excitonic pairing strength is determined by inter-exciton spacing $\ell_e$, instead of magnetic length $\ell_B$.
}
\end{figure*} 

\begin{figure*}[!t]
\includegraphics[width=1\linewidth]{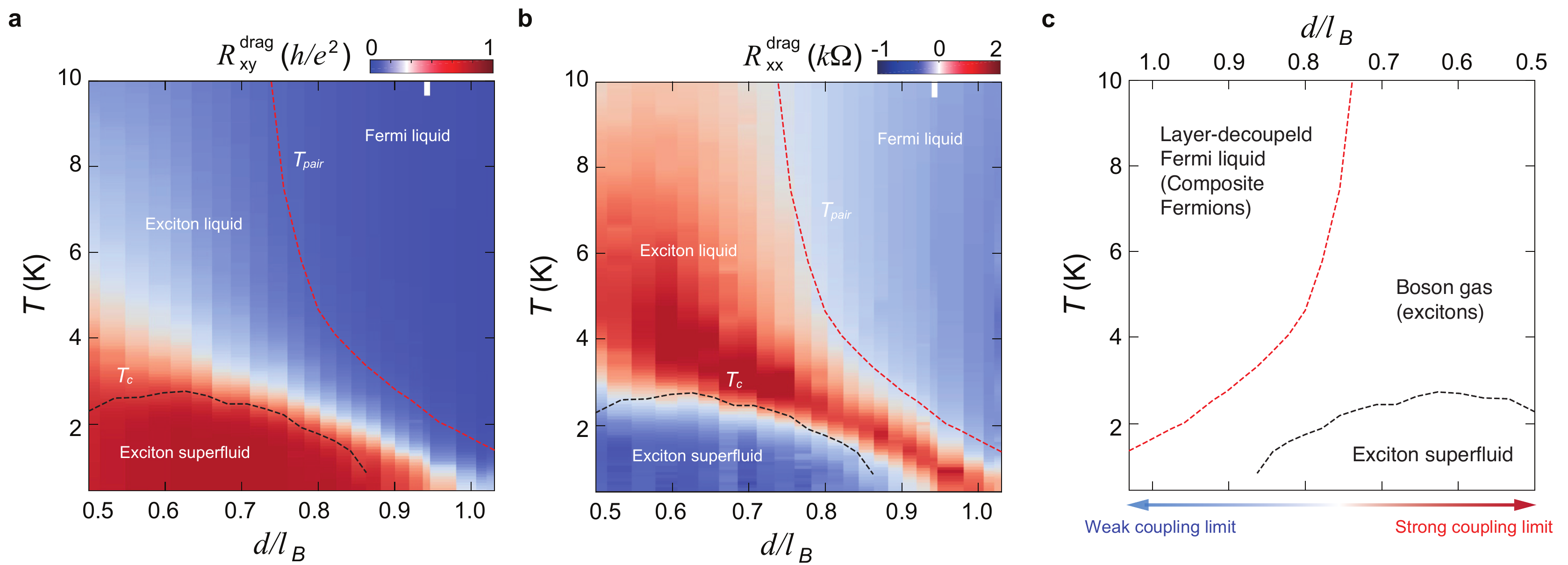}
\caption{\label{SIDT1} {\bf{The crossover phenomenon with varying \dlb\ at \dnu = 0.}} (a-b) Drag counterflow measurement from a graphene double-layer sample shaped into a Hall bar geometry.  (a) Hall drag  and (b) longitudinal drag response as a function of temperature and \dlb when each layer is tuned to occupy half a Landau level at $\nu_1=\nu_2=-1/2$. The transition temperature of the condensate phase $T_c$ is defined as the temperature where Hall drag response deviates from the quantized plateau, $R_{xy}^{drag}=h/e^2$. The dashed black line is operationally defined as the contour of constant Hall drag response of $R_{xy}^{drag}=0.98h/e^2$. Along the same vein, the pairing temperature $T_{pair}$ denotes the temperature where exciton pairing disappears. In this measurement, $T_{pair}$ is operationally defined as the temperature where the Hall drag response becomes less than $8\%$ of the quantization value. Red dashed line denotes $T_{pair}$, which is the contour line at a constant value of Hall drag response at , $R_{xy}^{drag}=0.08 h/e^2$. (c) Schematic diagram demonstrating the crossover between the weak coupling and strong coupling limit by tuning \dlb\ (the value of \dlb\ is marked near the top axis).
}
\end{figure*} 

\begin{figure*}
\includegraphics[width=0.98\linewidth]{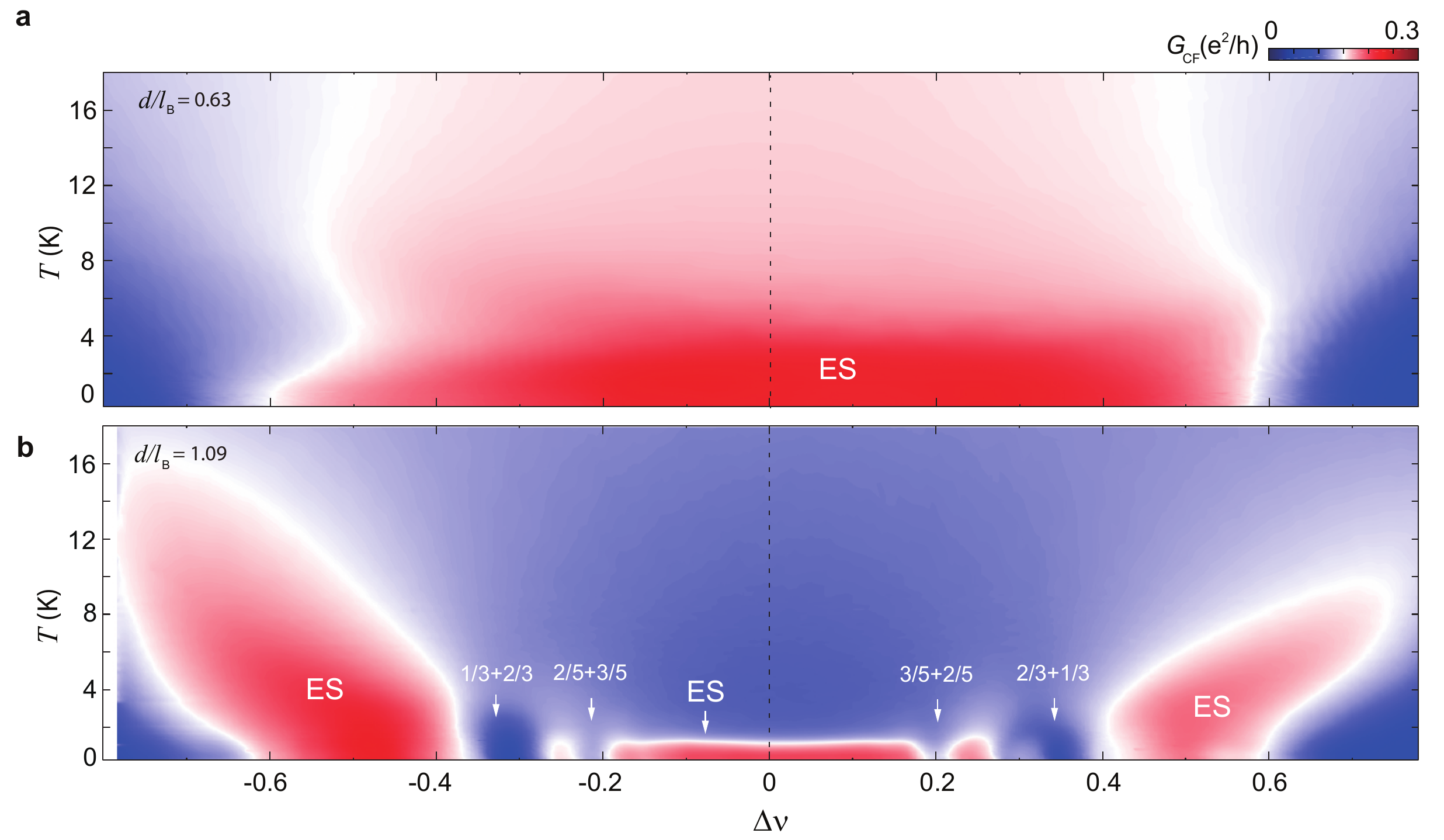}
\caption{\label{SIDT2} {\bf{The influence of \dnu\ on the exciton pairing strength.}} The counterflow conductance $G_{CF}$ as a function of \dnu\ and T measured at (a) \dlb $=0.63$ and (b) \dlb $=1.09$.  
}
\end{figure*}

The strength of interlayer coupling is reflected by the onset temperature of exciton pairing, $T_{pair}$, relative to the superfluid transition $T_c$ (Fig.~\ref{Tc}a) ~\cite{Liu2022crossover}. In the strong coupling regime, exciton pairing survives to high temperature in the absence of the condensate. As such, $T_{pair}$ is much larger than $T_c$ in the presence of strong exciton pairing. In the weak coupling limit, exciton pairing and the superfluid condensate phase occurs simultaneously with decreasing temperature, $T_{pair} \sim T_c$. As such, the pairing strength can be characterized based on the size of the temperature window $\Delta T = T_c - T_{pair}$. $T_{pair}$ is operationally defined as the temperature where Hall drag response exceeds a threshold of $20\%$ the plateau value (Fig.~\ref{Tc}a).

Such a crossover behavior is shown in Fig.~\ref{SIDT1}(a-b), which plots the Hall drag and longitudinal drag responses as a function of \dlb\ and $T$. 
At small \dlb, $T_{pair}$ diverges to high temperature, whereas $T_c$ remains around $3$ K. This gives rise to a large \DT, which is consistent with strong interlayer correlation in the strong coupling regime. With increasing \dlb, both $T_c$ and $T_{pair}$ trend towards zero, which corresponds to a decrease in \DT. 
A schematic phase diagram of $T_c$ and $T_{pair}$ is shown in Fig.~~\ref{SIDT1}c, where the bottom axis marks the evolution of excitonic pairing strength. The trend of $T_c$ and $T_{pair}$ 
resembles the BEC-BCS crossover phenomenon proposed by past theoretical works studying bosonic cold atom systems ~\cite{Randeria2014}.

Similar to Fig.~\ref{fig2}d-g, the value of \DT\ is dependent on layer imbalance $\Delta\nu$. as shown in Fig.~\ref{FigPairing}a, \DT\ increases with increasing |\dnu| At a given \dlb. This suggests that \dlb\ does not fully capture the excitonic pairing strength. On the other hand, replotting \DT\ as a function of \dnu\ and \dle\ reveals that \DT\ is independent of \dnu\ at a fixed value of \dle. As such, the excitonic pairing strength is fully determined by the ratio between $d$ and $\ell_e$. The critical role  of $\ell_e$, instead of $\ell_B$, provides an unambiguous evidence for a system of interacting excitons.

\begin{figure*}[!t]
\includegraphics[width=0.8\linewidth]{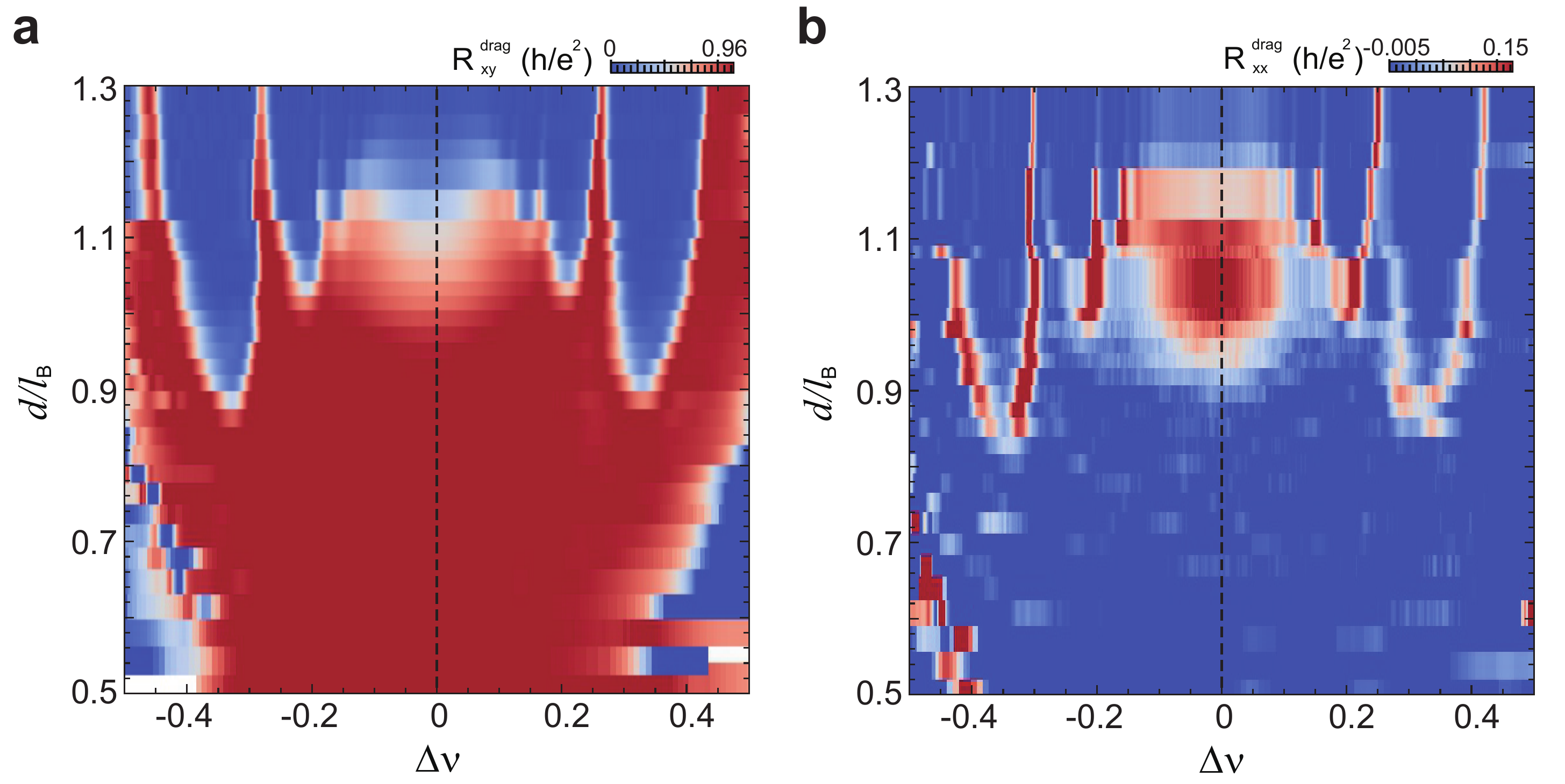}
\caption{\label{SIdragmap} {\bf{Drag counterflow measurement in Hall bar at \nutotal=1}} (a),(b) shows Hall resistance and longitudinal resistance in the drag layer as a function of \dnu and \dlb respectively. Both are measured at T = 0.3 K. The exciton condensate phase marked by unity Hall resistance and vanishing longitudinal resistance roughly occupies the same phase space as measured in corbino geometry shown in Fig.2a in the main text.} 
\end{figure*} 

\begin{figure*}
\includegraphics[width=0.7\linewidth]{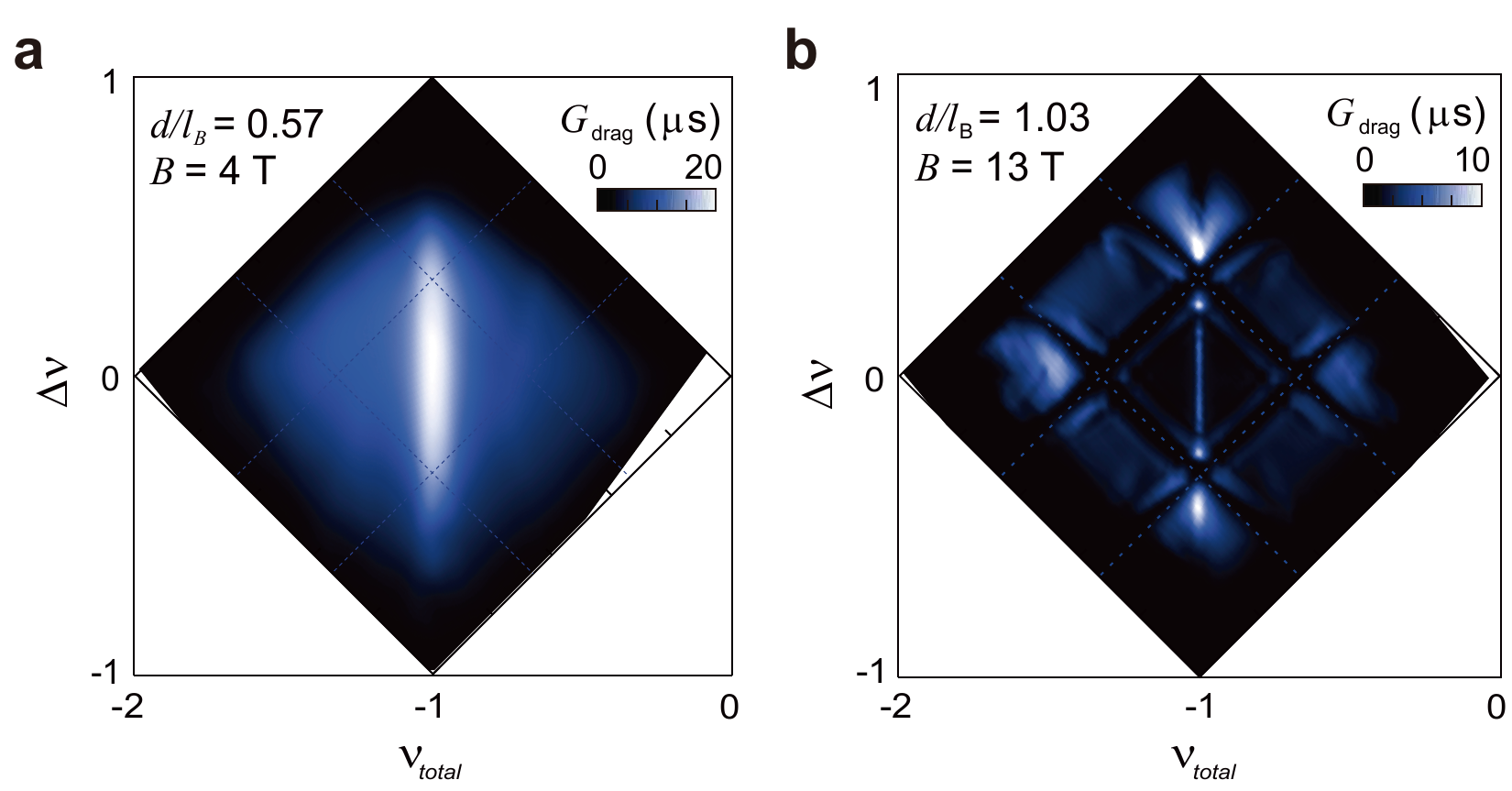}
\caption{\label{fig:gategate} {\bf{\nutotal - \dnu map of counterflow conductance at \dlb =0.57, 1.03}} The conductance measured in the drag counterflow configuration (analogous to the counterflow conductance) at \dlb = 0.57 (a) and 1.03 (b) at T = 0.3 K. In (a) the white stripe along \nutotal=-1 correspond to the exciton condensate phase where the conductance is maximum and bound by the contact resistance. (b) The dark checker board feature is the pattern of decoupled singke layer fractional quantum Hall states, which is discussed in Ref\cite{Li2019pairing,Liu2019interlayer,Zhang2025exciton, Nguyen2024fractionalexciton}. The ground state at \dnu = $\pm 1/3$, \nutotal=1 has zero drag conductance ,suggesting the ground state is decoupled FQHS. The dim color between \dnu=-1/3 to 1/3 along \nutotal=1 suggests the absence of exciton condensate which agree with the drag ratio results shown in figure 1 in the main text.}
\end{figure*}

\begin{figure*}
\includegraphics[width=0.6\linewidth]{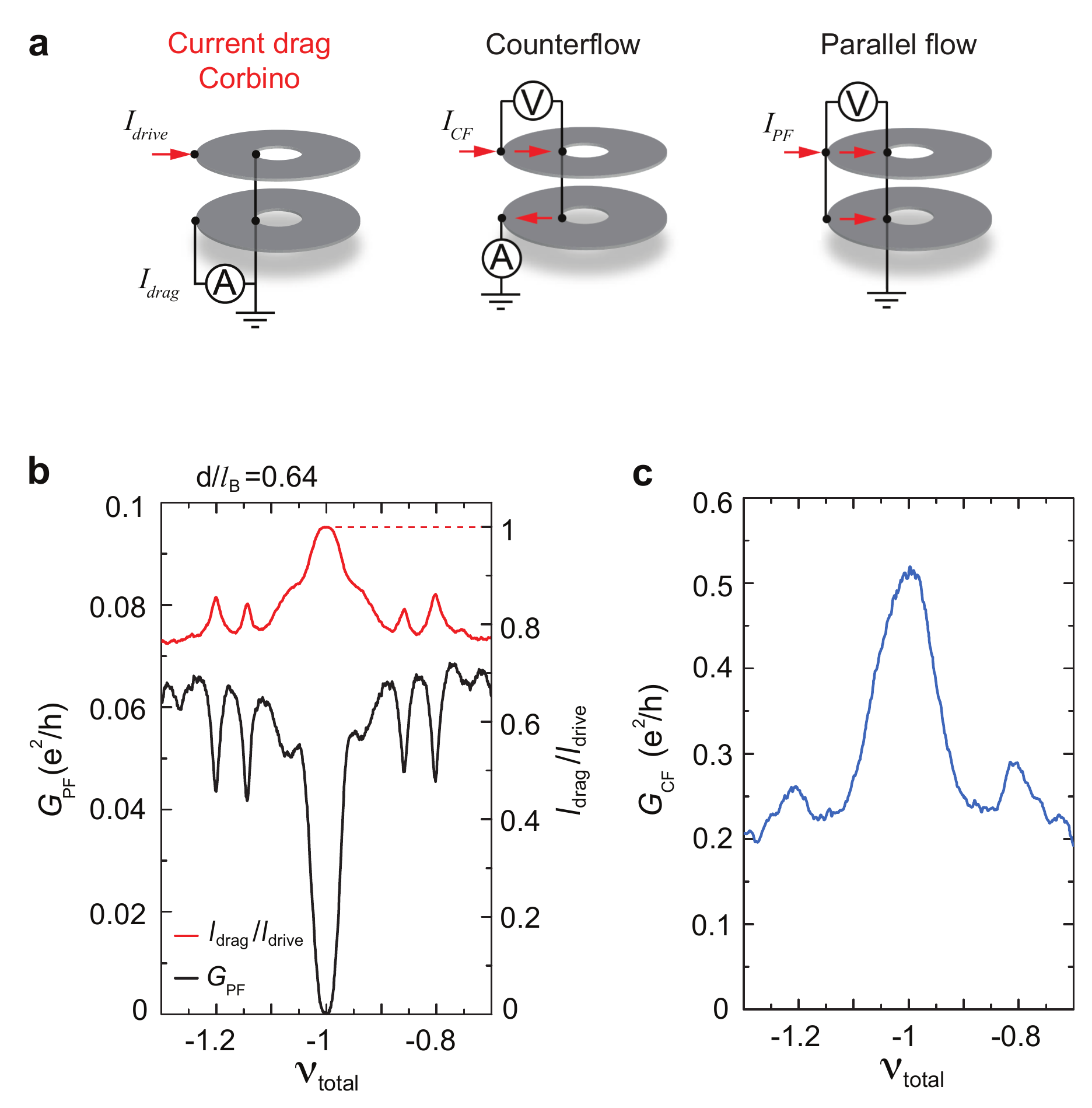}
\caption{\label{SI1} {\bf{Transport measurement in the Corbino-shaped sample.}}  (a) Schematic diagram for different  measurement configurations in Corbino-shaped samples.   (b) Parallel flow conductance $G_{PF}$ (black line) and the current drag ratio $I_{drag}/I_{drive}$ (red line) as a function of total Landau level filling factor $\nu_{total}$. The sample is under the layer-balanced condition throughout the measurement with \dnu=0.  (c) Counterflow conductance as a function of $\nu_{total}$. The existence of exciton condensate phase at $\nu_{total}=-1$ is evidenced by a combination of transport response:  vanishing parallel flow conductance, perfect current drag with $I_{drag}/I_{drive} = 1$ and an enhanced counterflow conductance. 
}
\end{figure*}

\end{widetext}

\end{document}